\documentclass[aps,prd,twocolumn,amsmath,amssymb,superscriptaddress]{revtex4-1}

\usepackage{graphicx}
\usepackage{dcolumn}
\usepackage{bm}
\usepackage{color}
\usepackage{multirow}
\usepackage[modulo]{lineno,xcolor}
\usepackage{hyperref}
\usepackage{amsmath}
\usepackage{braket}
\usepackage[caption=false]{subfig}
\usepackage{xspace}
\usepackage{microtype}
\usepackage{rotating}

\graphicspath{{./}{./figures/}}

\begin{document}
%
%
%
%

\newcommand{\vlvnt}{VLV\ensuremath{\nu}T}
\newcommand{\nue}{\ensuremath{\nu_e}}
\newcommand{\numu}{\ensuremath{\nu_\mu}}
\newcommand{\nutau}{\ensuremath{\nu_\tau}}
\newcommand{\nuebar}{\ensuremath{\bar{\nu}_e}}
\newcommand{\numubar}{\ensuremath{\bar{\nu}_\mu}}
\newcommand{\nutaubar}{\ensuremath{\bar{\nu}_\tau}}
\newcommand{\nuecc}{\ensuremath{\nue\rm{\ CC}}}
\newcommand{\numucc}{\ensuremath{\numu\rm{\ CC}}}
\newcommand{\nutaucc}{\ensuremath{\nutau\rm{\ CC}}}
\newcommand{\nuebarcc}{\ensuremath{\nuebar\rm{\ CC}}}
\newcommand{\numubarcc}{\ensuremath{\numubar\rm{\ CC}}}
\newcommand{\nutaubarcc}{\ensuremath{\nutaubar\rm{\ CC}}}
\newcommand{\nunc}{\ensuremath{\nu\rm{\ NC}}}
\newcommand{\nubarnc}{\ensuremath{\bar{\nu}\rm{\ NC}}}
\newcommand{\Enu}{\ensuremath{E_{\nu}}}
\newcommand{\Ereco}{\ensuremath{E_{\rm reco}}}
\newcommand{\CZtrue}{\ensuremath{\cos{\vartheta}_{\rm true}}}
\newcommand{\CZreco}{\ensuremath{\cos{\vartheta}_{\rm reco}}}
\newcommand{\PID}{\ensuremath{\rm PID}}

\providecommand{\dmOT}{\Delta m^2_{\rm 12}}
\providecommand{\dmTO}{\Delta m^2_{\rm 31}}
\providecommand{\dmsolar}{\Delta m^2_{\rm solar}}
\providecommand{\dmatm}{\Delta m^2_{\rm atm}}
\providecommand{\dmTT}{\Delta m^2_{\rm 32}}
\providecommand{\thatm}{\Delta m^2_{\rm atm}}
\providecommand{\thOT}{\theta_{\rm 13}}
\providecommand{\thTT}{\theta_{\rm 23}}
\providecommand{\sinsqOT}{\sin^{2}(\theta_{\rm 12})}
\providecommand{\sinsqsolar}{\sin^{2}(\theta_{\rm solar})}
\providecommand{\sinsqTT}{\sin^{2}(\theta_{\rm 23})}
\providecommand{\sinsqTTtrue}{\sin^{2}(\theta_{\rm 23}^{\rm true})}
\providecommand{\sinsqTwoTT}{\sin^{2}(2\theta_{\rm 23})}
\providecommand{\dcp}{\delta_{\rm CP}}

\providecommand{\gsim}{\gtrsim}
\providecommand{\lsim}{\lesssim}
\providecommand{\Enutau}{\rm{E}_{\nu_\tau}}
\providecommand{\Emu}{\rm{E}_\mu}
\providecommand{\Ecasc}{\rm{E}_{\rm casc}}
\providecommand{\Lmu}{\rm{L}_\mu}
\providecommand{\Lnu}{\rm{L}_\nu}
\providecommand{\Thetamu}{\theta_\mu}
\providecommand{\cosTheta}{\cos{\theta}}
\providecommand{\cosThetamu}{\cos{\theta_\mu}}
\providecommand{\cosThetanu}{\cos{\theta_\nu}}
\providecommand{\Thetanu}{\theta_\nu}
\providecommand{\nux}{\nu_{\rm x}}

\providecommand{\mares}{{M_A^{RES}}}
\providecommand{\maccqe}{M_A^{CCQE}}
\providecommand{\ahtby}{A^{BY}_{HT}}
\providecommand{\bhtby}{B^{BY}_{HT}}
\providecommand{\cvouby}{C^{BY}_{V1u}}
\providecommand{\cvtuby}{C^{BY}_{V2u}}

\providecommand{\ket}[1]{|#1\rangle}

\providecommand{\ue}[1]{|U_{e #1}|}
\providecommand{\uesq}[1]{|U_{e #1}|^{2}}

\providecommand{\Nch}{${\rm N}_{\rm ch}\,$}
\providecommand{\Ndir}{${\rm N}_{\rm dir}\,$}
\providecommand{\Nstr}{${\rm N}_{\rm str}\,$}
\providecommand{\Aeff}{${\rm A}_{\rm eff}\,$}
\providecommand{\Veff}{${\rm V}_{\rm eff}\,$}
\providecommand{\VeffNS}{${\rm V}_{\rm eff}$}
\providecommand{\Meff}{${\rm M}_{\rm eff}\,$}

\providecommand{\Cerenkov}{Cherenkov }
\providecommand{\deltaChiSq}{\Delta \chi^{2}}
\providecommand{\deltaChiSqBar}{\overline{\Delta \chi^{2}}}

\providecommand{\nubar}{\overline{\nu}\xspace}
\providecommand{\nue}{\nu_{\rm e}\xspace}
\providecommand{\numu}{\nu_\mu\xspace}
\providecommand{\nutau}{\nu_\tau\xspace}
\providecommand{\nuebar}{\overline{\nu}_{\rm e}\xspace}
\providecommand{\numubar}{\overline{\nu}_\mu\xspace}
\providecommand{\nutaubar}{\overline{\nu}_\tau\xspace}

\providecommand{\ket}[1]{|#1\rangle}

\providecommand{\ue}[1]{\ensuremath{|U_{e #1}|}}
\providecommand{\umu}[1]{\ensuremath{|U_{\mu #1}|}}
\providecommand{\utau}[1]{\ensuremath{|U_{\tau #1}|}}
\providecommand{\uesq}[1]{\ensuremath{|U_{e #1}|^{2}}}
\providecommand{\umusq}[1]{\ensuremath{|U_{\mu #1}|^{2}}}
\providecommand{\utausq}[1]{\ensuremath{|U_{\tau #1}|^{2}}}

\newcommand{\dragon}{\ensuremath{\mathcal{B}}\xspace}
\newcommand{\DRAGON}{\dragon}

\newcommand{\greco}{\ensuremath{\mathcal{A}}\xspace}
\newcommand{\GRECO}{\greco}

\newcommand{\nanos}{$\rm{ns}\ $}
\newcommand{\mus}{$\rm{\mu s}\ $}

\newcommand{\hyperplane}{Hyperplane\xspace}

\newcommand{\GRECOnorm}{$0.73^{+0.30}_{-0.24}$}
\newcommand{\GRECOCCnorm}{$0.57^{+0.36}_{-0.30}$}
\newcommand{\GRECOsignifCCNC}{$3.2\sigma$}
\newcommand{\GRECOsignifCC}{$2.0\sigma$}
\newcommand{\DRAGONnorm}{$0.59^{+0.31}_{-0.25}$}
\newcommand{\DRAGONCCnorm}{$0.43^{+0.36}_{-0.31}$}
\newcommand{\DRAGONsignifCCNC}{$2.5\sigma$}
\newcommand{\DRAGONsignifCC}{$1.4\sigma$}

\newcommand{\missing}{\textbf{\textcolor{red}{XXX}}\xspace}

\setlength\linenumbersep{5pt}
\renewcommand\linenumberfont{\normalfont\tiny\sffamily\color{gray}}



\title{Measurement of Atmospheric Tau Neutrino Appearance with IceCube DeepCore}

\affiliation{III. Physikalisches Institut, RWTH Aachen University, D-52056 Aachen, Germany}
\affiliation{Department of Physics, University of Adelaide, Adelaide, 5005, Australia}
\affiliation{Dept. of Physics and Astronomy, University of Alaska Anchorage, 3211 Providence Dr., Anchorage, AK 99508, USA}
\affiliation{Dept. of Physics, University of Texas at Arlington, 502 Yates St., Science Hall Rm 108, Box 19059, Arlington, TX 76019, USA}
\affiliation{CTSPS, Clark-Atlanta University, Atlanta, GA 30314, USA}
\affiliation{School of Physics and Center for Relativistic Astrophysics, Georgia Institute of Technology, Atlanta, GA 30332, USA}
\affiliation{Dept. of Physics, Southern University, Baton Rouge, LA 70813, USA}
\affiliation{Dept. of Physics, University of California, Berkeley, CA 94720, USA}
\affiliation{Lawrence Berkeley National Laboratory, Berkeley, CA 94720, USA}
\affiliation{Institut f\"ur Physik, Humboldt-Universit\"at zu Berlin, D-12489 Berlin, Germany}
\affiliation{Fakult\"at f\"ur Physik \& Astronomie, Ruhr-Universit\"at Bochum, D-44780 Bochum, Germany}
\affiliation{Universit\'e Libre de Bruxelles, Science Faculty CP230, B-1050 Brussels, Belgium}
\affiliation{Vrije Universiteit Brussel (VUB), Dienst ELEM, B-1050 Brussels, Belgium}
\affiliation{Dept. of Physics, Massachusetts Institute of Technology, Cambridge, MA 02139, USA}
\affiliation{Dept. of Physics and Institute for Global Prominent Research, Chiba University, Chiba 263-8522, Japan}
\affiliation{Dept. of Physics and Astronomy, University of Canterbury, Private Bag 4800, Christchurch, New Zealand}
\affiliation{Dept. of Physics, University of Maryland, College Park, MD 20742, USA}
\affiliation{Dept. of Astronomy, Ohio State University, Columbus, OH 43210, USA}
\affiliation{Dept. of Physics and Center for Cosmology and Astro-Particle Physics, Ohio State University, Columbus, OH 43210, USA}
\affiliation{Niels Bohr Institute, University of Copenhagen, DK-2100 Copenhagen, Denmark}
\affiliation{Dept. of Physics, TU Dortmund University, D-44221 Dortmund, Germany}
\affiliation{Dept. of Physics and Astronomy, Michigan State University, East Lansing, MI 48824, USA}
\affiliation{Dept. of Physics, University of Alberta, Edmonton, Alberta, Canada T6G 2E1}
\affiliation{Erlangen Centre for Astroparticle Physics, Friedrich-Alexander-Universit\"at Erlangen-N\"urnberg, D-91058 Erlangen, Germany}
\affiliation{Physik-department, Technische Universit\"at M\"unchen, D-85748 Garching, Germany}
\affiliation{D\'epartement de physique nucl\'eaire et corpusculaire, Universit\'e de Gen\`eve, CH-1211 Gen\`eve, Switzerland}
\affiliation{Dept. of Physics and Astronomy, University of Gent, B-9000 Gent, Belgium}
\affiliation{Dept. of Physics and Astronomy, University of California, Irvine, CA 92697, USA}
\affiliation{Dept. of Physics and Astronomy, University of Kansas, Lawrence, KS 66045, USA}
\affiliation{SNOLAB, 1039 Regional Road 24, Creighton Mine 9, Lively, ON, Canada P3Y 1N2}
\affiliation{Department of Physics and Astronomy, UCLA, Los Angeles, CA 90095, USA}
\affiliation{Dept. of Astronomy, University of Wisconsin, Madison, WI 53706, USA}
\affiliation{Dept. of Physics and Wisconsin IceCube Particle Astrophysics Center, University of Wisconsin, Madison, WI 53706, USA}
\affiliation{Institute of Physics, University of Mainz, Staudinger Weg 7, D-55099 Mainz, Germany}
\affiliation{Department of Physics, Marquette University, Milwaukee, WI, 53201, USA}
\affiliation{Institut f\"ur Kernphysik, Westf\"alische Wilhelms-Universit\"at M\"unster, D-48149 M\"unster, Germany}
\affiliation{Bartol Research Institute and Dept. of Physics and Astronomy, University of Delaware, Newark, DE 19716, USA}
\affiliation{Dept. of Physics, Yale University, New Haven, CT 06520, USA}
\affiliation{Dept. of Physics, University of Oxford, 1 Keble Road, Oxford OX1 3NP, UK}
\affiliation{Dept. of Physics, Drexel University, 3141 Chestnut Street, Philadelphia, PA 19104, USA}
\affiliation{Physics Department, South Dakota School of Mines and Technology, Rapid City, SD 57701, USA}
\affiliation{Dept. of Physics, University of Wisconsin, River Falls, WI 54022, USA}
\affiliation{Dept. of Physics and Astronomy, University of Rochester, Rochester, NY 14627, USA}
\affiliation{Oskar Klein Centre and Dept. of Physics, Stockholm University, SE-10691 Stockholm, Sweden}
\affiliation{Dept. of Physics and Astronomy, Stony Brook University, Stony Brook, NY 11794-3800, USA}
\affiliation{Dept. of Physics, Sungkyunkwan University, Suwon 440-746, Korea}
\affiliation{Dept. of Physics and Astronomy, University of Alabama, Tuscaloosa, AL 35487, USA}
\affiliation{Dept. of Astronomy and Astrophysics, Pennsylvania State University, University Park, PA 16802, USA}
\affiliation{Dept. of Physics, Pennsylvania State University, University Park, PA 16802, USA}
\affiliation{Dept. of Physics and Astronomy, Uppsala University, Box 516, S-75120 Uppsala, Sweden}
\affiliation{Dept. of Physics, University of Wuppertal, D-42119 Wuppertal, Germany}
\affiliation{DESY, D-15738 Zeuthen, Germany}

\author{M. G. Aartsen}
\affiliation{Dept. of Physics and Astronomy, University of Canterbury, Private Bag 4800, Christchurch, New Zealand}
\author{M. Ackermann}
\affiliation{DESY, D-15738 Zeuthen, Germany}
\author{J. Adams}
\affiliation{Dept. of Physics and Astronomy, University of Canterbury, Private Bag 4800, Christchurch, New Zealand}
\author{J. A. Aguilar}
\affiliation{Universit\'e Libre de Bruxelles, Science Faculty CP230, B-1050 Brussels, Belgium}
\author{M. Ahlers}
\affiliation{Niels Bohr Institute, University of Copenhagen, DK-2100 Copenhagen, Denmark}
\author{M. Ahrens}
\affiliation{Oskar Klein Centre and Dept. of Physics, Stockholm University, SE-10691 Stockholm, Sweden}
\author{D. Altmann}
\affiliation{Erlangen Centre for Astroparticle Physics, Friedrich-Alexander-Universit\"at Erlangen-N\"urnberg, D-91058 Erlangen, Germany}
\author{K. Andeen}
\affiliation{Department of Physics, Marquette University, Milwaukee, WI, 53201, USA}
\author{T. Anderson}
\affiliation{Dept. of Physics, Pennsylvania State University, University Park, PA 16802, USA}
\author{I. Ansseau}
\affiliation{Universit\'e Libre de Bruxelles, Science Faculty CP230, B-1050 Brussels, Belgium}
% ...
\date{\today}

\collaboration{IceCube Collaboration}
\email{analysis@icecube.wisc.edu}
\noaffiliation

\date{\today -- revision 5.1}

\begin{abstract}
We present a measurement of atmospheric tau neutrino appearance from oscillations with three years of data from the DeepCore sub-array of the IceCube Neutrino Observatory. This analysis uses atmospheric neutrinos from the full sky with reconstructed energies between $5.6$\,GeV and $56$\,GeV to search for a statistical excess of cascade-like neutrino events which are the signature of $\nu_\tau$ interactions. For CC+NC (CC-only) interactions, we measure the tau neutrino normalization to be \GRECOnorm\ (\GRECOCCnorm) and exclude the absence of tau neutrino oscillations at a significance of \GRECOsignifCCNC\ (\GRECOsignifCC)
These results are consistent with, and of similar precision to, a confirmatory IceCube analysis also presented, as well as measurements performed by other experiments.

\end{abstract}

\maketitle

\section{Introduction}
\label{sec:introduction}
The phenomenon of neutrino flavor oscillations is now well-established experimentally, building on the discoveries of atmospheric neutrino oscillations by the Super-Kamiokande (SK) experiment~\cite{Fukuda:1998mi} and solar neutrino oscillations by the Sudbury Neutrino Observatory (SNO) experiment~\cite{Ahmad:2001an,Ahmad:2002jz}.

These and most other neutrino oscillation experiments~\cite{Patrignani:2016xqp} are based on measuring the appearance or disappearance of electron neutrinos or muon neutrinos.
In contrast, there are only two experiments with measurements of the appearance of tau neutrinos through neutrino oscillations, leaving the {\nutau} sector relatively underexplored.
With the \nutau\ appearance measurement of the OPERA~\cite{Agafonova:2018auq} experiment, using an accelerator-based beam of \numu, the null hypothesis of no-\nutau\ appearance has been effectively ruled out. 
Additionally, a small excess of \nutau\ events has been measured by both OPERA (0.25$\sigma$) and SK~\cite{Li:2017dbe} (1.47$\sigma$) relative to what is expected under the standard three-flavor oscillation paradigm.

The measured excess may be interpreted in a number of ways. 
The tau neutrino charged current cross section directly contributes to the total number of detected $\nutau$, with theoretical uncertainties~\cite{Jeong:2010nt} of $\mathcal{O}(10)\%$ and much larger experimental uncertainties~\cite{Kodama:2007aa}.
These uncertainties can lead to an overall scaling of the number of observed $\nutau$ interactions which can be measured by atmospheric oscillation experiments sensitive to tau neutrinos.
This interpretation has been adopted in recent results from the SK collaboration, which recasts the excess in terms of a modification of the averaged tau neutrino charged current cross section.

Another potential interpretation for the observed excess in OPERA and SK would be the observation of non-unitarity in the neutrino sector.
In the standard oscillation picture, the dominant appearance mode of $\numu \rightarrow \nutau$ is given by 
\begin{align}
    P_{\nu_\mu \rightarrow \nu_\tau} &= \sum_{j,k} U_{\mu j}U^\ast_{\tau j}U^\ast_{\mu k}U_{\tau k} \exp{\left(i\frac{\Delta m^2_{jk}L}{2E_\nu}\right)} \label{eq:nutau_appearance} \\
    &\approx \cos^4{\theta_{13}} \sin^2{2\theta_{23}} \sin^2{\left(\frac{\Delta m^2_{31}L}{4E_\nu}\right)} \label{eq:nutau_appearance_approx}
\end{align}
with $U$ denoting the elements of the Pontecorvo-Maki-Nakagawa-Sakata (PMNS)~\cite{Pontecorvo:1957cp,Maki:1962mu} mixing matrix (see also Eq.~\ref{eq:pmns}), $\Delta m^2_{jk} = m_j^2-m_k^2$ the mass-squared splittings, $L$ the oscillation baseline, and $E_\nu$ the neutrino energy.
The angles $\theta_{13}$ and $\theta_{23}$ govern the amplitude of the mixing, while $\Delta m^2_{31}$ drives the oscillations on the length and energy scales. The benefit of using trigonometric angles is that they conveniently preserve oscillation probabilities to be within 0 and 1 while also reducing the number of physics parameters to fit. But not all measurements of the same angle probe the same individual elements of the underlying PMNS mixing matrix.

Measurements of $\theta_{23}$ from long baseline $\numu\rightarrow\numu$ disappearance probe $\umu{3}^2$, whereas measurements of $\theta_{23}$ from ${\numu\rightarrow\nutau}$ appearance probe $\umu{3}^2$ and $\utau{3}^2$, for further information see Supplementary Material of Ref.\cite{Parke:2015goa}. Not only do different experimental measurements of the same angle probe different underlying elements, but the relation between the angles and the nine canonical matrix elements is only preserved if the PMNS matrix is $3\times3$ unitary.

A core aspect of any theoretically consistent neutrino mixing matrix is that the individual rows and columns preserve norms and rational probabilities, {\it e.g.\@} ${\ue{3}^2+\umu{3}^2+\utau{3}^2=1}$. While checks of unitarity across the entire matrix are important, the mixing elements for the third mass eigenstate are particularly interesting because it has been experimentally established that ${\ue{3}^2+\umu{3}^2\simeq0.5}$, but it has only recently been confirmed by OPERA that ${\utau{3}^2>0}$ at $6.1\sigma$ significance~\cite{Agafonova:2018auq}. The only other evidence of $\utau{3}^2>0$ is from SK and reaches $4.6\sigma$ significance~\cite{Li:2017dbe}. Even with these two measurements, a global fit of leading oscillations results~\cite{Parke:2015goa} illustrates that the current constraint of ${0.2<\utau{3}^2<0.61}$ at $3\sigma$ lacks the precision necessary to probe unitarity of the third mass eigenstate at even $\mathcal{O}(10)\%$ precision. Unsurprisingly, the range of $\utau{3}^2$ from a global fit is not driven by the direct measurements of $\utau{3}^2$, but rather that values outside that range would induce small deviations in the $\nue$ and $\numu$ sectors that would exceed their ${3\times3}$ unitarity constraints. Using only current direct measurements, the allowed region is $\utau{3}^2 >0$ but is otherwise weakly constrained. 

\begin{widetext}
\begin{equation}
\left(
\begin{array}{c c c}
U_{e1} & U_{e2} & U_{e3}\\
U_{\mu1} & U_{\mu2} & U_{\mu3}\\
U_{\tau1} & U_{\tau2} & U_{\tau3}\\
\end{array}
\right)
=
\left(
\begin{array}{c c c}
1 & 0 & 0 \\
0 & c_{23} & s_{23} \\
0 & -s_{23} & c_{23} \\
\end{array}
\right)
\cdot
\left(
\begin{array}{c c c}
c_{13} & 0 & s_{13}e^{-i\delta_{CP}}\\
0 & 1 & 0 \\
-s_{13}e^{i\delta_{CP}} & 0 & c_{13}\\
\end{array}
\right)
\cdot
\left(
\begin{array}{c c c}
c_{12} & s_{12} & 0 \\
-s_{12} & c_{12} & 0 \\
0 & 0 & 1 \\
\end{array} \\
\right)
\textrm{, where}
\begin{cases}
s_{ij} = \sin{\theta_{ij}}\\
c_{ij} = \cos{\theta_{ij}}\\
\end{cases}
\label{eq:pmns}
\end{equation}
\end{widetext}

A measurement of $\utau{3}^2$ differing from $\simeq0.5$ would be further evidence for new physics beyond the Standard Model (SM), and would imply non-unitarity in the $\nu_3$ mass eigenstate, {\it i.e.,} $\ue{3}^2 + \umu{3}^2 + \utau{3}^2 \neq 1$. The impact of such a deviation could indicate the existence of:
\begin{itemize}
\itemsep-5pt
\item{ Non-standard interactions with the three active neutrinos in the SM.}
\item{ At least one new neutrino (sterile) which has no SM gauge interaction with normal matter.}
\end{itemize}
Conversely, a measurement of $\utau{3}^2\simeq0.5$ would demonstrate that the mixing matrix is (close to) unitary and further constrain interpretations of experimental neutrino oscillation anomalies in terms of $N$ admixed sterile neutrinos~\cite{Athanassopoulos:1996jb,AguilarArevalo:2009yj,Hampel:1998xg, Abdurashitov:2009tn,Abazajian:2012ys}.

In principle, the three channels to measure $\utau{3}^2$ are $\nue\rightarrow\nutau$, $\nutau\rightarrow\nutau$, and $\numu\rightarrow\nutau$. But, the ${\nue\rightarrow\nutau}$ channel is unfavorable because 1) experimentally a $\nue$ and $\nutau$ interaction produce a similar signature in most detectors, and 2) the magnitude of the oscillation is low due to the flavor composition of the third mass eigenstate.
The $\nutau\rightarrow\nutau$ channel probes $\utau{3}^2$ directly, but is also unfavorable because it requires a hitherto unrealized and experimentally challenging high-statistics focused $\nutau$ beam.

In practice, only the $\numu\rightarrow\nutau$ channel is feasible. This channel probes a combination of $\umu{3}^2$ and $\utau{3}^2$, where any degeneracy between $\umu{3}^2$ and $\utau{3}^2$ can be broken by either external constraints on $\umu{3}^2$ or by conducting a simultaneous measurement of ${\numu\rightarrow\numu}$ and ${\numu\rightarrow\nutau}$. 


Earlier IceCube neutrino oscillation measurements~\cite{Aartsen:2013jza,Aartsen:2014yll,Aartsen:2017nmd}, and the measurement presented here, use atmospheric neutrinos arising mainly from the decay of pions and kaons produced in cosmic ray air showers in the Earth's atmosphere.
The initial flux is dominated by \nue\ and \numu, and contains negligible numbers of {\nutau}~\cite{Bulmahn:2010pg}.
The atmospheric neutrinos interacting in the DeepCore subarray of IceCube travel distances ranging from $L\approx20$\,km (vertically downward-going) to $L\sim 1.3\times10^4$\,km (vertically upward-going; the full diameter of the Earth).
For vertically upward-going neutrinos, the first peak of maximal $\numu \rightarrow \nutau$ oscillation probability occurs at roughly 25\,GeV. This is comfortably above the $E_\nu = 5$\,GeV threshold for the DeepCore neutrino reconstruction used in this analysis~\cite{Leuermann:2018}.
The energy corresponding to the oscillation maximum is also above the kinematic energy threshold for charged current \nutau-nucleon interactions $E_{\nutau} = 3.5$\,GeV, where for lower energies there is a complete suppression of the cross section due to the relatively high $\tau$ lepton mass as compared to the other charged leptons. Even so, there is still a suppression to the CC-\nutau\ cross section compared to CC-$\nu_{e,\mu}$ up until $E_{\nutau}\approx10$\,TeV~\cite{Jeong:2010nt}.

The identification of individual \nutau\@ events at energies relevant for measuring atmospheric neutrino oscillation is precluded in DeepCore, as the outgoing tau lepton in CC interactions decays after $\approx 1$\,mm, far smaller than the meter-scale position resolution of DeepCore.  The \nutau\ CC interactions mainly manifest as ``cascades,'' similar to those from \nue\@ CC and neutral current (NC) interactions of all neutrino flavors.  Relative to the no-oscillation case, these $\nutau$-induced cascade events produce a distortion in the 2-D distribution of neutrino energy and direction (the zenith angle is directly related to the pathlength $L$ in eqs.~\ref{eq:nutau_appearance}-\ref{eq:nutau_appearance_approx}). This measurement is based on observing such oscillation-induced patterns between $5.6$\,GeV and $56$\,GeV in the atmospheric neutrino flux coming from all directions.

We present results based on two separate analyses that have different strategies for event selection and background estimation, but considerable overlap in their event selection variables and treatments of systematic uncertainties. The sample for our main analysis ``\GRECO'' targets a high acceptance of all-flavor neutrino events and its background estimation is simulation-driven. The sample for our confirmatory analysis ``\DRAGON'' is optimized for higher rejection of non-neutrino events and its atmospheric muon background estimation is data-driven.

\section{The IceCube DeepCore Detector}
\label{sec:detector}

\begin{figure}[htbp]
    \includegraphics[width=\linewidth]{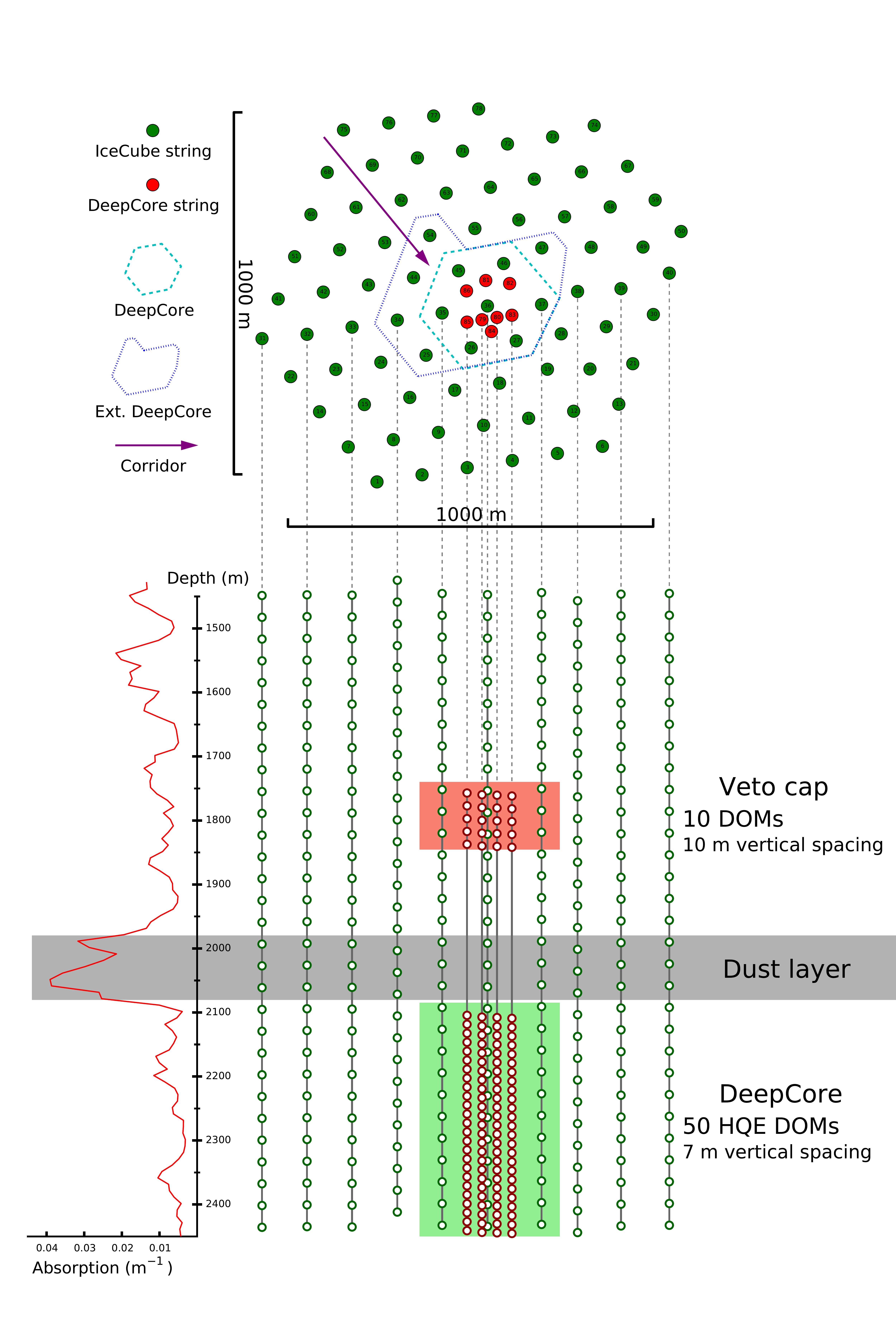}
    \caption{Top and side views of IceCube indicating the positions of DeepCore DOMs with red circles and surrounding IceCube DOMs with green circles. The DeepCore fiducial region is shown as a green box at the bottom center. The DeepCore DOMs were deployed mostly $>\!2100$\,m below the surface (shown highlighted in green) with some DeepCore DOMs also deployed around 1800\,m below the surface (shown highlighted in red) to aid in rejection of atmospheric muons.  The bottom left of the plot shows the absorption length for Cherenkov light vs.\@ depth. The purple arrow in the top view shows one example of a ``corridor'' path along which atmospheric muons can circumvent the simple veto cuts, as they may not leave a clearly detectable track signature (see Sec.~\ref{sec:selection} for details). The gray band indicates the {\it dust layer}, a region of higher scattering and absorption.
    }
    \label{fig:detector}
\end{figure}

The in-ice array of the IceCube detector~\cite{Aartsen:2016nxy}, buried in the South Pole glacial ice, comprises 5160 digital optical modules (DOMs).
Each DOM houses a downward-facing 10" photomultiplier tube (PMT) \cite{Abbasi:2010vc} and its associated electronics \cite{Abbasi:2008aa} in a glass pressure sphere~\cite{Abbasi:2008aa,Abbasi:2010vc}.
The modules are arranged along 86 vertical strings with 60 DOMs per string (see Fig.~\ref{fig:detector}).
Of these strings, 78 are deployed in a nearly regular grid, with an inter-string distance of about 125\,m and modules deployed between depths of 1.45\,km and 2.45\,km, instrumenting a total volume of roughly 1\,km$^3$.
This part of the detector is optimized for neutrinos from $0.1-10^5$\,TeV, and for the analysis presented here primarily serves as an active veto against the downward-going atmospheric muon background.
The remaining eight strings, situated at the bottom center of IceCube, form DeepCore~\cite{Collaboration:2011ym}. The PMTs on these strings have higher quantum efficiency and are primarily located below 2.1\,km in the clearest instrumented ice.
The DeepCore instrumented volume is roughly $10^7$\,m$^3$ with a module density about five times that of the surrounding IceCube array.

While the IceCube detector was fully commissioned in 2011, its noise rates were still stabilizing during the first year of operation. Therefore the data used here are limited to the period from April 2012 through May 2015.

\section{Event Sample}
\label{sec:sample}

\subsection{Simulation}
The simulation chain in IceCube involves three stages: generation, propagation, and detection in ice. Different software is involved at each stage depending on the particle type.

\subsubsection{Neutrinos}
\label{sec:nu_sim}
Neutrino interactions in IceCube are generated following the flux calculation of Honda {\it et al.}~\cite{Honda:2015fha} and using the interaction physics in GENIE~2.8.6~\cite{Andreopoulos:2009rq,Hoshina:GENIE_HE_patch}, which includes the nuclear model, cross sections, and hadronization process~\cite{Yang:2007zzt} based on KNO~\cite{Koba:1972ng} and PYTHIA~\cite{Sjostrand:2006za}.
The GRV98 \cite{Gluck:1998xa} parton distribution functions are used in the DIS cross sections calculations.
Muons created in \numu\ CC interactions are propagated through the ice using PROPOSAL~\cite{Koehne:2013gpa} for fast and precise modelling of the energy losses, while GEANT4~\cite{Agostinelli:2002hh} is used to handle the direct propagation of tau leptons and their decay products, including muons, hadrons, and electromagnetic (EM) showers below 100\,MeV. For events with EM showers above 100\,MeV, shower-to-shower variations are small enough to use parametrizations~\cite{Radel:2012kw} based on GEANT4 simulations.

The Cherenkov photons produced by the final state particles are then propagated through the ice using GPU-based software~\cite{CLSIM}. This simulation takes into account the optical properties (scattering and absorption) of the ice.
For the photons intersecting with a sensor module, the acceptance in terms of arrival angle and wavelength is then taken into account.
For analysis \DRAGON, a measure of the relative variation of optical efficiency among DOMs is included.
Additional hits caused by thermal noise, decaying radioactive isotopes in the PMT and DOM glass, and scintillation are added.
Finally, the PMT response and readout electronics are simulated and trigger algorithms are applied across the full detector in order to produce simulated neutrino events.

\subsubsection{Atmospheric Muon Background}

The generation of atmospheric muons is performed using a full CORSIKA~\cite{Heck:1998vt} air-shower simulation with a hadronic interaction model from~\cite{Ahn:2009wx}.
The propagation of these background muons and the detection of the Cherenkov radiation are the same as those due to a secondary muon in a neutrino interaction.

At the final level of the event selections (see Sec.~\ref{sec:selection}), the atmospheric muon background is reduced by roughly eight orders of magnitude.
The standard simulation tools are too computationally inefficient to produce sufficient amounts of muon background surviving all the selection criteria.
In order to estimate the muon background at final level, the two analyses use two distinct techniques:
\begin{itemize}
    \item Analysis \GRECO uses an atmospheric muon simulation employing a fast parametrized approach based on~\cite{Carminati:2009fj}. 
    This software targets the regions of the weakest background rejection: single low-energy muons aimed at the DeepCore fiducial volume, which make up approximately 75\% of the final simulated muon sample. 
    The simulation is approximately two orders of magnitude faster than one covering the entire IceCube detector. 
    Unsimulated regions in zenith and energy are augmented by simulation produced with the CORSIKA simulation package.
    All simulated atmospheric muons are weighted using the H4a cosmic ray flux model~\cite{Gaisser:2013bla}. 

    \item Analysis \dragon follows an alternative, data-driven approach to estimate the shape of the remaining muon background. 
    The method uses data side-bands consisting of events that would have been accepted in the final sample had they not included hits in DOMs in one of the corridor regions (see Fig.~\ref{fig:detector}). 
\end{itemize}

\subsection{Selection}
\label{sec:selection}

IceCube triggers on $\mathcal{O}(10^{11})$ downward-going atmospheric muons, $\mathcal{O}(10^5)$ atmospheric neutrinos, and $\mathcal{O}(10)$ high-energy astrophysical neutrinos per year, placing stringent demands on background rejection efficiency for IceCube analyses.
At neutrino energies above about 50~GeV, standard techniques to accept neutrinos and reject atmospheric muon background in IceCube include selecting events which reconstruct as upward-going, have a starting vertex deep within the detector fiducial volume, fall within a very narrow temporal or directional window, or have a very high energy.  
For lower-energy neutrinos, however, only DeepCore's higher density of DOMs allows accurate reconstruction of these dimmer events, as described in the next section.  The $\nutau$ appearance analyses therefore focus on events that are contained within the DeepCore fiducial volume.  Located at the bottom center of IceCube, DeepCore benefits from the exceptionally clear ice that has photon attenuation and absorption lengths of roughly 50~m and 150~m, respectively~\cite{Aartsen:2013rt, spice}. An important benefit of DeepCore's location is the use of over 4500 IceCube DOMs as an active ``veto region'' to identify background muons for removal.

The selection of the final event sample is implemented in a series of ``levels,'' the first three of which are very similar in Analyses \greco and \dragon, while the subsequent ones differ. These differences primarily reflect the looser Analysis \greco  selection criteria to prioritize the efficiency of selecting neutrino events, versus the tighter Analysis \dragon  criteria to prioritize the rejection of atmospheric muon background. Note that the Analysis \dragon selection criteria were originally optimized to measure $\numu$ disappearance and follow closely the criteria used for that measurement in~Ref.~\cite{Aartsen:2017nmd}.
Below we give a description of the selection criteria, highlighting important similarities and differences between the two analyses, and show distributions for some of the key variables central to the analyses.  We provide a more detailed description of the selection criteria in Appendices~\ref{app:common}--\ref{app:dragon}.

\subsubsection{Common Selection Criteria}

Analyses \GRECO and \DRAGON share the first three levels of selection criteria, starting with the online triggering at the South Pole. Detected photons or ``hits'' are labeled ``locally coincident'' and included in the trigger if they occur within 1\,$\mu$s of a hit on a nearby DOM on the same string. The trigger requires three or more locally coincident DeepCore DOMs to detect hits in a 2.5\,$\mu$s time window. When this condition is met, the data acquisition system reads out all available data in the full detector, in a time window that extends 6\,$\mu$s before and 6\,$\mu$s after the dynamic trigger window (see Sec.~6.4.2 of \cite{Aartsen:2016nxy} for more details).  In Level~2, a filtering algorithm is used to reject any events consistent with a muon traveling at $v \simeq c$ between the reconstructed interaction vertex within DeepCore and two or more more hit DOMs in the veto region~\cite{Collaboration:2011ym}.

After the application of the trigger and filter algorithms, a large number of background events are still present in the sample. Both analyses therefore perform a fast reconstruction at Level~2 that insures an adequate number of hit DOMs in IceCube consistent with either the track or cascade signature of a neutrino interaction.  Both analyses then define a slightly enlarged fiducial volume, and require $<7$~photoelectrons (p.e.) in the correspondingly smaller surrounding veto region.  A set of criteria is also applied to remove low quality events with too many noise hits, too few DOMs with multiple hits, too much deposited charge, a reconstructed vertex in the upper region of the fiducial volume, too small a fraction of the event's total p.e.\@ deposited at early times, or too large a fraction of DeepCore hits in the outer regions of the DeepCore fiducial volume.  
Detailed descriptions of these criteria, along with subtle differences between the two analyses, are discussed in Appendix~\ref{app:common}. In aggregate, these criteria remove events whose reconstruction is likely to be faulty, and those events that are likely to be downward-going atmospheric muon background.  The event rates after each of these first three levels of the common event selection for Analyses \greco and \dragon are shown in Table~\ref{tab:TableRate} of Appendix~\ref{app:common}.

\subsubsection{Additional Selection Criteria: Analysis \greco}

\begin{figure}[htbp]
    \includegraphics[width=\linewidth]{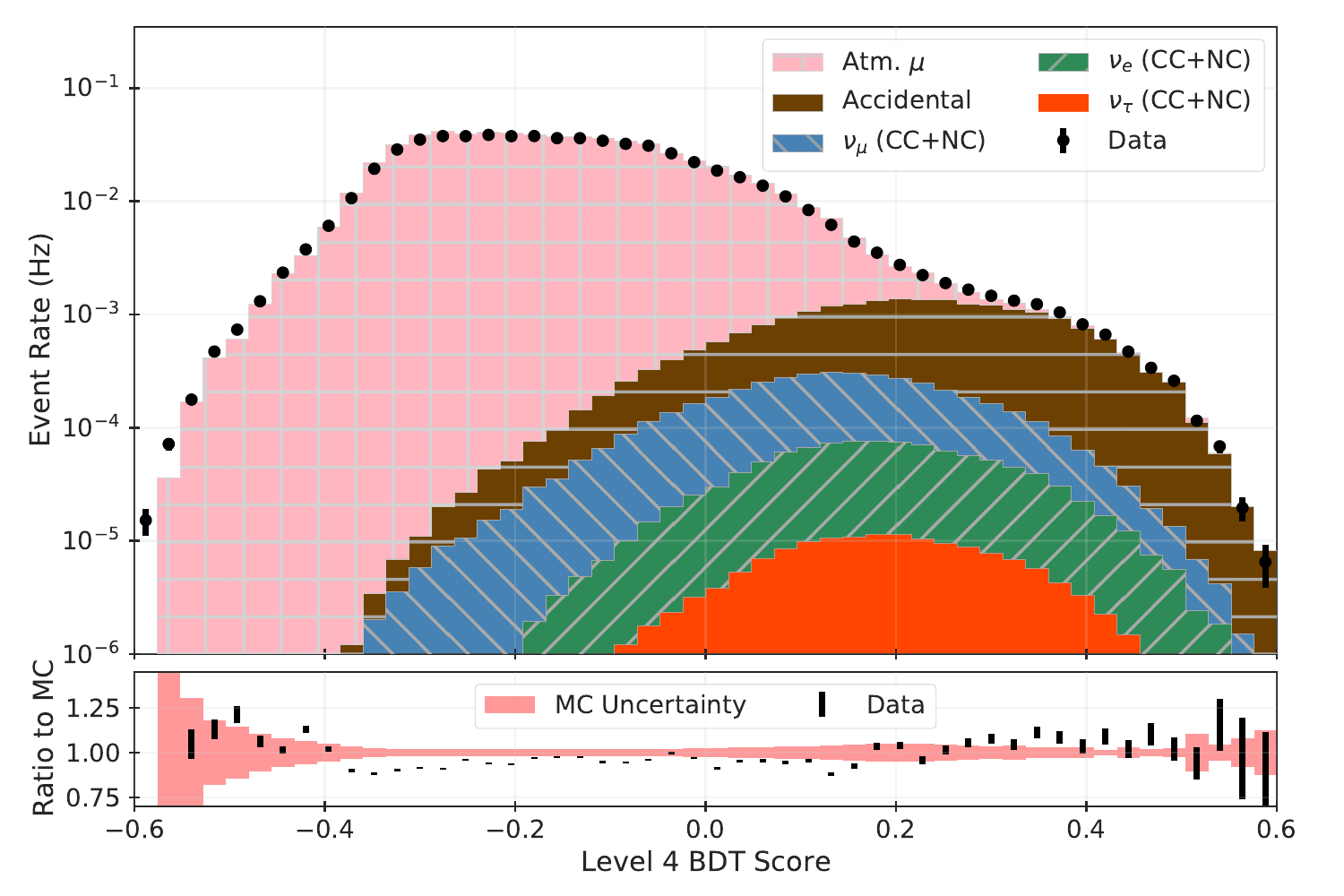}
    \caption{Top: BDT distribution at Level~4. Each shaded color represents the stacked histogram from Monte Carlo simulations for each event type. Black dots represent the data distribution. MC events are weighted by world averaged best fit oscillation parameters. Bottom: Ratio of distribution from data to that from MC. Black error bars are the statistical fluctuation from data, whereas shaded red areas are the uncertainties from limited MC statistics.
    At this stage of Analysis~\greco, atmospheric muons and accidental triggers due to random detector noise dominate both the signal and background regions. 
    Events below 0.04 are removed to reduce the fraction of atmospheric muon background events.
    }
    \label{fig:greco_l4_bdtscore}
\end{figure}

Event selection for Analysis \greco uses two boosted decision trees (BDTs)~\cite{FREUND1997119} to remove atmospheric muon background. The first BDT  (Level~4) uses six different input variables adapted from \cite{Aartsen:2012uu}: three related to the charge measured by the PMTs, a simple vertex estimator, an event speed estimator, and a calculation of event shape. The resulting BDT output is shown in Fig.~\ref{fig:greco_l4_bdtscore}.

Accidental triggers due to random detector noise occur primarily in the DeepCore fiducial volume with few hit DOMs, appearing neutrino-like for this selection level. In order to limit the impact of these events, dedicated selection criteria, detailed in Appendix~\ref{app:greco}, are introduced at later stages of the selection.

The second, subsequent BDT (Level~5) is used to further reduce the muon background based on six input variables: the time to accumulate charge, a vertex estimator, two variables using center-of-gravity calculations, a causal hit identifier, and a zenith angle estimation from a simple reconstruction. As an example, the distribution of this second BDT output for both simulation and data is shown in Fig.~\ref{fig:greco_l5_bdtscore}, and more distributions and information can be found in Appendix~\ref{app:greco}.

\begin{figure}[htbp]
    \includegraphics[width=\linewidth]{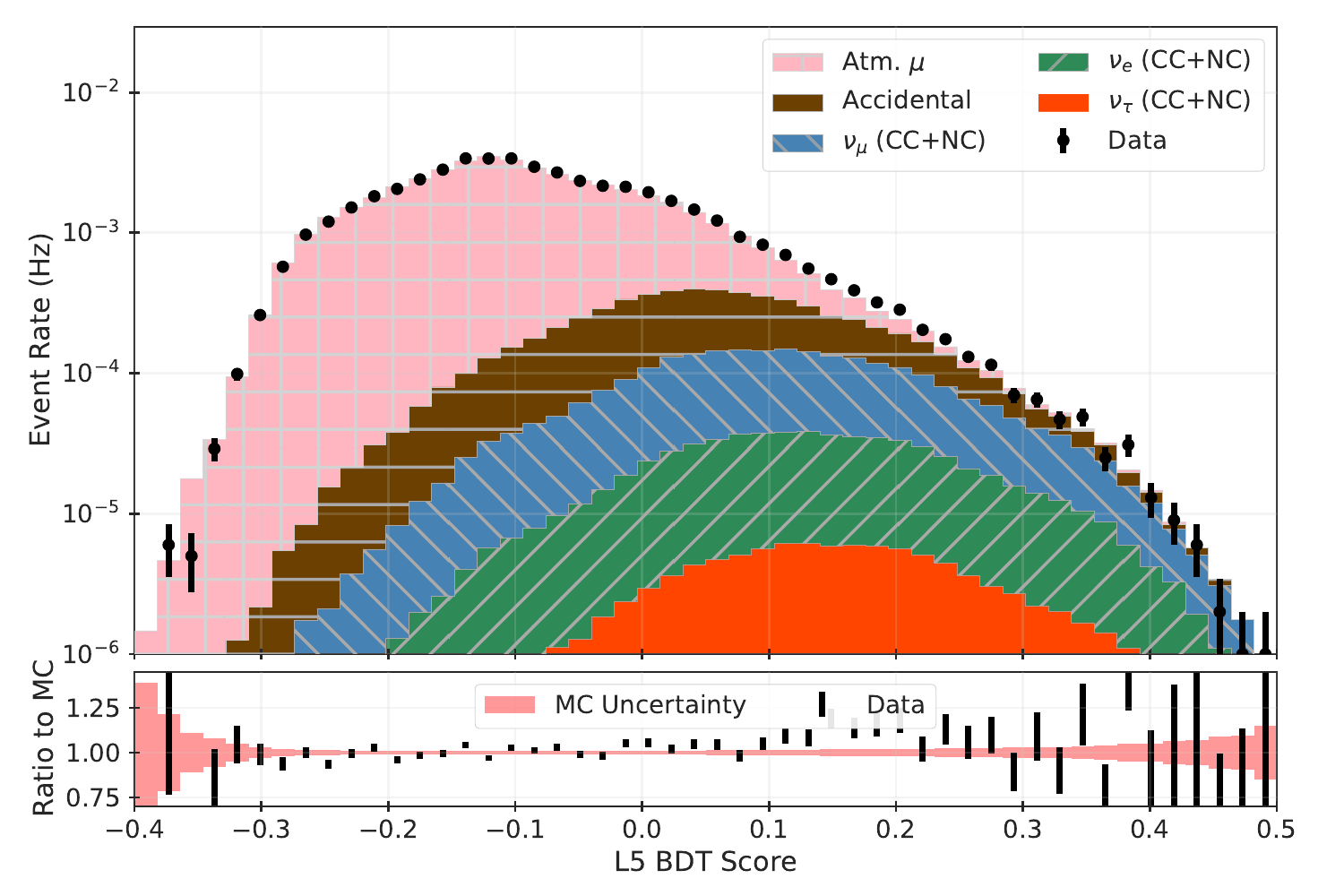}
    \caption{Top: BDT distribution at Level~5. Each shaded color represents the stacked histogram from each event type. Black dots represent the data distribution. MC events are weighted by world averaged best fit oscillation parameters. Bottom: Ratio of distribution from data to that from MC. Black error bars are the statistical fluctuation from data, whereas shaded red areas are the uncertainties from limited MC statistics.}
    \label{fig:greco_l5_bdtscore}
\end{figure}

The event rates after application of the Level~4 and~5 selection criteria are shown numerically in Table~\ref{tab:TableRate} of Appendix~\ref{app:common} and graphically in Fig.~\ref{fig:greco_rates} below.
After Level~5 the signal and background rates are roughly at parity.

\begin{figure}[htbp]
    \centering
    \includegraphics[width=\linewidth]{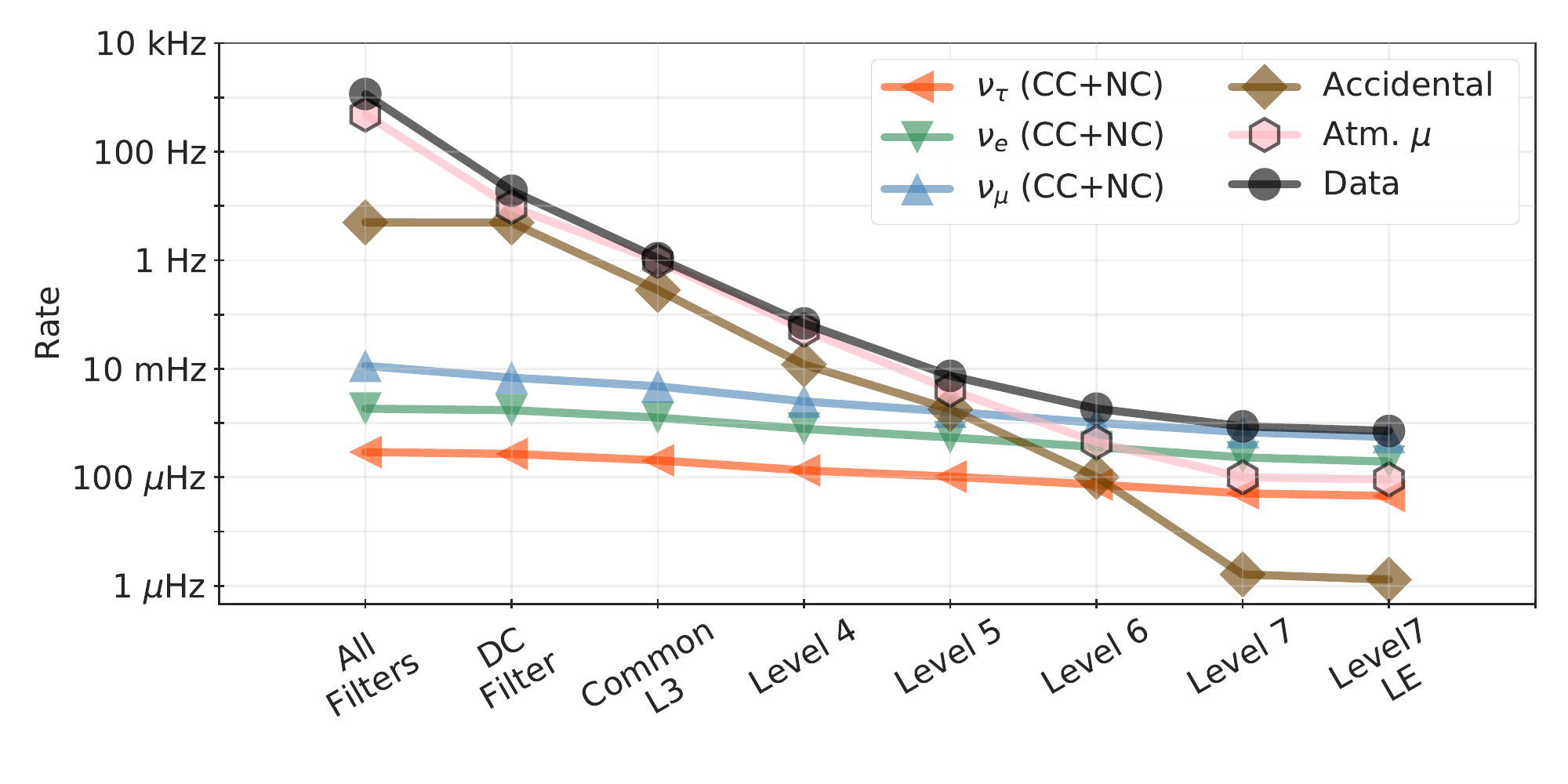}
    \caption{The event rates as a function of Analysis \greco cut level. The data is dominated by atmospheric muons and accidental triggers due to random detector noise until after Level~5, after which $\nu_\mu$ dominate the selection.}
    \label{fig:greco_rates}
\end{figure}

Following the application of the two BDT-based selections, a series of individual event selection criteria are applied (Level~6).  Requiring events to have a sufficient number of hits inconsistent with intrinsic DOM noise and to be spatially compact removes most remaining events caused purely by intrinsic noise hits.  Removal of many of the remaining atmospheric muons is accomplished by requiring a likelihood-based vertex estimate to be well contained in the DeepCore fiducial region, and by rejecting events with any hit DOMs along selected directions (``corridors'') through the surrounding IceCube veto volume.  Due to the regular hexagonal grid layout of the detector, these corridors have lower photosensor coverage  than other regions of the veto volume.

With a sufficiently low event rate, similar containment criteria are used as at Level~6, but with a more accurate and time-consuming reconstruction applied (Level~7).  In addition, events are required to have a reconstructed deposited energy between 5.6\,GeV and 56\,GeV.

\subsubsection{Additional Selection Criteria: Analysis \dragon }

The Analysis \dragon sample applies several selection criteria at Level~4.  These include requiring a sufficient number of p.e.\@ deposited in the largest cluster of hits in the fiducial region, a minimum number of non-isolated hits in the fiducial region, an event vertex contained in the fiducial volume, a space-time interval between the first and fourth temporal quartiles of the hits in DeepCore consistent with $v \le c$, no more than 5\,p.e.\@ in the surrounding veto region, and no more than two hits in the veto region consistent with speed-of-light travel to the hit in DeepCore whose time is closest to the event trigger time.  These criteria reject events caused by noise, reduce muon background, and favor the more cascade-like signature produced by most $\nutau$ interactions.  A BDT is then applied (Level~5), using 11 input variables, derived from the charge, time, and location of the hit DOMs, as well as reconstructed zenith angle and event speed using crude but fast track reconstructions.

Following the application of the BDT, events consistent with entering through corridor regions are rejected, and reconstructed events are further required to have starting and stopping positions in or near the DeepCore fiducial volume.  These Level~6 criteria further reject atmospheric muon background.  At this stage in the processing, the neutrino signal rate has been reduced by a factor of roughly 13 while the atmospheric muon background rate by a factor of $10^8$.  A more detailed breakdown is provided in Table~\ref{tab:event_rates} of Appendix~\ref{app:common}.

\subsection{Reconstruction}
\label{sec:reco}

The reconstruction used in both Analyses \greco and \dragon assumes that every event starts with an electromagnetic or hadronic shower followed by a finite, minimum ionizing muon at the same primary vertex. Due to the numerous charged particles in the shower, a cascade-like event is characterized by a localized Cherenkov light pattern centered at the interaction vertex. On the other hand, a track-like event involves a muon which deposits Cherenkov light uniformly along its trajectory, and travels much further than any non-muon particles produced in the primary shower. With the cascade plus track assumption, the reconstruction algorithm describes an event via eight parameters: the primary interaction vertex position ($x,y,z$) and time ($t$), the direction given by the zenith angle ($\theta_\nu$) and the azimuth angle ($\phi_\nu$) of the neutrino, the energy of the primary cascade ($E_{cscd}$), and the length of the track from the minimum ionizing muon ($L_\mu$).

Based on the above hypothesis, a likelihood-based reconstruction method compares the observed pattern of photon counts from all active DOMs in an event to that predicted. 
The PMT measures a charge linearly related to the number of Cherenkov photons arriving at a DOM.
Using the PMT charge as a proxy for photon counts, the number of photons arriving at the DOM is described by the time-binned PMT charge. 
The predicted pattern of charges from all DOMs in an event is then fitted to that of the observed event with the eight parameters in the event hypothesis allowed to vary freely.

To reduce computational complexity in running the reconstruction, energy deposition during an event is described using several independent light sources. 
In particular, the deposited energy from the primary cascade is treated independently of that from a muon track. 
Further, the energy deposition by the muon track is also discretized into segments with constant length. The total length of the track $L_\mu$ is directly related to the energy of the track via $E_\mathrm{trck} = L_\mu \frac{\mathrm{d}E_{\mu}}{\mathrm{d}x}$, where the differential energy loss of a minimum ionizing muon in ice, $\mathrm{d}E_{\mu} / \mathrm{d}x$, is fixed to $0.22\,{\mathrm{GeV}}/{\mathrm{m}}$.

The energy deposition of a muon along its track is not constant in reality nor in our simulation. 
This simplification is only used for reconstruction of low energy events and yields a good approximation at the $\mathcal{O}(10~\mathrm{GeV})$ scale.
The approximation begins to break down above about 50 GeV when stochastic losses along the muon track become non-negligible~\cite{Chirkin:2004hz}.

The expected charge $q_{i}(t)$ at the $i^{\text{th}}$ DOM at time $t$ is estimated by the charge due to energy depositions by the cascade $E_{cscd}$ and by the track $E_{trck}$ plus a time-independent noise term $n_i$. 
The expected charge can be expressed as,
\begin{equation} \label{eq:reponseMatrix}
    q_{i}(t) = \Lambda_{i}^{cscd}(t) \cdot E_{cscd}/\mathrm{GeV}  + \sum_{\mathrm{segments}\in L_\mu}\Lambda_{i}^{trck}(t) + n_i,
\end{equation}
where $\Lambda^{cscd}$ represents the charge expectation for a 1\,GeV cascade and $\Lambda^{track}$ for a minimal ionizing muon of one segment length.
A linear relation between $\Lambda^{cscd}$ in Eq.~\ref{eq:reponseMatrix} and the deposited cascade energy is assumed. 
To obtain the values of $\Lambda^{cscd}$ and $\Lambda^{trck}$, large sets of look-up tables are generated from simulations of photon propagation in the ice \cite{Aartsen:2013vja}.
These tables, used with the assumption that the number of Cherenkov photons emitted is directly proportional to the deposited energy of the particle, allow for the calculation of the expected charge from an arbitrary cascade or track.

The process of finding the maximum likelihood hypothesis for an event is an eight-dimensional optimization problem, and the likelihood space is typically non-convex,{\it i.e.\@} populated with local maxima. To cope with these challenges the MultiNest algorithm~\cite{Feroz:2008xx} is used to find the best-fit hypothesis.

Both presented analyses follow the above reconstruction algorithm but with two main differences. 
First, each track length segment in Analysis \greco  is 5\,m long, whereas analysis \dragon uses coarser 15\,m long segments. 
Second, the reconstruction used in analysis \greco ignores the observed charges, instead implementing a binary response of 0\,p.e\@ or 1\,p.e.\@ per 45\,ns in each DOM individually, while Analysis \dragon uses the observed charge in each DOM. 
The treatment of charge in Analysis \greco reduces the impact of observed discrepancies observed between the distributions of the average charge per DOM in data and simulation, which affect mainly the stochastic nature of charge depositions in events with a small number of hit DOMs.

\begin{figure}[htbp]
    \includegraphics[width=\linewidth]{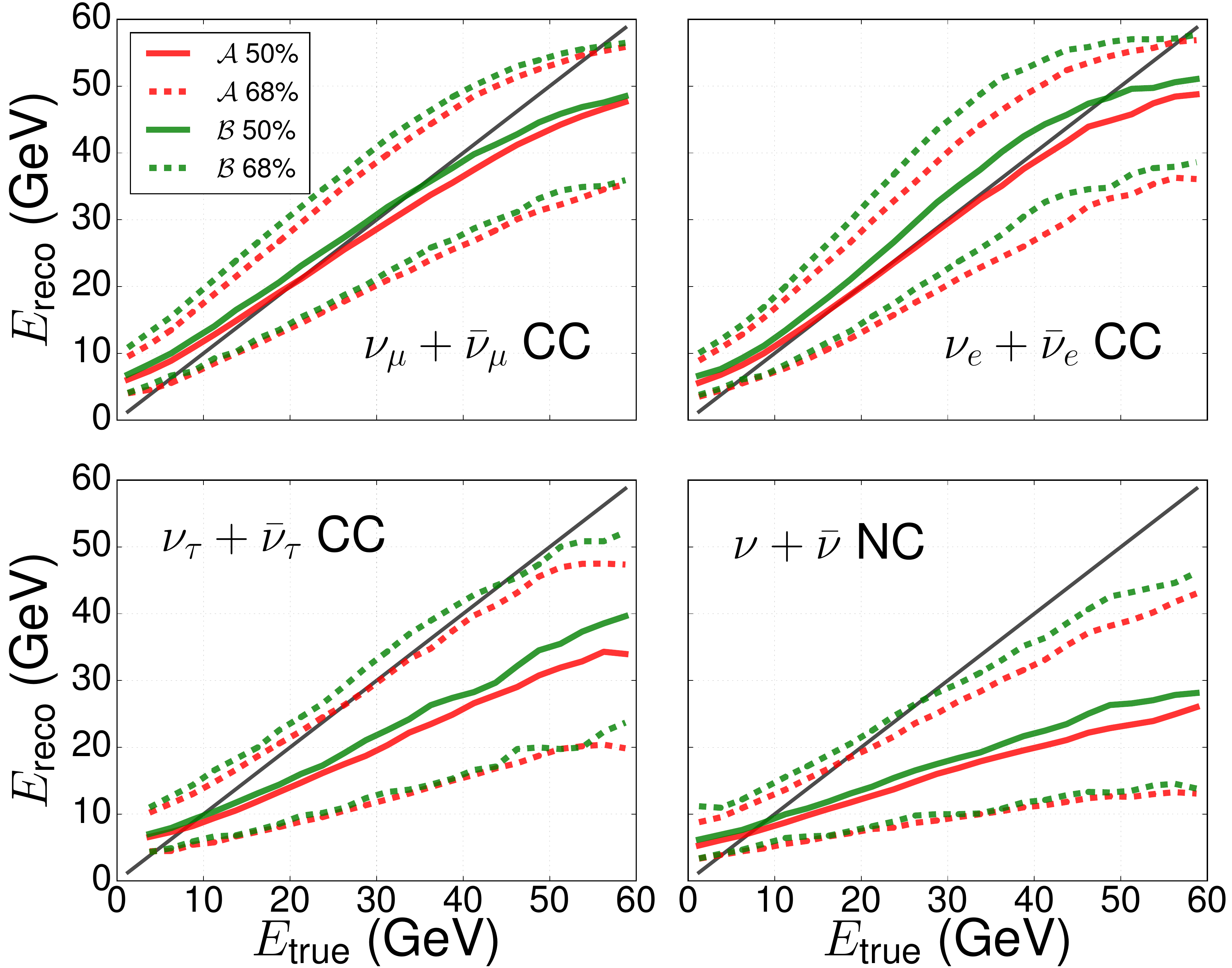}
    \caption{Reconstructed energy vs.\@ true energy for each neutrino flavor separately (CC interactions) and all flavors combined (NC interactions). The red and blue solid lines are the resolutions from Analyses \greco and \dragon respectively, and the dashed lines represent the 68\% ranges. The solid black lines are the references indicating perfect reconstruction.
    For \nutau\ CC and $\nu$ NC events the final state ensemble of out-going particles include at least one ``invisible'' neutrino which manifests as missing energy when comparing $E_{\rm true}$ to $E_{\rm reco}$.
    }
    \label{fig:energy_reco}
\end{figure}

\begin{figure}[htbp]
    \includegraphics[width=\linewidth]{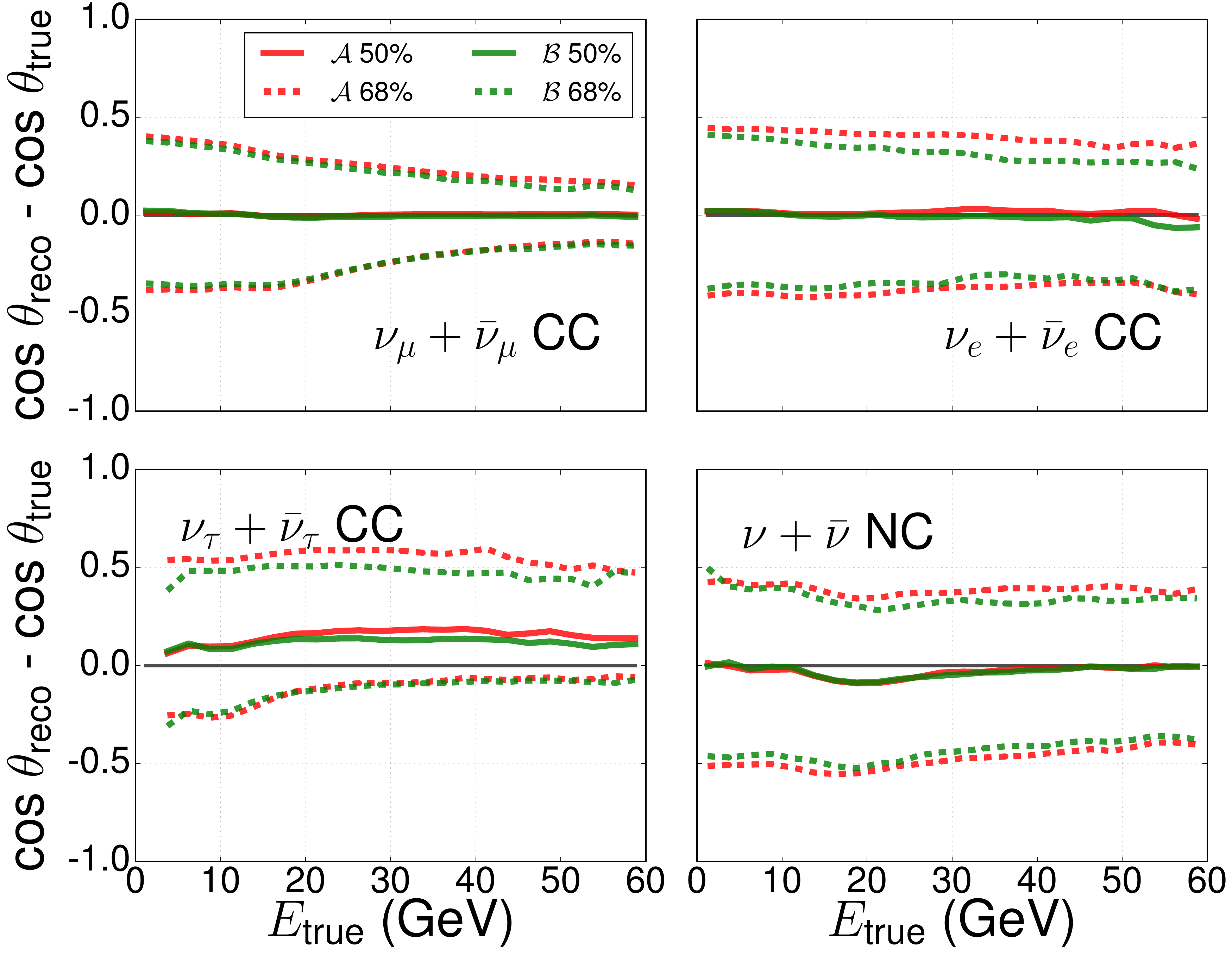}
    \caption{Difference between reconstructed and true cos $\theta$ vs.\@ true energy for each neutrino flavor separately (CC interactions) and all flavors combined (NC interactions). The red and blue solid lines are the resolutions from Analyses \greco and \dragon respectively, and the dashed lines represent the 68\% ranges. The solid black lines are the references indicating perfect reconstruction.}
    \label{fig:coszen_reco}
\end{figure}

Despite the differences between the two analyses, the energy and cosine of zenith angle resolutions of the two analyses are similar, as shown in Fig.~\ref{fig:energy_reco} and Fig.~\ref{fig:coszen_reco}.

\subsection{Classification}

A $\numu$ CC neutrino interaction often produces an event with an identifiable track, whereas events from $\nue$ CC and all-flavor $\nu_{\mu, \tau, e}$ NC have a cascade-like topology.
Most \nutau\ CC interactions also produce cascade-like events, with the short-lived $\tau$ lepton decaying roughly $83\%$ of the time to non-muon modes~\cite{Patrignani:2016xqp}. The $\approx 17\%$ muonic decay mode is $\tau^- \rightarrow \mu^- \numubar\ \nutau$ (and charge conjugate), where the daughter muon may have sufficient energy to create a visible track indistinguishable from a $\numu$ CC event causing it to be identified as a track-like.
To improve the sensitivity of $\nutau$ measurement, both analyses divide their samples into cascade- and track-like subsets, enhancing the purity of $\nutau$ events in the cascade channel.

\begin{figure}[htbp]
    \includegraphics[width=1\linewidth]{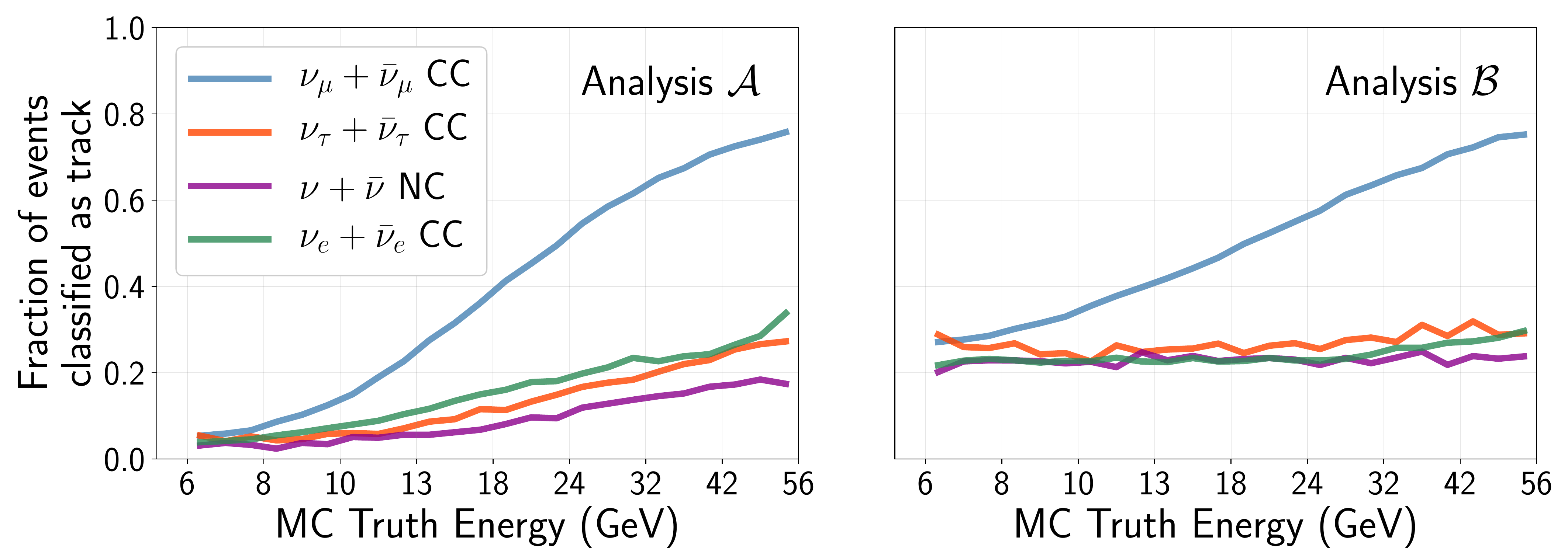}
    \caption{Fraction of track-like events as a function of true neutrino energy for each neutrino event type in Analyses \greco (left) and \dragon (right). Differences in particle classification lead to different fractions of track-like events at lower energies.}
    \label{fig:pid_reco}
\end{figure}

To determine if an event is cascade-like or track-like Analysis \greco relies on the reconstructed track length $L_\mu$. Events with a track length between 0\,m and 50\,m are considered cascade-like, and events with track lengths longer than 50\,m are considered track-like. For Analysis \dragon, an additional reconstruction is performed with the track length forced to 0\,m. Events are then classified based on the log-likelihood difference between the cascade-and-track hypothesis and that of cascade-only;
$\Delta\mathrm{LLH}_\mathrm{reco} = \ln{\mathcal{L_\text{cascade+track}}} - \ln{\mathcal{L_\text{cascade}}}$.
Events with $\Delta\mathrm{LLH}_\mathrm{reco} > 2$ are considered as track-like, while events with $-3 < \Delta\mathrm{LLH}_\mathrm{reco} < 2$ are cascade-like. The cascade only reconstruction should in principle never yield a likelihood that is better than the track+cascade one, but due to finite precision of the minimization process, negative $\Delta\mathrm{LLH}_\mathrm{reco}$ do occur. We allow events with a negative $\Delta\mathrm{LLH}_\mathrm{reco}$ as low as $-3$; the remaining events are removed from the analysis due to their bad reconstruction quality.
As shown in Fig.~\ref{fig:pid_reco}, the cascade and track separation powers from the two analyses are similar above 20\,GeV.

\section{Analysis}
\label{sec:analysis}

Our tau neutrino appearance analyses yield two distinct quantities: the level at which the null hypothesis of no $\nutau$ appearance is rejected and the measurement of the $\nutau$ normalization, which is defined as the ratio of the measured $\nutau$ flux to that expected assuming best-fit oscillation and other nuisance parameters for that $\nutau$ normalization.
These best-fit nuisance parameters are obtained simultaneously with the best-fit tau normalization during the optimization process, meaning that the expected distribution of tau neutrino events can be understood as: ($\nutau$ normalization) $\times$ ((baseline $\nutau$ expectation) + $\sum$(nuisance parameter)$\times$($\nutau$ systematic change)).

Since DeepCore cannot distinguish between $\nutau$ CC and NC interactions, our analyses benefit from treating them on an equal footing by applying the $\nutau$ normalization to both CC and NC tau neutrino interactions. 
However, to facilitate comparisons with results from other experiments, a second set of measurements are also performed applying the $\nutau$ normalization only to the $\nutau$ CC component.
In this second case, the $\nutau$ NC component is unaffected by the value of the $\nutau$ normalization.
In both the CC+NC and CC-only cases, there is a separate uncertainty assigned to all neutral current events.

\begin{figure}
    \includegraphics[width=\linewidth]{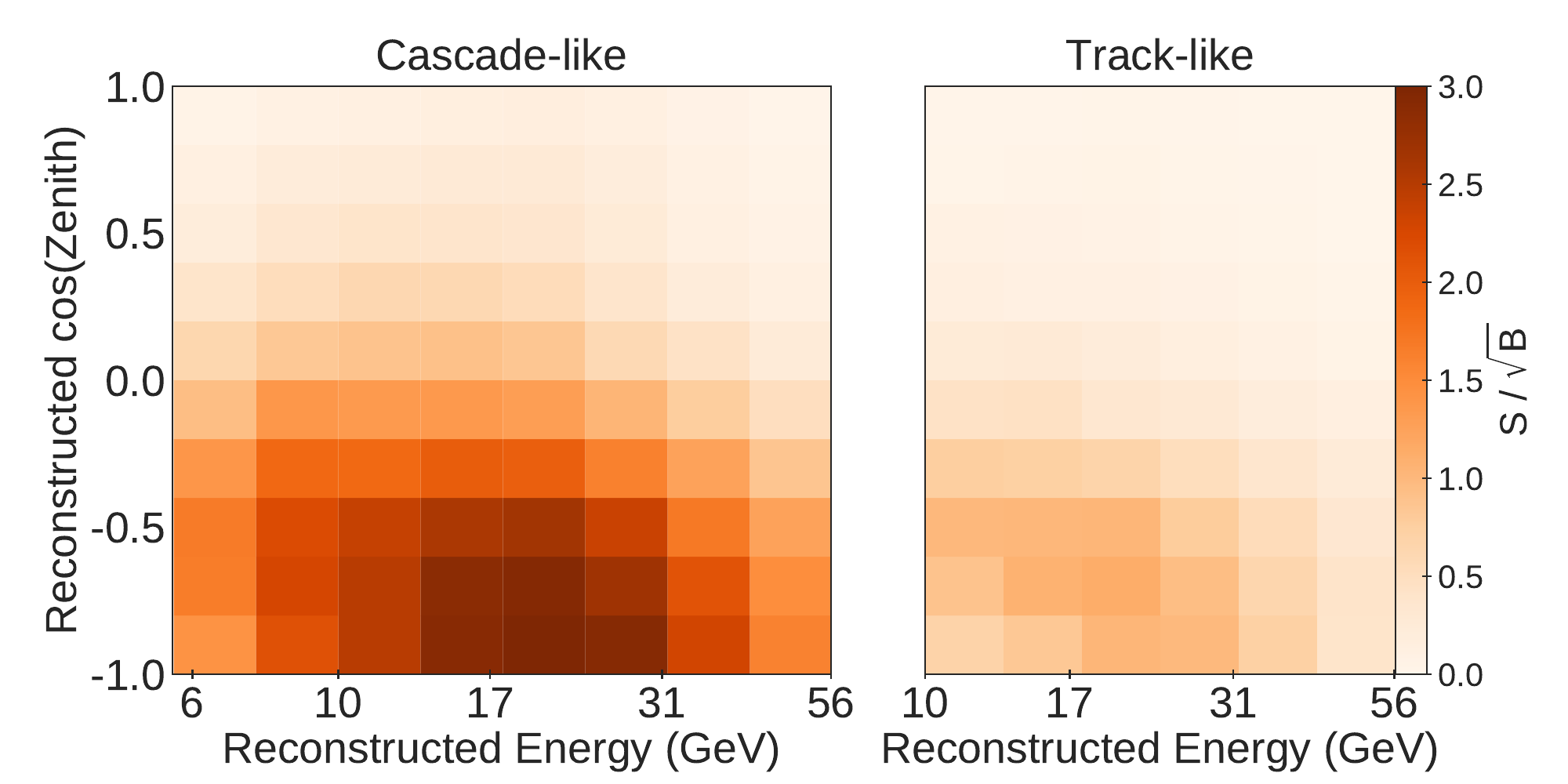}
    \caption{Expected signal $\nu_\tau$ (CC+NC) divided by the square-root of the expected background ($\nu_e$, $\nu_\mu$, atmospheric $\mu$, and noise-triggered) events as a function of reconstructed cosine of the zenith angle and reconstructed energy. Cascade-like events are shown on the left and track-like events on the right.  The plots include both neutrinos and anti-neutrinos.}
    \label{fig:soverb}
\end{figure}

In Analysis \greco, data is binned into a 3-d histogram with eight reconstructed energy bins spaced logarithmically between 5.6~GeV and 56~GeV, 10 reconstructed cosine zenith bins spaced linearly between $-1$ and 1, and two reconstructed track length bins for particle identification (PID). Track-like events in Analysis \greco have a reconstructed energy of at least 10~GeV associated with the minimum track length of 50~m. Therefore, the first two energy bins for track events are empty by construction and not included in the analysis. Figure~\ref{fig:soverb} shows the $S/\sqrt{B}$ as a figure of merit, where $S$ and $B$ are the number of signal and background events, respectively. The figure indicates that upward-going cascade events with reconstructed energies around 20~GeV dominate the measurement. With the same energy binning as Analysis \greco, Analysis \dragon covers the same cosine zenith range with eight bins instead of 10 and uses  $\Delta\mathrm{LLH}_\mathrm{reco}$ for PID instead of reconstructed track length. 

In each of Analyses \greco and \dragon, a $\chi^2$ minimization is performed on the binned data as a function of the $\nutau$ normalization and nuisance parameters associated with the relevant systematic uncertainties, see Sec.~\ref{sec:sys}. The $\chi^2$ function is defined as
\begin{equation}
    \chi^2 = \sum_{i \in \{\text{bins}\}} \frac{(N^{\text{exp}}_i - N^{\text{obs}}_i)^2}{N^{\text{exp}}_i + (\sigma^{\text{exp}}_i)^2} + \sum_{j \in \{\text{syst}\}} \frac{(s_j - \hat{s}_j)^2}{\sigma_{s_j}^2},
\label{eq:chi2}
\end{equation}
where $N^{\text{exp}}_i$ is the number of total events expected from the signal and all background events in the $i^{\rm th}$ bin, and $N^{\text{obs}}_i$ is the number of events observed in the $i^{\rm th}$ bin. For both Analyses \greco and \dragon, the denominator consists of the standard Poisson variance $N^{\text{exp}}_i$ and the uncertainty in the prediction of the number of expected events $\sigma^{\text{exp}}_i$ of the $i^{\rm th}$ bin. Analysis \greco uses Monte Carlo simulation for the prediction of all event types, and the term $\sigma^{\text{exp}}$ is the sum of uncertainties due to finite statistics of MC simulation from each event type. In Analysis \dragon, the term $\sigma^{\text{exp}}$ encompasses both the uncertainty due to finite MC statistics as well as the uncertainty in the data-driven muon background estimate described in the Sec.~\ref{sec:atmmu}. The second term of Eq.~\eqref{eq:chi2} is the sum of penalty terms for nuisance parameters that have prior constraints imposed, where $s_j$ is the central value of $j^{\rm th}$ systematic parameter, $\hat{s}_j$ is its maximum likelihood estimator, and $\sigma_{s_j}^2$ is the prior's Gaussian standard deviation. 

For both analyses the uncertainty due to limited MC statistics is small for signal neutrinos, as the effective livetime for simulation is an order of magnitude higher than that of the acquired data.
The situation is different for the muon background predictions: for Analysis \greco the uncertainty arises from simulation with {\it less} effective livetime than the actual data and for Analysis~\dragon from a data side-band, in both cases resulting in larger uncertainties than for signal neutrinos.
However, any ensuing variations are predominantly constrained to the track-like and downward-going region of the event sample which is away from the cascade-like and upward-region region associated with our targeted signal events.

While both analyses use data from the same operating period of April 2012 through May 2015, minor differences in the event selection criteria lead to a total livetime of 1006 days for Analysis \greco and 1022 days for Analysis \dragon. Table~\ref{tab:events} shows the expected number of events at the best fit point for each neutrino flavor and interaction type, and for atmospheric muons and noise-triggered backgrounds. 


\begin{table}[htbp]
\centering
\vspace{0.2cm}
\caption{Expected number of events at the NC+CC best fit point, grouped by flavor and interaction type, and including atmospheric muons. The observed counts from the data are shown in the last row. Associated $\pm 1\sigma$ uncertainties due to limited simulation statistics are also shown (the uncertainty showed on the observed count is just the Poisson error).}
\label{tab:events}
\vspace{0.2cm}
\begin{tabular}{l|cc|cc}
                                     & \multicolumn{2}{c|}{Analysis \greco}             & \multicolumn{2}{c}{Analysis \dragon}             \\
Type                                 & Events             & $\pm 1\sigma$      & Events             & $\pm 1\sigma$      \\
\hline
$\nu_e + \bar{\nu}_e\ \rm{CC}$       & 13462              &  29                 & 9545               & 23                 \\
$\nu_e + \bar{\nu}_e\ \rm{NC}$       & 1096               &  9                  & 923                & 8                  \\
$\nu_\mu + \bar{\nu}_\mu\ \rm{CC}$   & 35706              &  48                 & 23852              & 39                 \\
$\nu_\mu + \bar{\nu}_\mu\ \rm{NC}$   & 4463               &  19                 & 3368               & 17                 \\
$\nu_\tau + \bar{\nu}_\tau\ \rm{CC}$ & 1804               &  9                  & 934                & 5                  \\
$\nu_\tau + \bar{\nu}_\tau\ \rm{NC}$ & 556                &  3                  & 445                & 4                 \\
Atmospheric $\mu$                    & 5022               &  167                & 1889               & 45                 \\
Noise Triggers                       & 93                 &  27                 & $<$ 25                &  $<$ 5                  \\
\textbf{total (best fit)}            & \textbf{62203}         &  \textbf{180}        & \textbf{40959}     & \textbf{68}        \\
\textbf{observed}                    & \textbf{62112}         &  \textbf{249}        & \textbf{40902}     & \textbf{202}       \\
\end{tabular}
\end{table}

\section{Systematic Uncertainties}
\label{sec:sys}

The effect of systematic uncertainties is included in the analyses with nuisance parameters that impact the shape and normalization of the expected event distributions. The uncertainties considered can be broadly grouped in categories according to their origin: the initial unoscillated flux of atmospheric neutrinos, neutrino-nucleon cross sections, neutrino flavor oscillation parameters, detector response, and atmospheric muon background estimates. The associated parameters, together with their best-fit values, are summarized in Table~\ref{tab:syst}.
Each category of uncertainties will be discussed in turn.

\begin{figure}
    \centering
    \includegraphics[width=\linewidth]{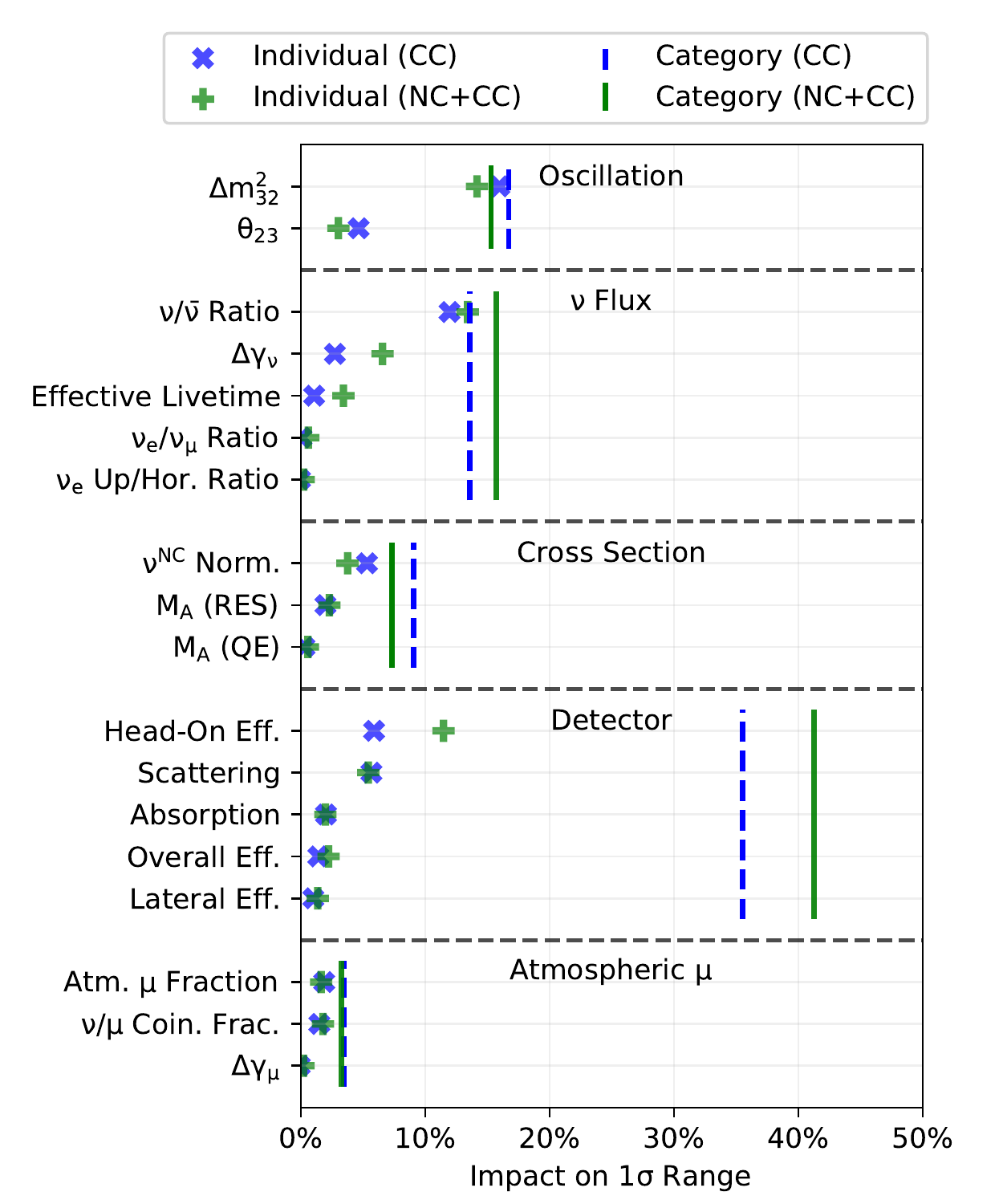}
    \caption{The relative impact from each systematic uncertainty and each group on the final 1$\sigma$ confidence interval width in Analysis \greco. Each systematic uncertainty is fixed to the best-fit value in turn and the change in the interval is measured. The most important systematic uncertainty is $\Delta m^2_{31}$, with a 14\% (16\%) impact on the NC+CC (CC) measurement. The detector uncertainties show degeneracies that limit the impact of individual parameters, but together account for 41\% (36\%) of the uncertainty in the NC+CC (CC) measurement in Analysis \greco. }
    \label{fig:sys_impact}
\end{figure}

To quantify the impact of each systematic uncertainty, the 1$\sigma$ confidence interval of the expected tau neutrino normalization measurement was calculated while fixing one parameter at a time. The resulting change in the confidence interval is shown in Fig.~\ref{fig:sys_impact}.
Of the fitted systematic uncertainties, the neutrino mass splitting provides the strongest impact on the final confidence interval. The impact of each category of systematic uncertainties was also tested in a similar way. When entire categories of systematic uncertainties are fixed at the same time, the largest impact comes from the detector uncertainties, which account for 41\% (36\%) of the NC+CC (CC) measurement in Analysis \greco. This is due to individual systematic variations being correlated, especially the ones in the detector uncertainty group.

In addition to the systematic uncertainties mentioned above and included in the analysis, we have studied different optical models for the glacial ice as well as a newly available charge calibration for the detector.
In both cases, the impact on the final result was found to be negligible, and they were thus omitted from the fit and the error calculation.

\subsection{Atmospheric neutrino flux}
\label{sec:flux}

The measurement presented in this work is extracted from an observed distortion of the flux of atmospheric neutrinos. Our nominal model is the calculation of Honda {\it et al.}~\cite{Honda:2015fha}. 
The calculation covers the energy range 100~MeV to 10~TeV, and was produced specifically for a detector situated at the geographic South Pole, so local geomagnetic effects are included.
The cosmic rays that contribute the most to the neutrino production at the energies of interest, between 5.6--56~GeV, are protons and helium.
Honda {\it et al.\@} model the energy spectrum of each of these incident particles using a single power law, fitting the flux to data from satellite and balloon experiments.
In this calculation, interactions of cosmic rays with the Earth's atmosphere are simulated using a combination of the JAM interaction model~\cite{jam_model} and a modified version of DPMJET-III~\cite{Roesler:2000he}.
The modifications, discussed in~\cite{Honda:2011nf}, are changes to the yields of $\pi$ and $K$ mesons to reach a better agreement with muon measurements from the BESS experiment~\cite{Abe:2003cd}.
The atmospheric conditions such as temperature and column density are taken from the NRLMSISE-00 model~\cite{JGRA:JGRA16630}, whose authors estimate the resulting calculation has an uncertainty on the neutrino flux of $\leq 15$\%. Seasonal variations are included in the flux calculations, one-year averaged values are used in our analyses.

In both Analyses \GRECO and \DRAGON, a detailed modification of the neutrino flux prediction as a function of energy, zenith angle, and particle species has been used.
The basis of this modification is the work of Barr {\it et al.\@}~\cite{Barr:2006it}, who have performed a detailed study of the uncertainties on neutrino flux predictions by systematically modifying the inputs required to perform the calculation.
Their work suggests that, for the energies that are of interest here, the flux calculation is mostly affected by the uncertainties on the spectral index assumed when modeling the cosmic ray fluxes, and the lack of measurements on the production of $\pi$ and $K$ mesons with energies above 500\,GeV and 30\,GeV, respectively, and where the secondary particle contains $>10$\% of the incident particle energy.

A modification to the spectral index on the cosmic rays translates into a very similar modification of the neutrino flux.
We therefore account for this uncertainty by modifying the neutrino flux using the function $E^{\Delta\gamma_{\nu}}$, which only depends on neutrino energy.
Modifying the yields of pions and kaons in hadronic interactions produces changes in the neutrino flux, not only as a function of energy, but also incoming zenith angle for each of the particle species in it.
In~\cite{Barr:2006it} a summary of these modifications is shown for the $\nu/\bar{\nu}$ flux ratio as function of energy and as function of zenith angle for three energy regions, and the upward-going to horizontal $\nu$ ratio as a function of neutrino energy.
We use that information to build a model able to reproduce the effects described as function of both energy and zenith angle.

In summary, four effective parameters account for the uncertainties considered on the atmospheric neutrino flux.  These are a modification of the spectral index ($\Delta\gamma_{\nu}$), the ratio of $\nue$ to $\numu$ fluxes (``$\nue/\numu$ ratio''), the ratio of the $\nu$ to $\bar{\nu}$ fluxes as function of zenith angle and energy (``$\nu/\bar{\nu}$ ratio''), and an additional parameter for the remaining uncertainty in the upward-going vs. horizontal flux of electron neutrinos (``up/hor ratio''). 
All parameters are introduced assuming that they are uncorrelated. 
A 5\% uncertainty is assumed for the $\nue / \numu$ flux ratio.
The two parameters that modify $\nu/\bar{\nu}$ and \nue\ up/hor receive an uncertainty such that, when both are evaluated at $1\sigma$, the results from~\cite{Barr:2006it} are reproduced.
They roughly correspond to a 10\% energy-dependent change to the neutrino flux with a 3\% zenith-dependent modulation.
The top two panels of Fig.~\ref{fig:syst_effect_on_event} demonstrate the effect of these parameters in the reconstructed final sample.
The error assigned to $\Delta\gamma$ is discussed in the next section.

Sources of uncertainty that result in a global scaling of the neutrino flux, independent of energy or zenith angle, are not considered in this work as the normalization is left free in the fit (scaled by the effective livetime parameter).

\subsection{Atmospheric muon flux}
\label{sec:muflux}

While the sources of uncertainties discussed above are included in both Analyses \GRECO and \DRAGON, an additional uncertainty related to neutrino-muon coincidence is taken into account in Analysis \GRECO.
An extra simulation set was produced, in which every neutrino event is contaminated by an atmospheric muon resulting from an independent air shower.
Together with the baseline neutrino sets with no muon contamination, the event count is parametrized per bin as a function of coincident fraction.
Because previous high-energy analyses using the IceCube volume found less than 10\% contamination due to coincident muons, a one-sided Gaussian prior centered at 0 with a width of 10\% is applied to the coincident fraction for Analysis \GRECO.
The effect from neutrino-muon coincidences is normalized to leave the total event rate unchanged.

Analysis \GRECO also considers an uncertainty related to the cosmic ray spectral index in the atmospheric muon flux.
Atmospheric background muons in Analysis \greco are produced in air showers of energies 1~TeV to 1~PeV.
These shower energies are higher than the expected energies from the atmospheric neutrinos making it into the final analysis.
To be conservative, the effect of a change in the cosmic ray spectral index is treated independently between neutrinos and muons to account for the separate energy regimes probed.

Measurement uncertainties from a fit to cosmic ray experimental data~\cite{Evans:2016obt} are used to obtain an estimate for the uncertainty on the spectral indices associated with proton and helium cosmic ray primaries. 
Based on the error bars from the experiments, the deviation from the central fit value is determined as a function of primary energy using CORSIKA simulations. 
This change in the flux weighting for atmospheric muons is parametrized as a function of true energy and zenith angle and applied to the final simulated atmospheric muon sample.
A Gaussian prior is applied to the spectral index uncertainty, with a 1$\sigma$ deviation in the parameter corresponding to a 1$\sigma$ change in the cosmic ray spectral index.

\subsection{Neutrino--nucleon interactions}
\label{sec:xsec}

Deep-inelastic-scattering (DIS) interactions make up the bulk of neutrino interactions visible in DeepCore.
The uncertainties associated with these interactions were investigated in the final samples.

The first studies were on the parameters used in the Bodek-Yang model to allow the parton distribution functions used in the calculation of cross sections to be extended to the lower $Q^2$ region~\cite{Bodek:2002ps}.
These DIS events were re-weighted on an event-by-event basis in response to changes in the higher-twist parameters and valence quark corrections using the reweighting scheme included in the GENIE generator \cite{Andreopoulos:2009rq}.
Though this did have a small impact on the final analysis, they were fully degenerate with either the overall neutrino scaling provided by the neutrino event rate (via the ``effective livetime" parameter) or the energy dependent scaling provided by the spectral index parameter $\Delta\gamma_\nu$.
Since these two systematics fully absorb the effect of the uncertainty in the Bodek-Yang model, no additional parameter was included in the final analysis.

We further investigated the impact of both high- and low-W averaged charged hadronization multiplicity, a systematic uncertainty also related to DIS interactions \cite{Katori:2014fxa}.
These studies were done by modifying PYTHIA to change the multiplicity of outgoing charged particles to be within the range observed by bubble chamber experiments~\cite{Zieminska:1983bs,PhysRevD.25.624,ALLEN1981385}. 
These changes were then propagated through GENIE to evaluate the effect on the final sample.
It was found this has less than 0.1\% impact on events at the final level, with the change being energy dependent.
Due to the small size of this effect and its shape being degenerate with that of spectral index changes ($\Delta\gamma_\nu)$, we did not include this as an additional parameter in the final fit.

The final DIS uncertainty studied was its differential cross section.
The approach here was to modify the structure function as a function of the Bjorken-$x$ within the uncertainties measured by NuTeV~\cite{Fleming:2000bg}.
This resulted in a change at final level of less than 1\% up to 3\% at 200~GeV.
As with the studies on hadron multiplicity, these changes are degenerate with a change in the spectral index uncertainty and so are not included in the final fit.

Many cross section systematic uncertainties were tested, but the only two which were not already degenerate with other systematic uncertainties were the axial mass form factors for charged current quasi-elastic ($M_A^{CCQE}$) and resonant ($M_A^{res}$) events.
Both of these are included in the final analysis and change the expected number of CCQE or resonant events seen in the event sample.
The systematic is implemented so that a change in $M_A^{CCQE (res)}$ will result in a change to each CCQE (resonant) event weight on an event-by-event basis using GENIE's re-weighting capabilities.

\begin{figure}[htbp]
\vspace*{2mm}
\includegraphics[width=\linewidth,height=4cm]{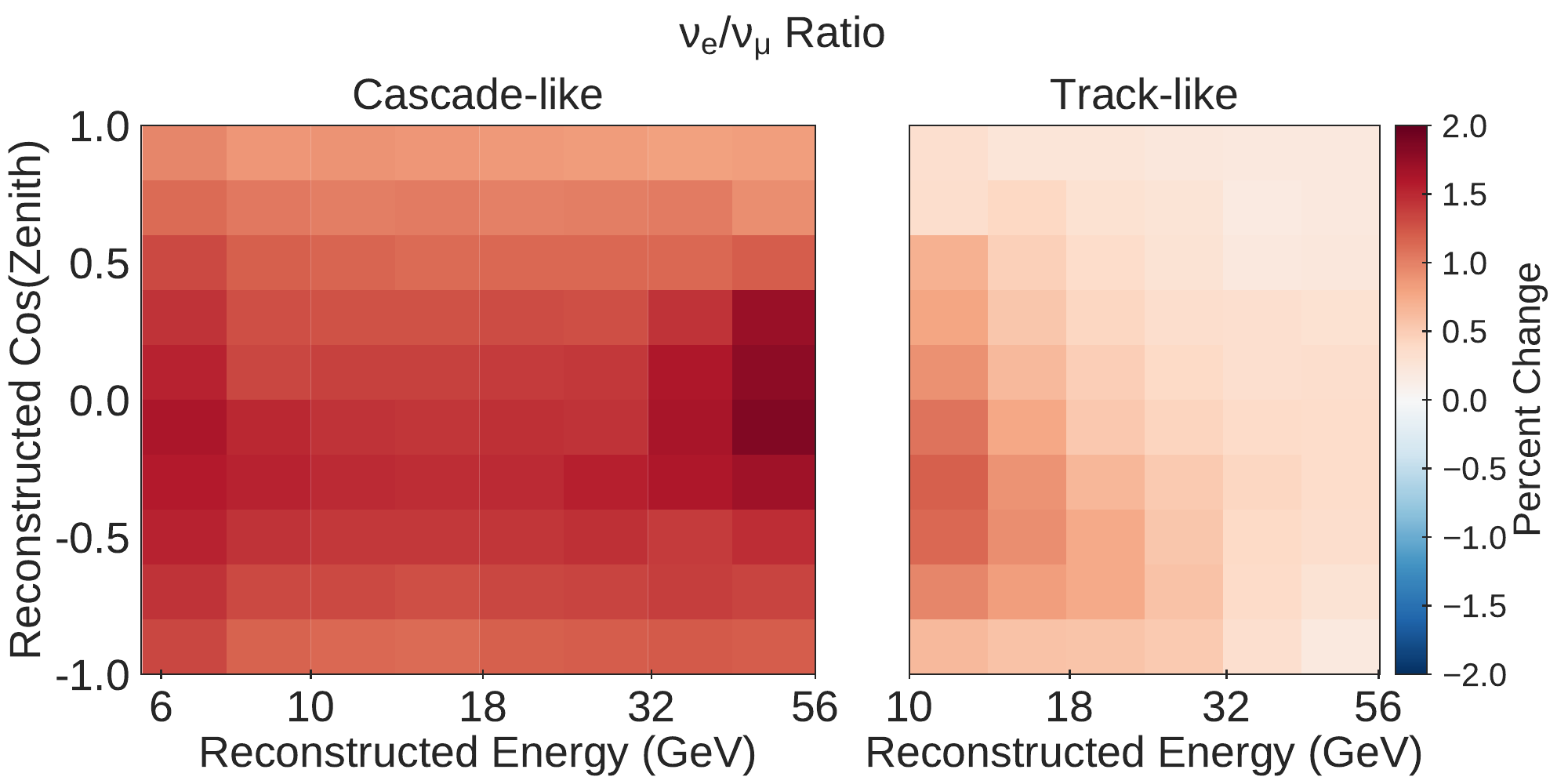}
\vspace*{2mm}
\includegraphics[width=\linewidth,height=4cm]{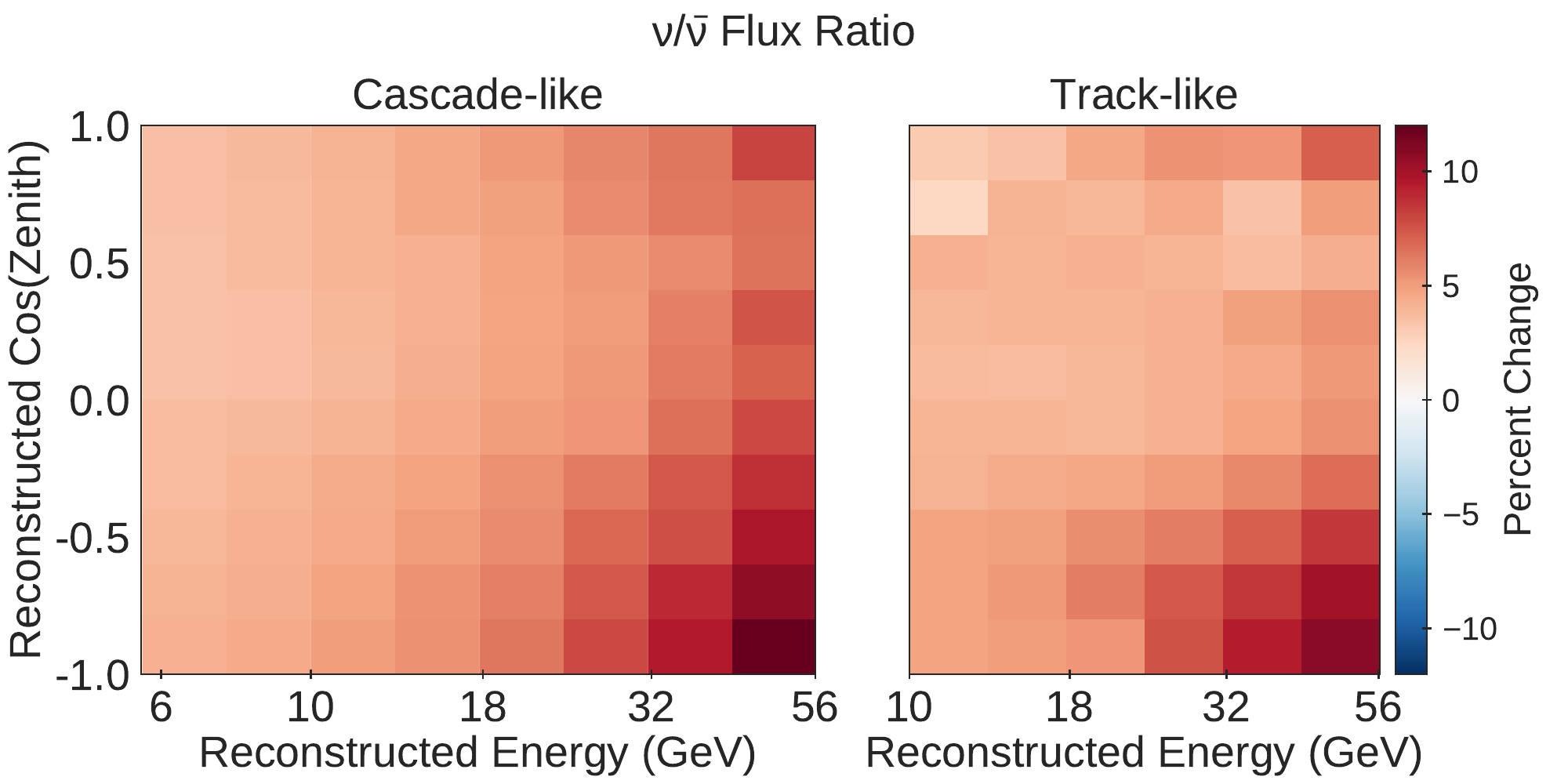}
\vspace*{2mm}
\includegraphics[width=\linewidth,height=4cm]{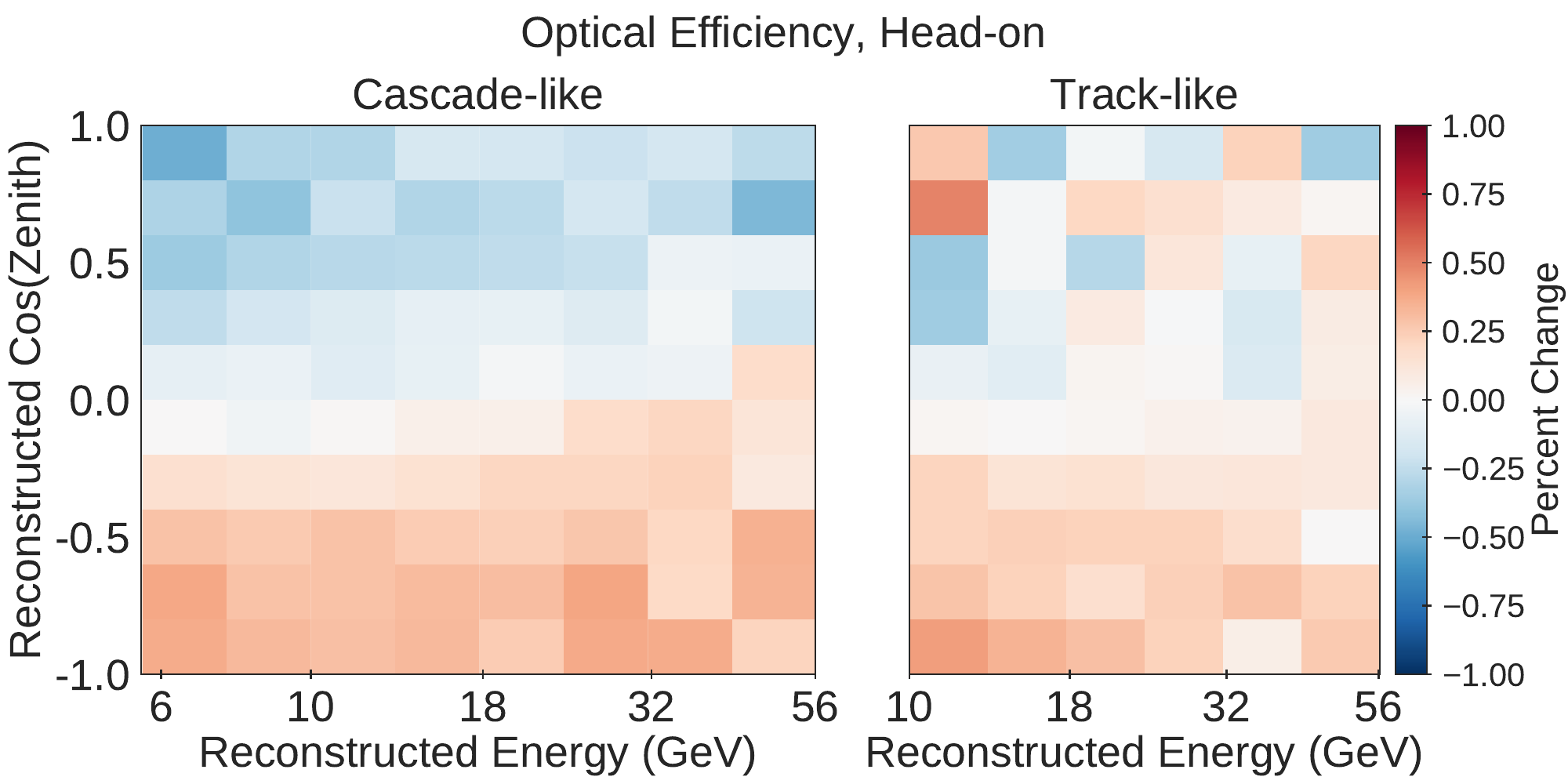}
\vspace*{2mm}
\includegraphics[width=\linewidth,height=4cm]{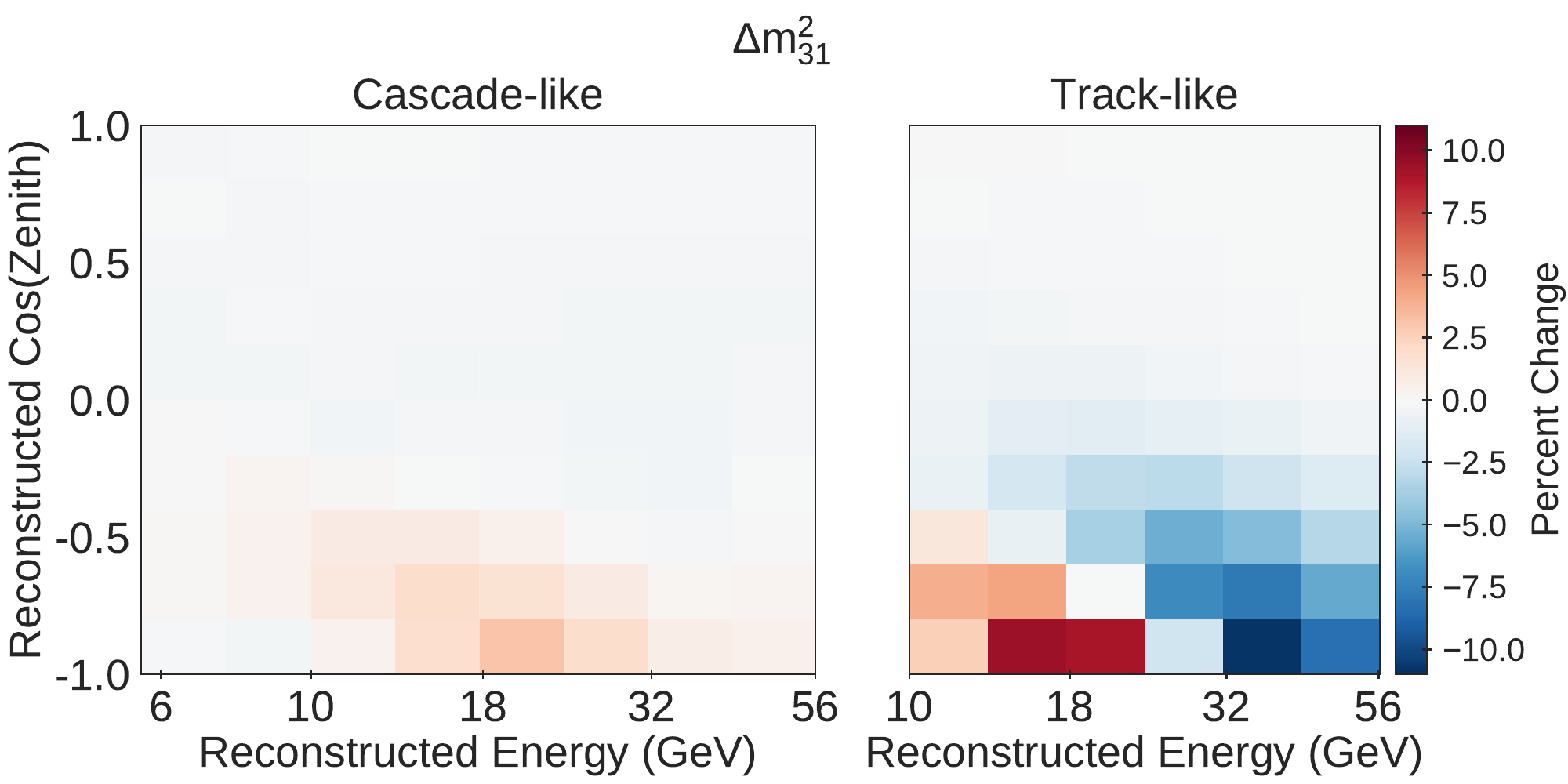}
\vspace*{2mm}
\includegraphics[width=\linewidth,height=4cm]{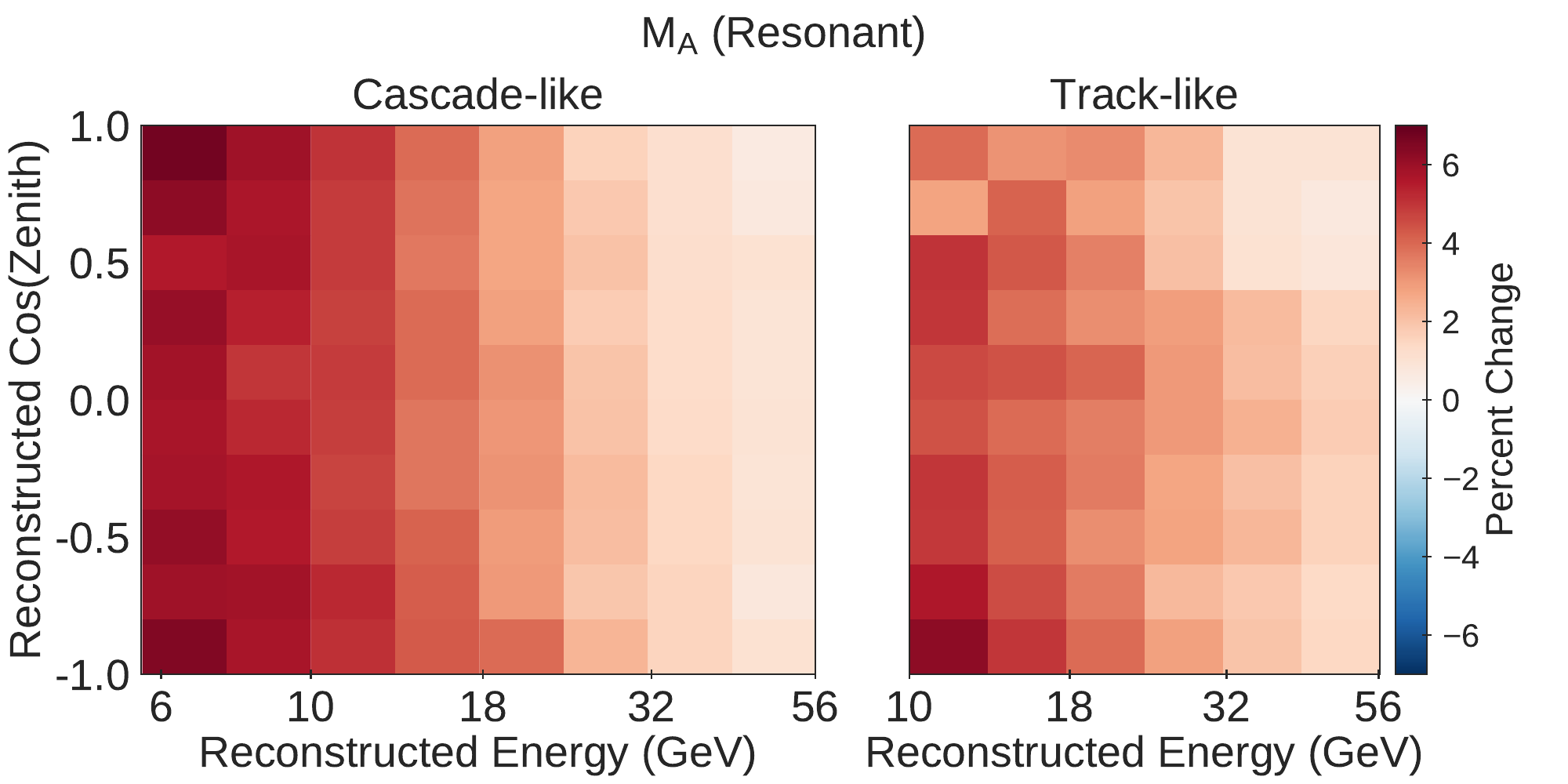}

\caption{Effect of selected systematic uncertainties on the nominal event distribution shown as a percentage change of the expectation per bin. With cascade-like events on the left and track-like events on the right, shown from top to bottom are: $\nu_e/\nu_\mu$ flux ratio at $+1\sigma$, $\nu/\bar{\nu}$ flux ratio at $+1\sigma$, head-on optical efficiency at $+1$, $\Delta m^2_{32}$ at $2.778 \times 10^{-3}$\, eV$^2$ instead of $2.526\times 10^{-3}$\,eV$^2$, and $M_A^{res}$ at $+1\sigma$.  
(See text for definitions of these parameters.)
A complete collection of plots is provided in Appendix~\ref{app:systematics}.
}
\label{fig:syst_effect_on_event}
\end{figure}

The nominal value used for $M_A^{CCQE}$ is 0.99\,GeV, with an uncertainty of $(-0.1485, +0.2475)$\,GeV used as a prior; for $M_A^{res}$ we used 1.12\,GeV and $\pm0.22$\,GeV.
These are the same values used as GENIE's default model and re-weighting scheme, respectively.
The last row of Fig.~\ref{fig:syst_effect_on_event} shows that the impact of the $M_A^{res}$ uncertainty on the event distribution is energy dependent, with the largest impact at lower energies where the majority of resonant events are expected.
The effect of $M_A^{CCQE}$ follows a similar shape with even smaller changes, as the quasi-elastic events are peaked at lower energies.
The axial mass uncertainties have little impact as a function of $\cos{\theta_\nu}$.

Measurements of the $\nutau$ cross section exist from only a few experiments, with DONUT providing a ratio of $\sigma(\nutau)/\sigma(\nu_{e,\mu})$ of $1.37\pm0.35\pm0.77$~\cite{Kodama:2007aa}.
Uncertainties on the $\nutau$ CC cross section in the energy range of interest in this analysis~\cite{Jeong:2010nt} differ primarily by a factor degenerate with the $\nutau$ CC normalization tested.
Indeed, this degeneracy is used by SK to reinterpret their best-fit $\nutau$ normalization as a modification to the $\nutau$ neutrino CC cross section~\cite{Li:2017dbe}.
Due to this degeneracy, we do not include any nuisance parameters specifically modifying the $\nutau$ CC cross section.

\subsection{Oscillation Parameters}
\label{sec:osc}

The model in this analysis assumes three-flavor oscillations and hence relies on three mixing angles, two mass-squared splittings, and a CP violating phase.
We use the Prob3++~\cite{prob3pp} software which incorporates matter effects for full three-flavor oscillations calculations.
The earth is approximated with 12 radial layers of constant density~\cite{Dziewonski:1981x}.
For earth crossing neutrinos, matter effects start to significantly alter the $\nue\leftrightarrow\numu$ transition probabilities only at energies of around 6~GeV and below, hence the effect is very small for these analyses.

With atmospheric neutrinos we are not sensitive to the solar parameters, so we fix the mass splitting $\Delta m^2_{21}$ to $7.5\times 10^{-5}$\,eV$^2$ and the mixing angle $\theta_{12}$ to $33.48^\circ$.
The reactor angle $\theta_{13}$ is treated as a systematic uncertainty in Analysis \dragon and is assigned a Gaussian prior with a central value of $8.5^{\circ}$ and an uncertainty of $\pm0.21^\circ$.
All of the above values are taken from~\cite{Gonzalez-Garcia:2014bfa}.

No prior constraints are used for the two atmospheric parameters $\Delta m^2_{31}$ and $\theta_{23}$ which vary freely in the fit.
Since this analysis is insensitive to $\delta_{CP}$ it is fixed to $0^{\circ}$.
Also, since the neutrino mass ordering is not yet known, we check both normal and inverted orderings in the fit and accept the one yielding the better likelihood.
To avoid any bias in the fitted value of $\theta_{23}$, we fit its value in both octants ($\sin^2{\theta_{23}} <0.5$ and $>0.5$) and accept the value yielding the maximum likelihood.



\begin{table*}[tbh]
\centering
\caption{Nuisance parameters along with their associated priors where applicable and the best fit values from Analysis~\greco when fitting the charged and neutral current \nutau\ normalization combined (NC+CC) and the charged current alone (CC), and the same for Analysis~\dragon.
Priors are given as central value together with the $\pm1\sigma$ ranges when a Gaussian prior is imposed, while "-" denotes that no external prior constraint (i.e. flat prior) is used.
}
\label{tab:syst}
\begin{tabular}{lc|cc|cc}
                                                          &                               &  \multicolumn{2}{c|}{Analysis \greco}              &  \multicolumn{2}{c}{Analysis \dragon}  \\
Parameter                                                 & Prior                         &  \begin{tabular}[c]{@{}c@{}}Best fit\\ (CC+NC)\end{tabular}   &  \begin{tabular}[c]{@{}c@{}}Best fit \\ (CC)\end{tabular}  &  \begin{tabular}[c]{@{}c@{}}Best fit\\ (CC+NC)\end{tabular}   &  \begin{tabular}[c]{@{}c@{}}Best fit \\ (CC)\end{tabular}         \\
\hline
\multicolumn{2}{l|}{\textbf{Neutrino Flux \& Cross Section:}}                                      & & & & \\
$\nu_e/\nu_\mu$ Ratio                                     & $1.0\pm 0.05$                 & 1.03              & 1.03              & 1.03               & 1.03     \\
\nue\ Up/Hor. Flux Ratio ($\sigma$)                       & $0.0 \pm 1.0$                 & $-0.19$             & $-0.18$              & $-0.25$              & $-0.24$    \\
$\nu/\bar{\nu}$ Ratio ($\sigma$)                          & $0.0 \pm 1.0$                 & $-0.42$              & $-0.33$               & 0.01               & 0.04     \\
$\Delta\gamma_\nu$ (Spectral Index)                     & $0.0 \pm 0.1$                 & 0.03              & 0.03               & $-0.05$              & $-0.04$   \\
Effective Livetime (years)                                & -                             & 2.21              & 2.24               & 2.45               & 2.46     \\
$M_A^{CCQE}$ (Quasi-Elastic) (GeV)                               & $0.99 ^{+ 0.248}_{- 0.149}$   & 1.05              & 1.05               & 0.88               & 0.88     \\
$M_A^{res}$ (Resonance) (GeV)                                   & $1.12 \pm 0.22$               & 1.00              & 0.99               & 0.85               & 0.85     \\
NC Normalization                                          & $1.0 \pm 0.2$                 & 1.05              & 1.06               & 1.25               & 1.26     \\
&&&&&\\
\multicolumn{2}{l|}{\textbf{Oscillation:}}                                      & & & & \\
$\theta_{13}$ ($^\circ$)                                  & $8.5 \pm 0.21$                & -                 & -                  & 8.5                & 8.5      \\
$\theta_{23}$ ($^\circ$)                                  & -                             & 49.8              & 50.2               & 46.1               & 45.9     \\
$\Delta m^2_{32}$ ($10^{-3}\rm{eV}^2$)                  & -                             & 2.53              & 2.56               & 2.38               & 2.34     \\
&&&&&\\
\multicolumn{2}{l|}{\textbf{Detector:}}                                         & & & & \\
Optical Eff., Overall (\%)                                & $100 \pm 10$                  & 98.4              & 98.4               & 105                & 104      \\
Optical Eff., Lateral ($\sigma$)                          & $0.0\pm1.0$                   & 0.49              & 0.48               & $-0.25$              & $-0.27$    \\
Optical Eff., Head-on (a.u.)                              & -                             & $-0.63$             & $-0.64$              & $-1.15$              & $-1.22$    \\
Local Ice Model                                           & -                             & -                 & -                  & 0.02               & 0.07     \\
Bulk Ice, Scattering (\%)                                 & $100.0 \pm 10$                & 103.0             & 102.8              & 97.4               & 97.3     \\
Bulk Ice, Absorption (\%)                                 & $100.0 \pm 10$                & 101.5             & 101.7              & 102.1                & 101.9      \\
&&&&&\\
\multicolumn{2}{l|}{\textbf{Atmospheric Muons:}}                                      & & & & \\
Atm. $\mu$ Fraction (\%)                                  & -                             & 8.1               & 8.0                & 4.6                & 4.6      \\
$\Delta\gamma_\mu$ ($\mu$ Spectral Index, $\sigma$)       & $0.0\pm1.0$                       & 0.15              & 0.15               & -                  & -        \\
Coincident $\nu+\mu$ Fraction                             & $0.0+0.1$                     & 0.01              & 0.01               & -                  & -        \\
&&&&&\\
\multicolumn{2}{l|}{\textbf{Measurement:}}                                      & & & & \\
$\nu_\tau$ Normalization                                & -                             & 0.73              & 0.57               & 0.59               & 0.43     \\ %
\end{tabular}
\end{table*}

\subsection{Detector Uncertainties}
\label{sec:holeice}

\begin{figure}[htbp]
    \includegraphics[width=0.8\linewidth]{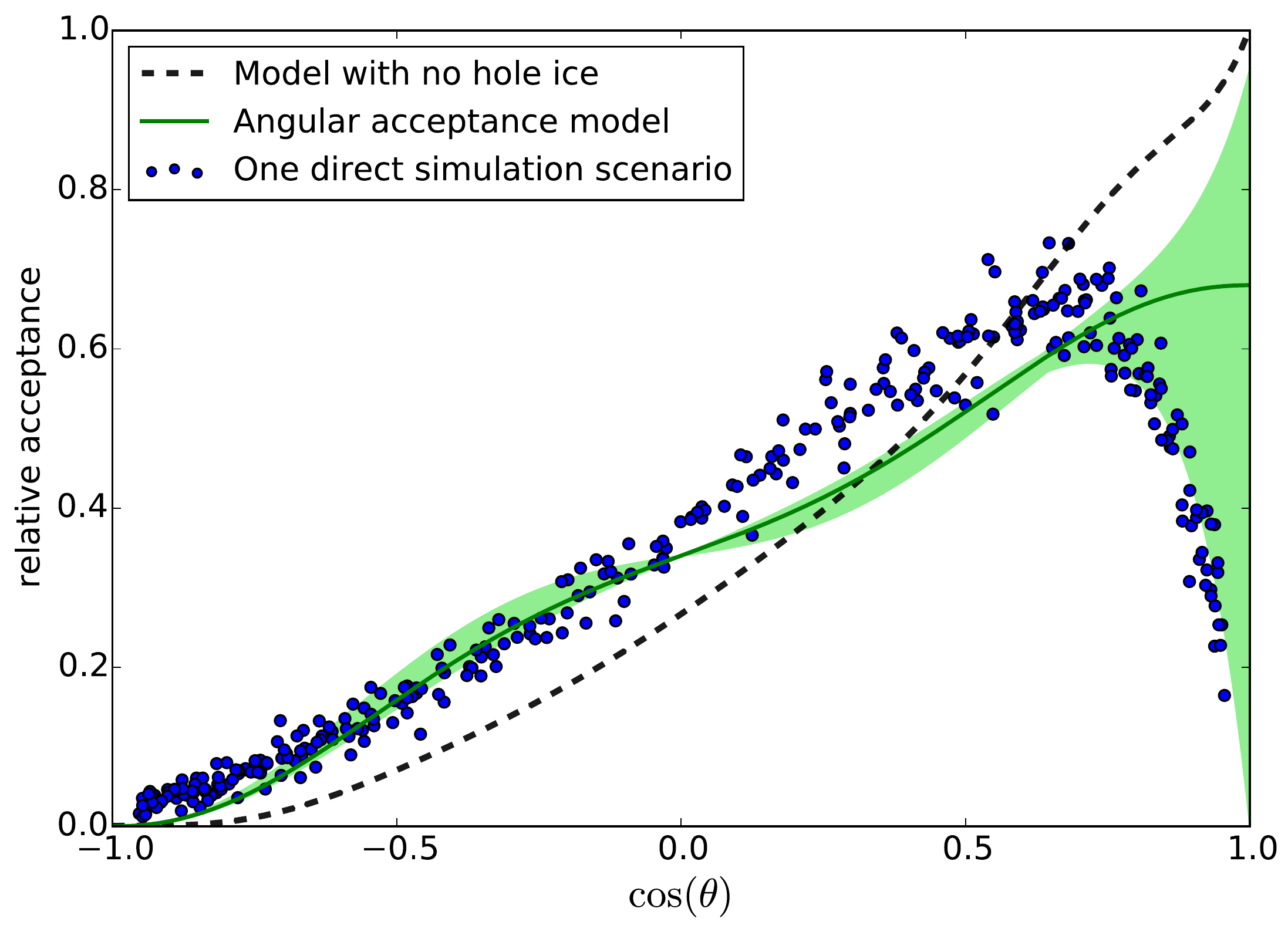}
    \caption{Relative acceptance of photons versus photon arrival angle for different optical models of the ice.  Zenith angles $\theta$ with $\cos{\theta} =-1.0$ indicate vertically downward-going photons (hitting the top of a DOM), $\cos{\theta} = 0.0$ horizontal photons, and $\cos{\theta} = -1.0$ vertically upward-going photons.    The black line shows the angular photon sensitivity of a module as measured in the laboratory.
    The green line and surrounding green band show the angular acceptance used and its uncertainty, respectively, and is based on two parameters, lateral and head-on sensitivity. The head-on area has a large associated uncertainty.
    Data points obtained from the direct simulation of a bubble column (not based on angular acceptance) are overlaid in blue.
    }
    \label{fig:holeice}
\end{figure}

Systematic uncertainties related to the response of the detector itself play an important role in the analyses.
The impact of these uncertainties is complex, depending upon the properties of the detector, on the impacts in event selection, and on the effect in the reconstruction used to estimate particle properties.
In order to account for these complexities, separate simulations for different settings of the detector response were produced and propagated through each step of the event selection and reconstruction as described in Sec.~\ref{sec:sample}.
Each simulation set includes a change to at least one detector uncertainty parameter.
The change in the number of expected events for each of the analysis bins relative to the baseline simulation set is used to estimate the effective impact of each systematic uncertainty for each simulated discrete point of parameter settings.

To arrive at a continuous description, the effects are approximated using a function with linear dependencies on the nuisance parameters. For $N$ linear parameters, we use $N$-dimensional ``hyperplanes" as given in the following equation for each bin $k$ in the analysis histogram:
\begin{equation}
    f^{\nu}_k(p_1, p_2,..., p_N) = \sum_{i=1}^N{a_{ik} p_i} + b_k,
\end{equation}
with the nuisance parameters $p_i$, the fitted hyperplane slopes $a_i$, and the common offset $b$. Thus for $N$ parameters $N+1$ values are fitted.
Such parameterizations are obtained independently for every analysis bin, separately for each of the three neutrino flavors in CC interactions, and combined for all NC interactions. These relative changes of event rates are then applied as scale factors to the event weights during the analysis.

In Analysis \greco, detector response uncertainties of simulated atmospheric muons are also parametrized in a similar way to the neutrino uncertainties.
Variations in the overall efficiency of the optical modules and the absorption result in particularly strong changes in the observed light yields in the veto region from muon tracks, leading to large changes in the atmospheric muon event rates after selection.
These simulated muon rates are not well-modeled with linear parametrizations. 
In these two cases, an exponential form is instead used, giving the form for each bin $k$ as
\begin{equation}
    f^{\mu}_k(p_1, p_2,..., p_{N+M}) = \sum_{i=1}^N{a_{ik} p_i} + \sum_{j=1}^M{a_{jk} e^{-b_{jk} p_j}} + c_k
\end{equation}
where $N$ parameters describe the lateral and head-on optical efficiency as well as the scattering of the glacial ice and $M$ parameters cover the overall efficiency and the absorption.

The values $f_k$ give the fractional change for each histogram bin given the values of the detector nuisance parameters $\vec{p}$.
This is applied as a multiplicative reweighting factor for each bin of the analysis histogram.

Both analyses incorporate six nuisance parameters to account for detector uncertainties. 
Each nuisance parameter is modeled by 2--5 additional simulation sets for each neutrino flavor and, in the case of Analysis~\greco, atmospheric muons.
Using the obtained parametrizations, we obtain an average $\chi^2$/expected degrees of freedom, per flavor and bin, of 13.1/13 across the included neutrino simulation sets and 6.0/6 for background muon sets in Analysis \greco.
Similarly, a $\chi^2$ distribution with 24.0/25 degrees of freedom is obtained from the neutrino simulation sets in Analysis \dragon.

The transparency of the ice in our fiducial volume was calibrated using remotely-controlled light-emitting diodes (LEDs) inside every deployed DOM. The optical properties affect the light yield and temporal arrival distributions of photons that are produced from events seen by the DOMs.
The parameters in the model--scattering and absorption coefficients as a function of depth--were determined as a function of location within the detector as described in~\cite{Aartsen:2013rt, spice}.
Both coefficients have associated uncertainties of $\pm10\%$ and are included as systematic uncertainties in this measurement. Additional MC sets were produced with enhanced scattering (+10\%), enhanced absorption (+10\%), and diminished scattering and absorption ($-7$\%, $-7$\%) to estimate the effects.

The overall photon detection efficiency of the IceCube DOMs depends on both individual PMTs as well as properties of the glass housing and nearby cables. 
Dedicated measurements of the efficiency of the DOMs yield a relative uncertainty of 10\%~\cite{Aartsen:2016nxy}.
This effect is modeled by changing the light collection efficiency of the DOMs in simulation, with the efficiency of all modules scaled simultaneously by a common factor.
Simulated data sets ranging from 88\% to 112\% of the nominal optical efficiency were used to parametrize the effect of the DOM efficiency uncertainty and a Gaussian prior with a width of 10\% was applied to the overall photon collection efficiency for these analyses.

In addition to modifying the absolute efficiency, any bubbles in the refrozen ice in the borehole  (``hole ice'') near the DOMs can cause increased scattering of Cherenkov photons. 
The effect of the refrozen ice column is modeled by two effective parameters controlling the shape of the DOM angular acceptance curve (see Fig.~\ref{fig:holeice}).
The {\it lateral} parameter controls the relative sensitivity between photons traveling roughly 20$^\circ$ above and below the horizontal.
The uncertainty on this parameter is constrained by LED calibration data\cite{Aartsen:2013rt}.

Simulated data sets were generated covering the $\pm1\sigma$ uncertainty range and a Gaussian prior based on the calibration data is used for this parameter.
The {\it head-on} parameter modulates the sensitivity for photons traveling upwards and arriving near the DOM's lower face. 
This is a region that is poorly constrained by the string-to-string LED calibration because no bright, upward-pointing LEDs were deployed.
To account for this uncertainty, the acceptance curve is altered using a dimensionless parameter ranging from $-5$ (corresponding to a bubble column completely obscuring the DOM's lower face for vertically incident photons) to $2$ (no obscuration).  
Simulated data sets covering the range from $-5$ to 2 were used to parameterize this effect.  
No prior is imposed on this parameter due to lack of information from calibration data.  
Modelling the hole ice via the angular acceptance curve is an approximation, as it only truly holds in the far field. 
In addition, it can only model hole ice radii significantly larger than the DOM radius as no azimuthal dependence is incorporated.

An additional model of the hole ice has also been tested in Analysis \dragon, incorporating an explicit simulation of the bubble column consisting of ice with enhanced scattering located in the refrozen holes~\cite{POCAM}. 
In addition, photons arriving at a DOM are not accepted based on their incident angle, but by requiring that they impact the DOM's lower hemisphere. 
Although in principle more realistic than the angular acceptance model, the tuning of all parameters involved in such a simulation is a challenge.
Various MC sets for a range of different settings (optical properties of bubble column ice, column radius) and using the best knowledge of the position of the column with respect to each DOM were produced.
For comparison, the fraction of photons arriving at several DOMs as a function of the arrival direction is also shown in Fig.~\ref{fig:holeice}.
A fourth parameter (the local ice model) is introduced in Analysis~\dragon to account for differences not covered by the angular acceptance model.
A value of zero corresponds to the purely angular acceptance base simulations, while a value of one is assigned to the explicit bubble column simulations.
Since this model is disfavoured by the data of Analysis \dragon, Analysis \greco only incorporates the angular acceptance model.

\subsection{Atmospheric Muon Uncertainties}
\label{sec:atmmu}

The last nuisance parameter pertains to the amount of atmospheric muon contamination in the final data sample, where Analysis \GRECO is based on Monte Carlo simulation and parameterizations while Analysis \DRAGON is data-driven. For Analysis \GRECO, uncertainties due to atmospheric muons flux include the uncertainties associated with the cosmic ray spectral index in Section~\ref{sec:muflux} based on~\cite{Evans:2016obt}. Additional uncertainties due to detector response are treated the same way as the case of neutrinos, where additional sets are produced and a hyperplane fit is performed per bin. 

\begin{figure}[htbp]
    \centering
    \includegraphics[width=\linewidth]{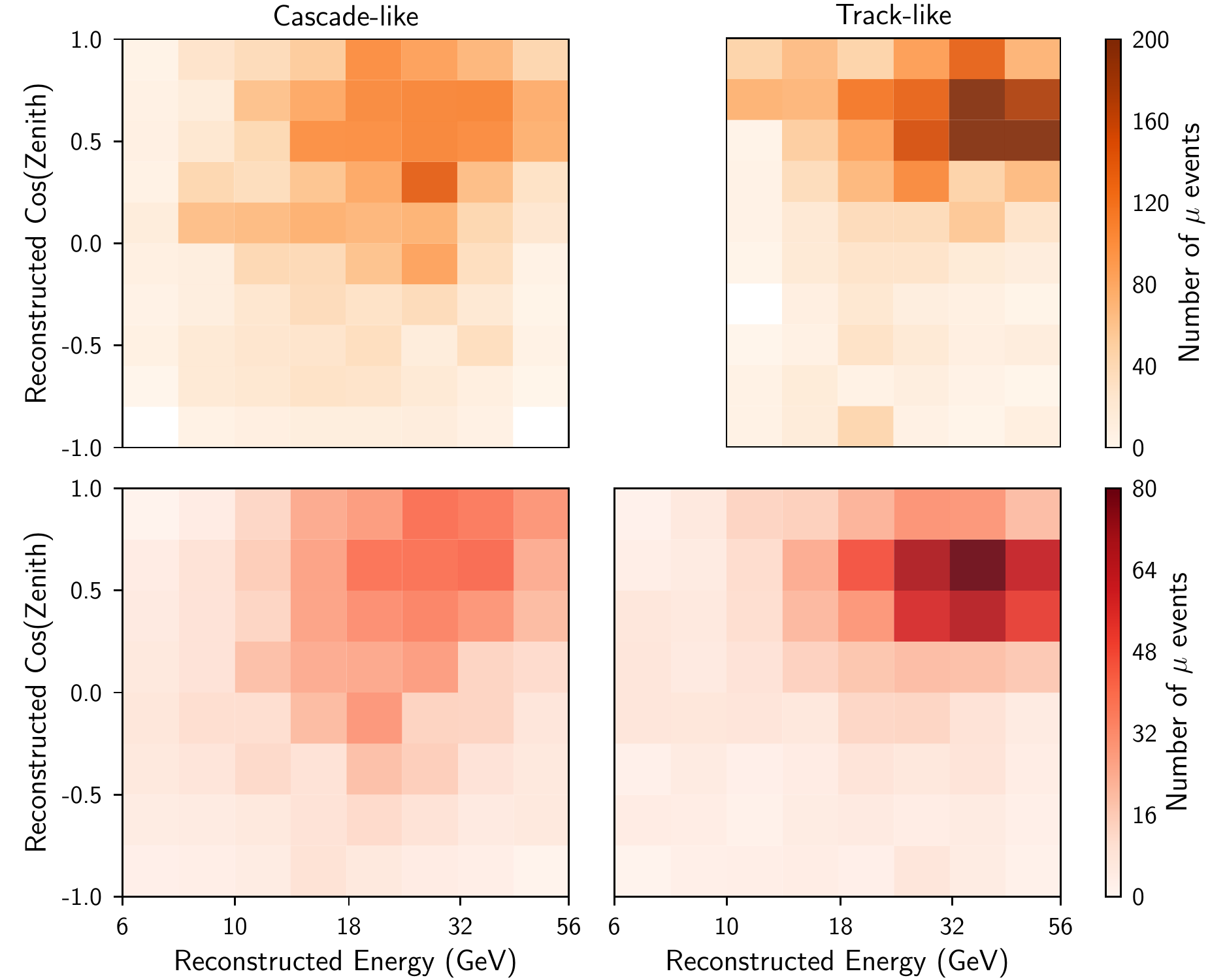}
    \caption{Event distributions of the atmospheric muon background for Analysis \greco (top row) obtained from the best-fit simulation, and for Analysis \dragon (bottom row) obtained from the data sideband.}
    \label{fig:icc_background}
\end{figure}

For Analysis \DRAGON, a data-driven method is used to estimate the shape of this background as described in Section~\ref{sec:selection} (see Fig.~\ref{fig:icc_background}).
With the absolute efficiency for tagging background events not amenable to direct measurement, the normalization of the muon contribution is left unconstrained in the fit.  
Its nominal value is set to match the expected rate from simulated atmospheric neutrinos, and error terms are calculated with respect to this nominal value.
In addition, we account for uncertainties in these background templates arising from shape changes when modifying the selection cuts. 
Two samples are obtained by requiring more than one hit and more than two hits in the muon veto regions, with the latter being a more muon-rich sample.
The difference in shape between the two (ignoring normalization differences) is added in quadrature, together with the limited statistics term, to the uncorrelated uncertainties $\sigma^{\text{exp}}$ in Eq.~\ref{eq:chi2}.
The output shape and uncertainty are in agreement with muon simulations.

\section{Confirmatory Measurement of Atmospheric Neutrino Oscillation Parameters}
\label{sec:greconumu}

Under the assumption of a unitary PMNS mixing matrix, the atmospheric neutrino oscillation parameters $\Delta$m$_{23}^2$ and sin$^2 \theta_{23}$ are measured as a cross-check of the validity of Analysis \GRECO presented earlier. 
With the $\nutau$ normalization fixed to 1, all sources of systematic uncertainties listed in Table~\ref{tab:syst} are taken into account. 
With 140 non-zero bins and 133 effective degrees of freedom, a $\chi^2$ defined in Eq.~\ref{eq:chi2} of 129.4 is obtained when letting all 16 nuisance and two oscillation parameters float. 
The best fit values of $\Delta$m$^2_{23}$ and sin$^2 \theta_{23}$ are $2.55^{+0.12}_{-0.11} \times 10^{-3}$ eV$^2$ and $0.58^{+0.04}_{-0.13}$, respectively.

\begin{figure}[htbp]
    \includegraphics[width=\linewidth]{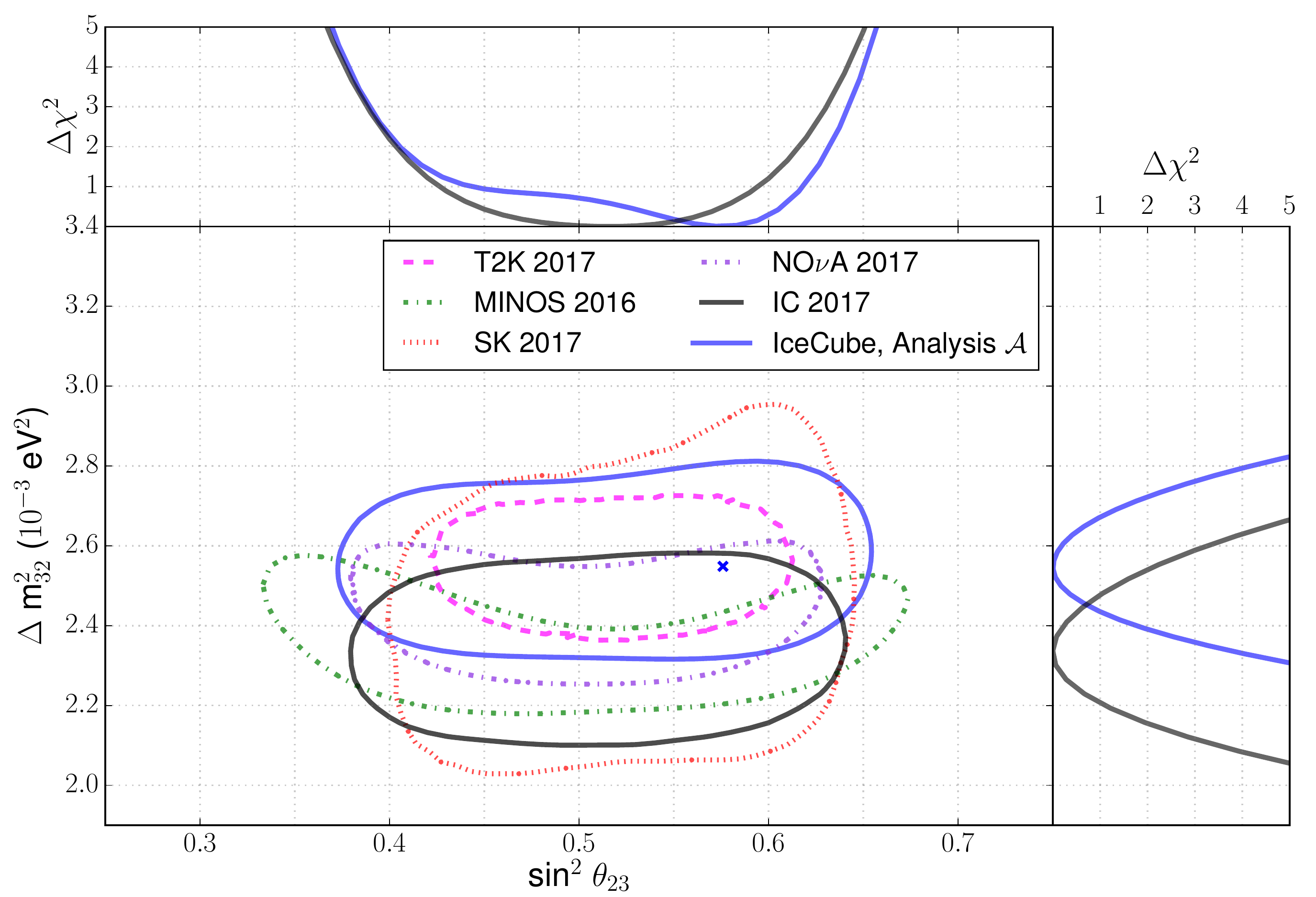}
    \caption{The 90\% allowed region using the data sample from Analysis \greco in blue compared to other experiments\cite{Singh:2017thk,Haegel:2017ofz,Whitehead:2016xud,Wendell:2014dka,Aartsen:2017nmd}. The best fit point from Analysis \greco is shown as the blue cross mark. The IceCube 2017 result~\cite{Aartsen:2017nmd}, represented in black, uses the data sample from Analysis \dragon. The top and right plots are the 1-d $\Delta \chi^2$ profiles of the measured oscillation parameters.
    }
    \label{fig:greconumu}
\end{figure}

Previous measurements of the atmospheric neutrino oscillation parameters have been performed using the IceCube detector, including a measurement of atmospheric muon neutrino disappearance performed using the event sample from Analysis~\DRAGON, here referred to as ``IC2017''~\cite{Aartsen:2017nmd}.
The IC2017 analysis included a subset of systematic uncertainties from Analysis~\DRAGON found to be significant for the disappearance measurement.
Detector systematics related to the optical efficiency were included, but used a different parametrization of the detector uncertainties than that described in Sec.~\ref{sec:holeice}.

Figure~\ref{fig:greconumu} shows the 90\% allowed region of atmospheric neutrino oscillation parameters for the analyses based on Analysis~\greco and IC2017, and the allowed regions reported by other experiments.
Overall, the 90\% allowed regions from Analysis \greco and IC2017 are statistically consistent, and both results compare favorably with the latest published 90\% contours from other neutrino experiments~\cite{Singh:2017thk,Haegel:2017ofz,Whitehead:2016xud,Wendell:2014dka}.

The shift between the two IceCube contours in both $\Delta$m$^2_{23}$ and $\sin^2 \theta_{23}$ can be explained by statistical fluctuations alone, as 65\% of events in Analysis~\greco are unique with respect to Analysis~\dragon (while 48\% of events in Analysis~\dragon are unique).
Furthermore, detailed investigation showed that differences in the analyses--namely differences in the parametrization of detector effects (hyperplane), inclusion of bulk ice uncertainties, and the differences in the event selection and reconstruction as described in Sec.~\ref{sec:sample}--can lead to small ($<0.5\sigma$) systematic shifts in the result as well.

Separate analyses based on the same event samples as presented here, but testing the neutrino mass orderings, were performed using only up-going events. Results including best-fit oscillation parameters are reported in \cite{NMO}, and are also compatible with the values reported in this section.
\section{Results and Conclusion}
\label{sec:results}

\begin{figure}[htbp]
    \includegraphics[width=\linewidth]{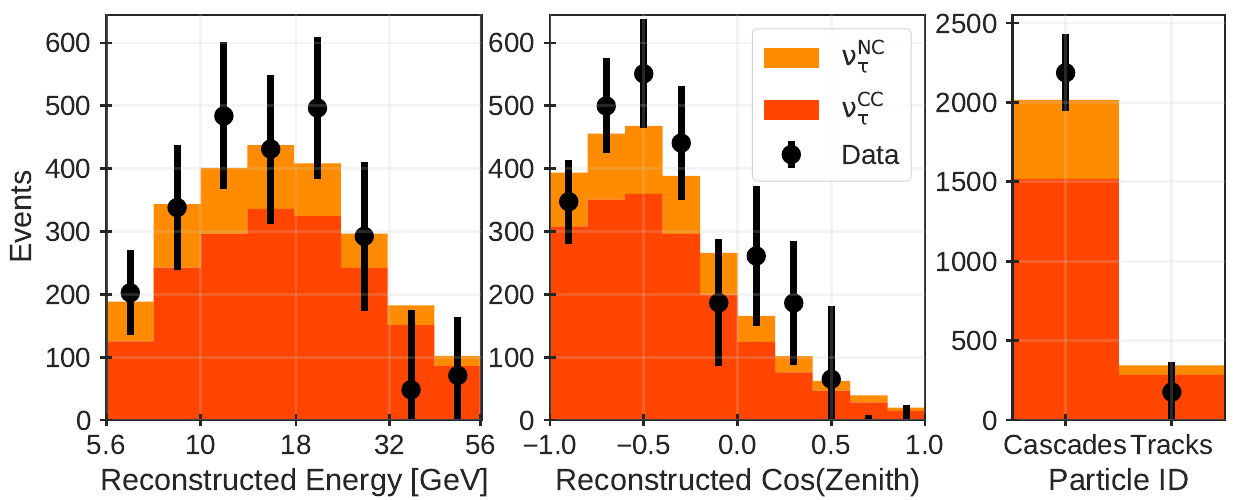}
    \caption{
    Distributions of the data with best-fit neutrino and muon backgrounds subtracted and signal simulation. Statistical errors are shown for the data. The best-fit hypothesis shows good agreement in the reconstructed energy axis (left), the cosine of the reconstructed zenith angle (middle) and PID categories (right) for Analysis \greco.
}
    \label{fig:E_plot}
\end{figure}

\begin{figure}[htbp]
    \includegraphics[width=\linewidth]{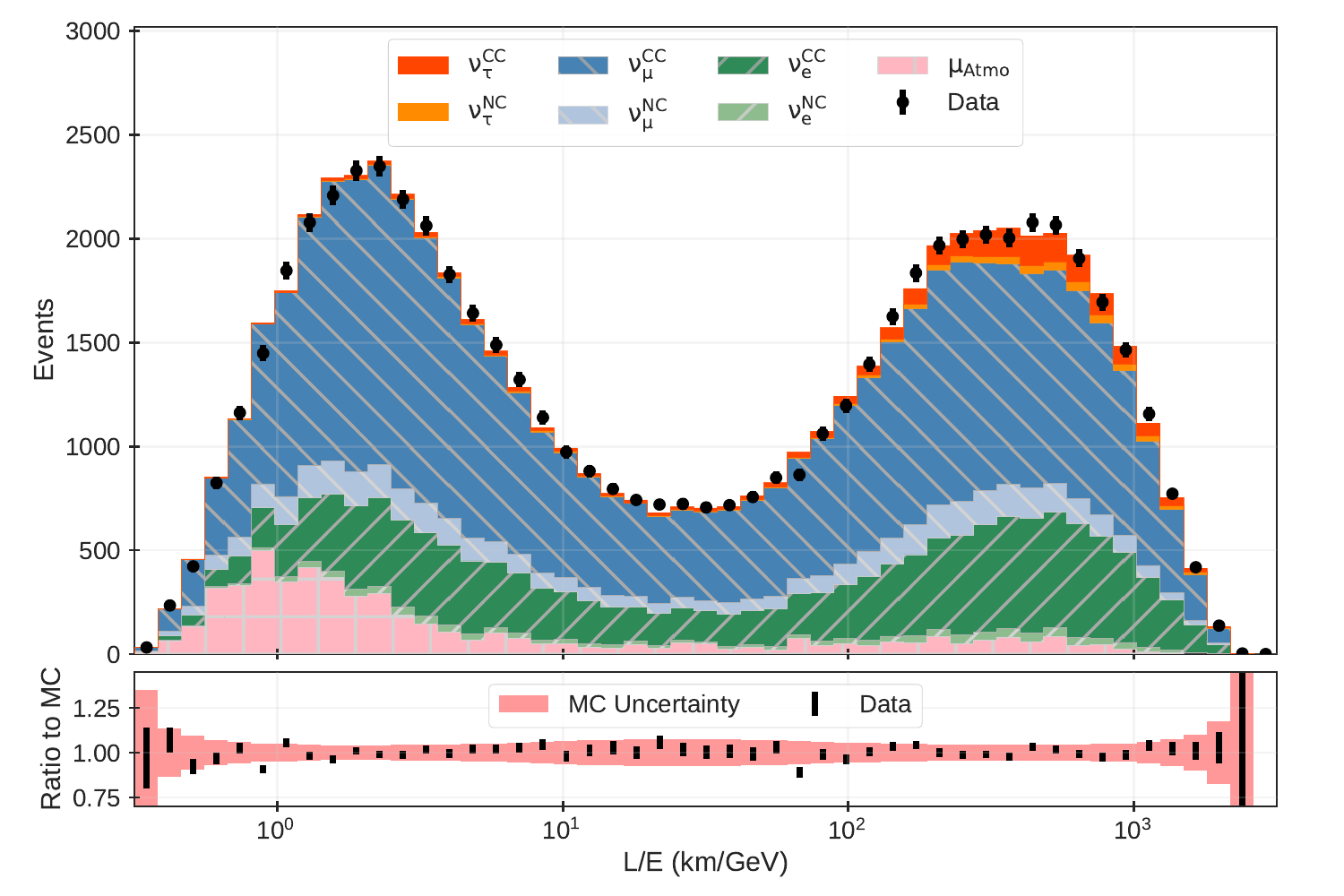}
    \includegraphics[width=\linewidth]{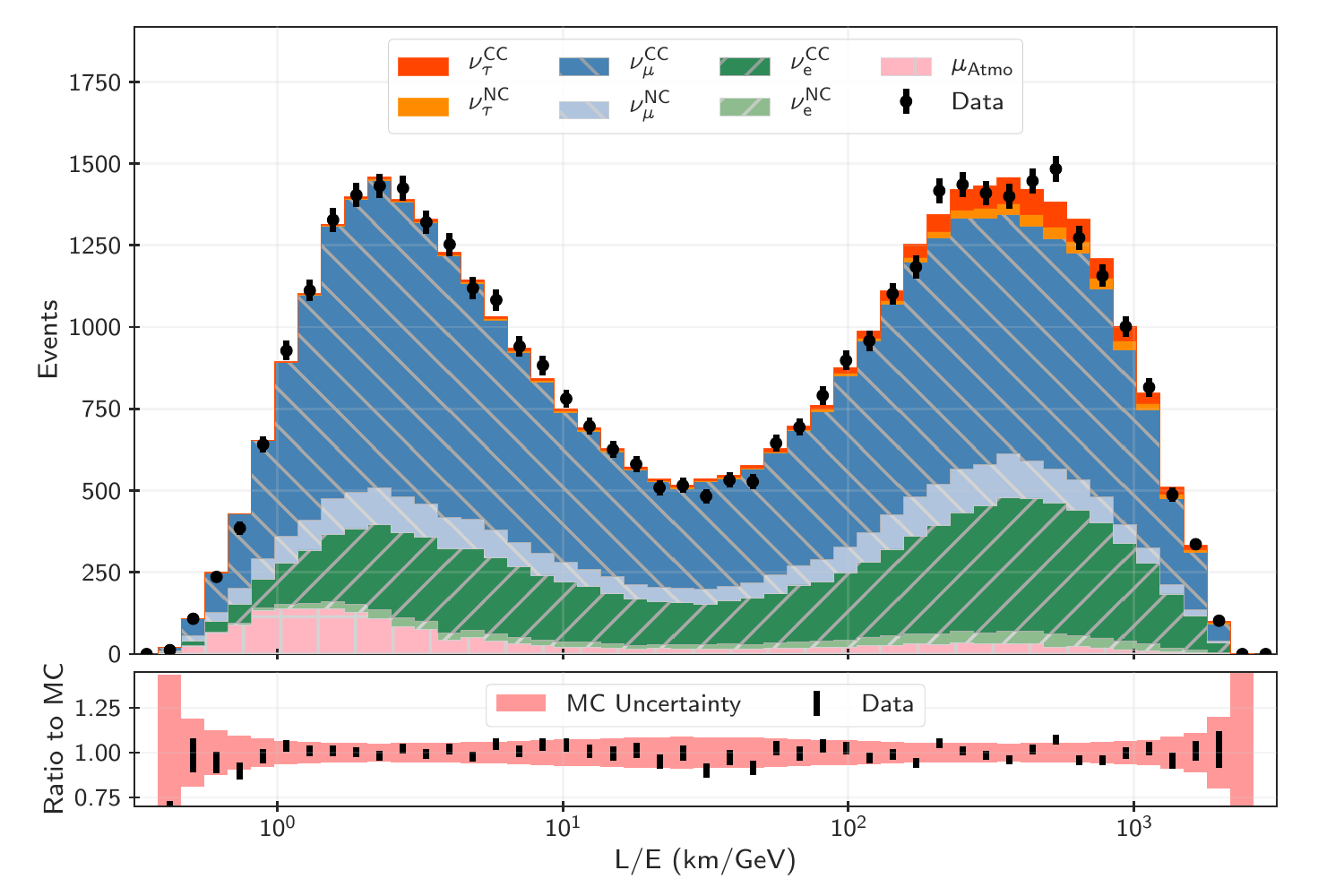}
    \caption{Distribution of the data as a function of reconstructed $L/E$, overlaid with the best fit neutrino and cosmic-ray muon histograms for Analysis \greco (top) and \dragon (bottom).
    The bottom portion of each shows the ratio of the data to the predicted distribution at the best fit point, with black points representing data and the height of the shaded band the uncertainty of the best fit (statistical errors only).
    }
    \label{fig:LE_plot}
\end{figure}

The distributions of the reconstructed neutrino energy, the reconstructed zenith angle, and the event class for the best fit tau neutrino hypothesis for Analysis \greco are shown in Fig.~\ref{fig:E_plot}, overlaid with background-subtracted data, {\it i.e.,} cosmic-ray muons and all non-\nutau\ neutrinos subtracted. Figure~\ref{fig:LE_plot} shows all events projected onto the $L/E$ axis for the best fit expectations overlaid with the observed data for both analyses separately. The excellent agreement of the model with the data can be seen qualitatively in the figure. Using the actual measurement bins and setting all parameters to their best fit values, the model agrees well quantitatively in Analysis \greco (\dragon) with the observed data with a total $\chi^2$ of 127.6 (113.3), corresponding to a p-value of 55\% (20.3\%), estimated via pseudo-data trials. The corresponding values for the nuisance parameters can be found in Table~\ref{tab:syst}.

\begin{figure}[tbp]
    \includegraphics[width=\linewidth]{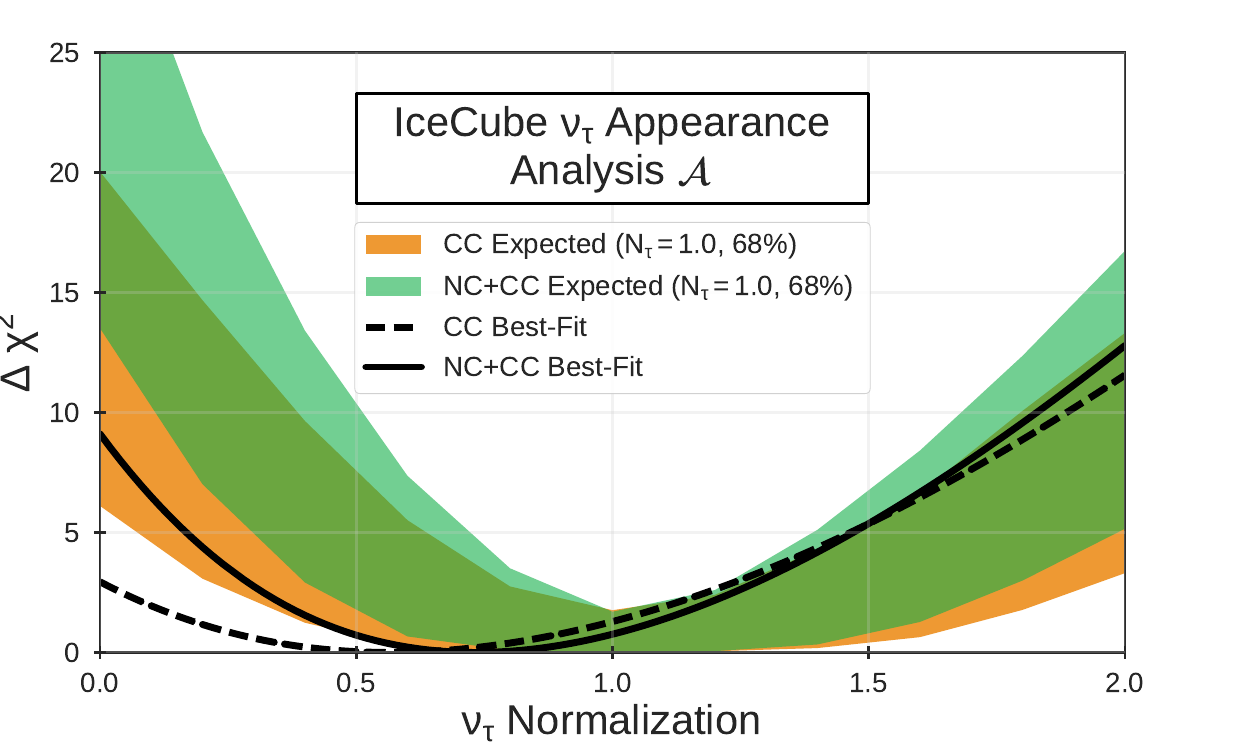}
    \caption{Observed $\Delta\chi^2$ from the best fit CC+NC (CC) $\nu_\tau$ normalization of 0.75 (0.62) as a function of the $\nutau$ normalization (black lines). Shaded bands show the 68\% ranges of the expected distribution of $\Delta\chi^2$ values obtained from pseudo-experiments assuming nominal values for oscillation parameters and a $\nutau$ normalization of 1.0.}
    \label{fig:brazil}
\end{figure}

Figure~\ref{fig:brazil} shows the expected and observed $\Delta\chi^2$ values for a \nutau\ normalization ranging from 0 to 2.0.
The band of expected values assumes standard oscillations with a \nutau\ normalization of 1.0.
Our main result for the CC+NC measurement has a best fit value of 0.73 with the 68\% confidence interval (C.I.) covering the range $(0.49, 1.07)$ and the 90\% C.I.\@ covering $(0.34, 1.30)$.
For the CC-only normalization, we observe the best fit at 0.57 with the 68\% C.I.\@ $(0.30, 0.98)$ and the 90\% C.I.\@ $(0.11, 1.25)$.

\begin{figure}[tbp]
    \includegraphics[width=\linewidth]{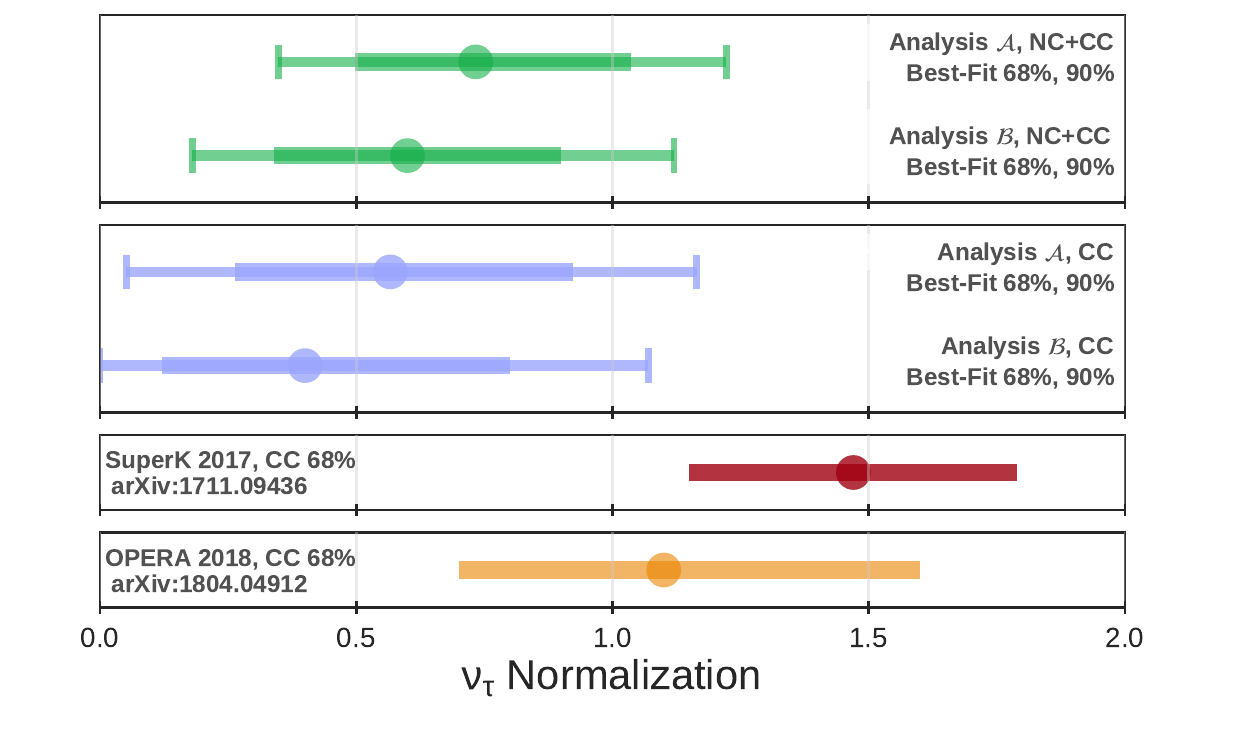}
    \caption{The measured values for CC+NC and CC-only results in both analyses. Also shown are previous best-fit values of the CC-only $\nutau$ normalization from OPERA and SK, which were performed with different energy ranges and fluxes and a different definition of the $\nutau$ normalization from those used in IceCube. All measurements of tau neutrinos are consistent with standard  oscillations ($\nutau$ normalization of 1.0), with the two analyses presented here showing excellent internal agreement.}
    \label{fig:intervals}
\end{figure}

These measured values are compatible with corresponding values obtained from Analysis \dragon within less than $1\sigma$ standard deviation.
These confirmatory results of Analysis \dragon are \DRAGONnorm (\DRAGONCCnorm) for the CC+NC (CC-only) measurement, also see Fig.~\ref{fig:intervals}.

All values are also compatible within the 90\% confidence interval with expectations assuming the three-flavor neutrino oscillation paradigm ({\it i.e.,} $\nu_\tau$ normalization = 1.0) and the assumed $\nutau$ CC cross sections.
The significance at which we can reject the null hypothesis of no \nutau\ appearance is 3.2~$\sigma$ and 2.1~$\sigma$ for the CC+NC and the CC-only case for Analysis \greco, respectively. The confirmatory Analysis \dragon yields slightly weaker limits of \DRAGONsignifCCNC~(\DRAGONsignifCC).

The confidence intervals for the measurements presented here, shown in Fig.~\ref{fig:intervals}, are calculated using the approach of Feldman and Cousins~\cite{Feldman:1997qc} to ensure proper coverage.

The presented results are of a comparable precision to those of SK and OPERA (see Fig.~\ref{fig:intervals}), and complementary to those measurements in terms of energy scale, $L/E$ range, systematic uncertainties, and statistics. Specifically, the SK measurement is based on lower-energy events where roughly 50\% interact via CC quasi-elastic or resonant scattering, while the IceCube data are dominated by higher-energy events that interact primarily via the deep inelastic scattering interaction and are thus subject to different sources of neutrino interaction uncertainties~\cite{Formaggio:2013kya}. Additionally, the event samples used here are considerably larger than both OPERA and SK, with an estimated 1804 CC and 556 NC $\nutau$ events for the final sample in Analysis \greco and 934 CC and 445 NC $\nutau$ events in the final sample in Analysis \dragon.

Determining the impact on tests of PMNS matrix unitarity requires global fits incorporating results from other experiments, as our result is only sensitive to the two elements $U_{\mu3}$ and $U_{\tau3}$ of the matrix, while unitarity tests involve elements from a full row or column of the matrix.  Also, as noted earlier, one could also use the measured $\nutau$ normalization reported here along with the previously reported results from OPERA and SK to better constrain the CC $\nutau$ cross section.

The measurement is limited by systematic uncertainties, in particular uncertainties in the initial flux of atmospheric neutrinos and uncertainties in our detector model.
Nevertheless, our result will improve with more statistics, as the aforementioned uncertainties are constrained by the data in the measurement itself---the increased sample size from more data allows us to control various detector effects and other sources of systematic uncertainties at a higher precision.

This defines a clear path forward towards a higher precision tau neutrino appearance measurement: more data, extended event selection and better control of detector uncertainties.
With ten years of DeepCore data we expect an analysis similar to the one presented here to attain a precision of 15\%.
Better reconstruction algorithms--currently under development--promise to improve the precision, as do approved detector upgrades~\cite{PINGU-LOI}.  The upgrades will include advanced calibration devices to improve our understanding of detector-related uncertainties, and the additional optical modules will be better and more efficient at identifying and reconstructing low energy neutrinos. These improvements will yield an anticipated precision of the tau neutrino normalization of better than 10\% with a single year of operation.

\begin{acknowledgments}
We acknowledge the support from the following agencies and institutions: 
USA -- U.S. National Science Foundation-Office of Polar Programs,
U.S. National Science Foundation-Physics Division,
Wisconsin Alumni Research Foundation,
Center for High Throughput Computing (CHTC) at the University of Wisconsin-Madison,
Open Science Grid (OSG),
Extreme Science and Engineering Discovery Environment (XSEDE),
U.S. Department of Energy-National Energy Research Scientific Computing Center,
Particle astrophysics research computing center at the University of Maryland,
Institute for Cyber-Enabled Research at Michigan State University,
and Astroparticle physics computational facility at Marquette University;
Belgium -- Funds for Scientific Research (FRS-FNRS and FWO),
FWO Odysseus and Big Science programmes,
and Belgian Federal Science Policy Office (Belspo);
Germany -- Bundesministerium f\"ur Bildung und Forschung (BMBF),
Deutsche Forschungsgemeinschaft (DFG),
Helmholtz Alliance for Astroparticle Physics (HAP),
Initiative and Networking Fund of the Helmholtz Association,
Deutsches Elektronen Synchrotron (DESY),
and High Performance Computing cluster of the RWTH Aachen;
Sweden -- Swedish Research Council,
Swedish Polar Research Secretariat,
Swedish National Infrastructure for Computing (SNIC),
and Knut and Alice Wallenberg Foundation;
Australia -- Australian Research Council;
Canada -- Natural Sciences and Engineering Research Council of Canada,
Calcul Qu\'ebec, Compute Ontario, Canada Foundation for Innovation, WestGrid, and Compute Canada;
Denmark -- Villum Fonden, Danish National Research Foundation (DNRF), Carlsberg Foundation;
New Zealand -- Marsden Fund;
Japan -- Japan Society for Promotion of Science (JSPS)
and Institute for Global Prominent Research (IGPR) of Chiba University;
Korea -- National Research Foundation of Korea (NRF);
Switzerland -- Swiss National Science Foundation (SNSF).

\end{acknowledgments}

\bibliographystyle{apsrev}
\bibliography{bib/osc}

\appendix
\label{app:appendices}
\section{Common Event Selection Variables}
\label{app:variables}
This section describes the technical details of the selection variables used in both Analyses \GRECO and \DRAGON. 

\subsection{Interaction Vertex}
\label{app:variables_vertex}
In IceCube coordinates ($x, y, z$), the DeepCore fiducial volume is centered on String 36 at $(x,y)=(x_{36},y_{36})$ in the middle of the detector 1950~m below the surface (see Fig.~\ref{fig:detector}). Along the $z$-axis, DeepCore DOMs are located between $-500$~m and $-150$~m, and the dust layer is between $-210$~m and $-135$~m. During an event selection, the radial position $\rho$, defined by
\begin{equation}
\rho=\sqrt{(x-x_{36})^2+(y-y_{36})^2},
\end{equation}
and depth position $z$ of the interaction vertex are often used. Both Analyses \GRECO and \DRAGON perform two simple guesses to roughly estimate the vertex position of an event without any fitting reconstruction algorithms.

The first guess is \textit{FirstHLC} which estimates the vertex position using the earliest hard local coincidence (HLC) hit DOM. Because an event is triggered when at least three HLC hits are recorded in the DeepCore fiducial volume, the $\rho$ and $z$ positions of the first HLC hit are likely to be near the interaction point.

The second method is \textit{VertexGuess}, which is the position of the first hit DOM in a cleaned hit series. For a neutrino signal event, the interaction happens in the DeepCore fiducial volume, whereas an atmospheric muon is expected to leave early hits in the veto region. Therefore, the variable \textit{VertexGuessZ} provides a quick guess at an early stage in the event selection process for the $z$ position of the interaction point.

\subsection{Charge Variables}
\label{app:variables_charge}
The event selection of Analyses \GRECO and \DRAGON use the following charge information from a cleaned hit series to identify downward-going atmospheric muon events.

First, \textit{NAbove200} is the integrated charge from hit DOMs above $z=-200$~m and $2~\mu$s before the DeepCore trigger. Compared to an upward-going neutrino, a downward-going atmospheric muon is more likely to hit DOMs in the upper part of the detector and deposit a larger total charge in the veto region before trigger time.

The second charge variable is the charge ratio (\textit{QR}) and has several distinct implementations. The variable \textit{QR6} is the fraction of accumulated charges from cleaned hits during the first 600\,ns after the DeepCore trigger with respect to the total accumulated charge from all cleaned hits; that is,
\begin{equation}
QR6 = \frac{1}{Q_{\text{total}}} \sum_i Q_i
\end{equation}
for $0\,\rm{ns}<t(Q_i)<600\,\rm{ns}$. A contained neutrino event deposits more charge in a shorter time scale than a through-going muon, so a \textit{QR6} value closer to zero indicates a potential atmospheric muon event. Similarly, the variable \textit{QR3} is calculated with a tighter time window of 300\,ns instead of 600\,ns. Further, to reduce the impact from noise hits, the charge ratio variables \textit{C2QR6} and \textit{C2QR3} are calculated as described above for \textit{QR6} and \textit{QR3}, respectively, but with the two initial hits in the cleaned hit series removed.

\subsection{Veto Regions}
\label{app:variables_veto}
Veto variables are used in both analyses to identify and remove atmospheric muon backgrounds. These variables define veto regions, event-by-event, based on the positions and/or photon arrival times of the hit DOMs.

\begin{figure}
    \includegraphics[width=\linewidth]{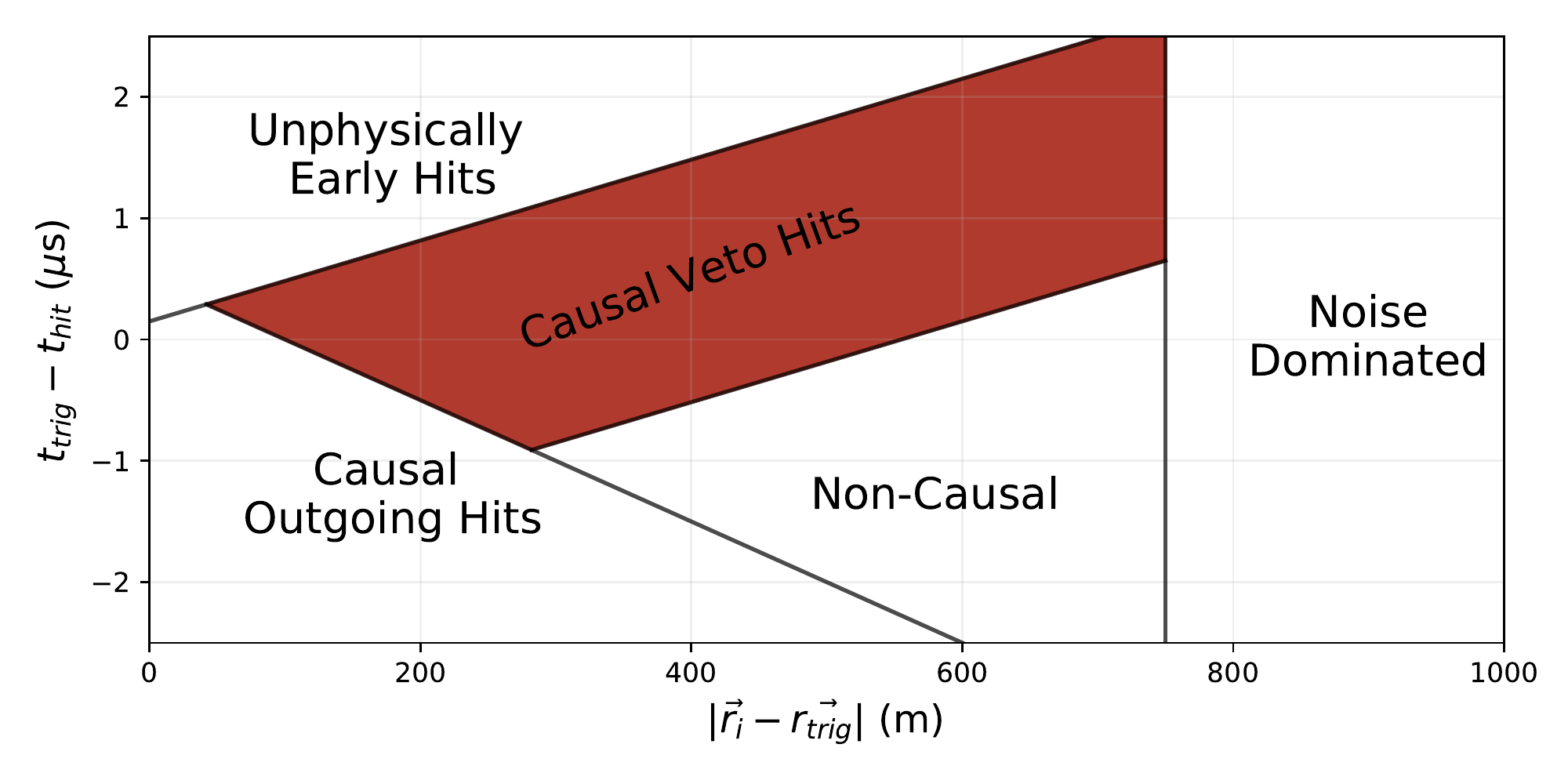}
    \caption{Definition of veto region based on the \textit{VICH} algorithm. The four lines surrounding the red region define the causal veto volume, event-by-event, based on the hit closest to the trigger time. The more total charge from cleaned hits deposited inside the veto region, the more likely that the event is caused by an atmospheric muon.}\label{appfig:vich_region}
\end{figure}

First used in~\cite{Ha:2011fvs} and optimized for DeepCore atmospheric oscillation searches in~\cite{Euler:2014}, the Veto Identified Causal Hits (\textit{VICH}) algorithm uses the cleaned hit closest to the trigger time as a reference hit and determines a veto region in which the hits may be causally connected with the reference hit. Fig.~\ref{appfig:vich_region} shows the causal veto region (in red), which is defined by
\begin{itemize}
\item{$\Delta r / c < 2.5\ \rm{\mu s}$},
\item{$\Delta r / c < - \frac{2}{3} \Delta t + \frac{1}{3}\ \rm{\mu s}$},
\item{$\Delta t - 0.15\ \rm{\mu s} < \Delta r / c < \Delta t + 1.85\ \rm{\mu s}$},
\end{itemize}
where $\Delta r$ and $\Delta t$ are the distance and photon arrival time between a given hit and the reference hit, respectively, and $c$ is the speed of light in vacuum. The width of the veto region accounts for a reasonable amount of scattering in the ice combined with the typical time scale of a GeV-scale neutrino event in IceCube (1\,$\mu$s). Restrictions of the red region close to the time and position of the reference hit are made to allow variations between the interaction time, the first trigger, and subsequent hits that could occur in a neutrino event. With the defined veto region, the \textit{VICH} variable is the integrated charge from cleaned hits that lie in the veto region. The more charge from cleaned hits deposited inside the veto region, the more likely that the event is caused by an atmospheric muon.

Another effective way to identify atmospheric muon background is to count the number of \textit{corridor DOMs}. As shown in Fig.~\ref{fig:detector}, the IceCube and DeepCore strings are arranged in a roughly triangular lattice in the horizontal plane. An atmospheric muon interacting in DeepCore can potentially come from a corridor (shown as the purple arrow in the figure) leaving no detected light in the veto region. To identify these muons, known corridor regions are studied. Given an interaction vertex, the corridor algorithm first finds direct hits from a cleaned hit series. Direct hits are hits due to minimally-scattered photons in the ice between their emission point and their detection in the DOM (the procedure to identify direct hits can be found in~\cite{Aartsen:2014yll}). The algorithm then looks through the closest IceCube strings along the known corridors and counts the number of direct hits on those strings, which is defined as the number of corridor DOMs.

\subsection{Center of Gravity}
\label{app:variables_CoG}
The Center of Gravity (CoG) is a parameter that measures space-time correlations between assumed signal hits in DeepCore and likely veto hits in the surrounding IceCube DOMs. For a total of $N$ hits, the average position $\vec{x}_{\text{CoG}}$ is given by
\begin{equation}
\vec{x}_{\text{CoG}}=\frac{1}{N}\sum_i^N \vec{x}_i,
\end{equation}
where $\vec{x}_i$ is the position of the $i^{\text{th}}$ hit DOM relative to the center of IceCube. Then, an average CoG time $t_{\text{CoG}}$ is calculated by assuming a simple cascade hypothesis with light propagating from the CoG without scattering;
\begin{equation}
t_{\text{CoG}}=\frac{1}{N} \sum_i^N \bigg ( t_i - \frac{|\vec{x}_i - \vec{x}_{\text{CoG}}|}{c_{\text{ice}}}\bigg ).
\end{equation}
Here, $t_i$ is the photon arrival time at the $i^{\text{th}}$ hit DOM, and $c_{\text{ice}}$ is the speed of light in ice. The same calculations can be applied to obtain the average position and time from a specific group of cleaned hits.

CoG is often used to check for causality. The CoG position and time are calculated from the cleaned hits inside the DeepCore fiducial volume. Then, for each cleaned hit in the veto region, a causally connected veto hit is identified if its $\vec{x}$ and $t$ satisfy
\begin{equation}
    0.25 \text{ m/ns} \leq \frac{|\vec{x}_{\text{CoG}} - \vec{x}|}{t_{\text{CoG}} - t} \leq 0.4 \text{ m/ns}.
\end{equation}
This requirement is used to reject muon tracks entering the fiducial volume after leaving hits in the veto region.

CoG is also used to evaluate event topology by dividing the time-sorted and cleaned hit series equally into four quartiles. Quartile 1 (Q1) consists of the earliest hits; quartile 4 (Q4) the latest hits. The CoG position and time are calculated for the hits in each quartile. The following three variables, based on these CoG quartiles, are used to identify background events.

First, to identify atmospheric muons, a \textit{z-travel} variable is defined as the vertical distance between the CoG $z$ position of Q1 and that from all cleaned hits in the fiducial volume; that is, \textit{z-travel} $\equiv z_{\text{all}} - z_{\text{Q1}}$. For a downward-going atmospheric muon event, its earlier hits tend to have a $z_{\text{Q1}}$ position above the average $z_{\text{all}}$, resulting in a negative \textit{z-travel}. Similarly, an upward-going neutrino event usually has a positive \textit{z-travel}.

Second, the spatial separation between Q1 and Q4 is also used for rejecting muon background events. The \textit{Q1-Q4 separation} is defined as $|\vec{x}_{\text{Q4}} - \vec{x}_{\text{Q1}}|$. Because atmospheric muons usually travel long distances across the detector, they often have a larger spatial separation between their earlier and later hits compared to neutrino events.

Third, to discriminate noise triggers from physics triggers, the space-time interval $\Delta s^2$ between Q1 and Q4 is also used. By definition, $\Delta s^2 \equiv |\vec{x}_{\text{Q4}} - \vec{x}_{\text{Q1}}|^2 - (ct_{\text{Q4}} - ct_{\text{Q1}})^2$. For an event caused by random detector noise, its $\Delta s^2$ will appear either too time-like or too space-like compared to an event due to a neutrino interaction.

In addition to hit causality and event topology, a charge-weighted CoG can also estimate the size of an event by determining the standard deviations of $z$ position ($\sigma_z$) and photon arrival time ($\sigma_t$) from all cleaned hits. An atmospheric muon tends to produce hits across the detector for a longer period of time, so its $\sigma_z$ and $\sigma_t$ are typically larger than for a neutrino event.

\subsection{Quick Track Reconstructions}
\label{app:variables_track}
At lower selection levels with high event rates, computationally-inexpensive reconstruction algorithms are often used to provide a rough estimate of the event parameters related to the interacting particle. These parameters include the particle's speed, direction, and point of interaction. The following two quick algorithms assume a track event hypothesis and are used in both Analyses \GRECO and~\DRAGON.

The improved LineFit, or \textit{iLineFit}, is based on the LineFit reconstruction. Assuming an infinitely-long muon track, the simple LineFit algorithm analytically minimizes a least-squares fit of cleaned hits with respect to the event parameters. The \textit{iLineFit} then takes into account effects such as random detector noise and the scattering and absorption properties of the ice. The fitting procedure is described in~\cite{Aartsen:2013bfa}.

The second reconstruction is a likelihood-based single photoelectron fit with eleven seeds (\textit{SPEFit11}). Given a track-like event with a set of event parameters, the likelihood between the expected and observed photon arrival times is determined for each DOM. The total likelihood from all DOMs is minimized with respect to the event parameters. This fit runs iteratively from eleven different starting orientations to avoid falling into local minima. More information is found in~\cite{Ahrens:2003fg}.

\section{Common Event Selection Criteria}
\label{app:common}
This section discusses the basic event filtering at the early stages of selection processes. These early selection cuts are mostly identical between Analyses \GRECO and \DRAGON. The first requirements (Levels~1 and~2) rely on trigger conditions, whereas the next selection criteria (Level~3) depends largely on veto algorithms.

\subsection{Common Level 1 and 2}
\label{app:common_L1L2}
Levels~1 and 2 include the standard online triggering and filtering, both of which rely on charges recorded by the PMTs in all DOMs. In IceCube, the charge that a DOM records is measured in effective photoelectron units (p.e.), and the calibration and characteristics of a PMT are described in~\cite{ABBASI2010139}. When a DOM's PMT exceeds a 0.25\,p.e.\@ threshold, an incident hit is detected, and the DOM is known as a ``hit DOM.''

When several nearby DOMs are hit, a local coincidence (LC) occurs, which indicates a potential neutrino signal event. In particular, two LC types are of interest: LC1 coincidences between two nearest-neighbor DOMs on a string and LC2 coincidences between two next-nearest-neighbor DOMs on a string. If either the LC1 or LC2 condition is met, the LC is called hard local coincidence (HLC), and the initial hit is a HLC hit. For each recorded HLC, a full digitization readout is performed~\cite{Aartsen:2016nxy}. If a DOM is hit with no coincidence from its neighbors, the hit only results in a charge and time stamp readout for the DOM instead of a full waveform readout.

Both the online trigger (Level~1) and online filter (Level~2) are performed at the South Pole~\cite{Aartsen:2016nxy}. Level~1 is a simple multiplicity trigger (SMT) that requires at least three HLC hits within 2.5~$\mu$s among the DeepCore DOMs; this trigger condition is known as SMT3. The Level~2 online filter looks for causally-connected hits in the veto and DeepCore regions using the CoG algorithm discussed in Appendix~\ref{app:variables_CoG}. For a given event, if the CoG algorithm identifies one or more causally connected veto hits, the event is likely caused by an atmospheric muon and is thereby rejected.

\subsection{Common Level 3}
\label{app:common_L3}
In general, the goal of the Level~3 selection is to remove events that are triggered by random detector noise and atmospheric muons. Most algorithms at Level~3 rely on hit information inside and outside the extended DeepCore volume. This region is defined to include DOMs that are 2100~m below the surface on all strings except the strings in the outermost three layers of IceCube. Compared to the standard DeepCore fiducial volume defined in Fig.~\ref{fig:detector}, the extended DeepCore volume contains five more IceCube strings.

Two algorithms are used to identify events triggered by random detector noise. First, the \textit{NoiseEngine} algorithm looks for directionality among hits. It starts by removing isolated hits and determines a time window that maximizes the number of cleaned hits. For each cleaned hit within the time window, it is connected to all other hits if the hit pair satisfies a space-time correlation window. A map of all possible hit pairs is produced, and they are projected onto a binned HEALPix sphere. If more than three pairs land in a single HEALPix bin, then hits are directional, and the event is unlikely to be noise-triggered. The second algorithm uses charge and hit information from cleaned hits within a dynamic time window. Noise-triggered events tend to have no more than two cleaned hits with a total charge less than 2~p.e.

Two selection criteria are applied to quickly identify candidate neutrino signal events in DeepCore. First, the total amount of charge recorded from a set of cleaned hits on DOMs inside the extended DeepCore fiducial volume must be greater than zero. The second selection variable is the \textit{VertexGuessZ} described in Appendix~\ref{app:variables_vertex}. An event passes if its first cleaned hit has a $z$ position below $-120$\,m, which is the top of the extended DeepCore fiducial volume. 

To identify obvious atmospheric muon events, two simple charge variables are defined to study light deposited in the veto region. As discussed in Appendix~\ref{app:variables_charge}, an atmospheric muon tends to have a higher value of \textit{NAbove200} than a neutrino signal event. Thus, an event is rejected if its \textit{NAbove200} is $\ge 12$~p.e. A second method uses the CoG algorithm discussed in Appendix~\ref{app:variables_CoG}. The algorithm is used to identify causally-connected hits in the veto region, and their charges are summed. If the total charge is $\ge 7$~p.e., the event is likely caused by an atmospheric muon and rejected.

The remaining two common Level~3 selection criteria are also charge-related. The first variable is the ratio of total charge outside the extended DeepCore fiducial region to that inside. The calculation is performed on a cleaned hit series, and events with a ratio smaller than 1.5 are kept. Second, the charge ratio \textit{C2QR6} discussed in Appendix~\ref{app:variables_charge} is used. A typical background muon event has a lower value of \textit{C2QR6}, so only events with \textit{C2QR6} greater than 0.4 are kept.

With these quick and simple Level~3 criteria applied, the event rate is diminished by a factor of 20 to roughly 1~Hz. Both Analyses \GRECO and \DRAGON share the same Level~3 selection criteria above, and an additional cut at Level~3 is applied in Analysis \GRECO as discussed in Appendix~\ref{app:greco_level3}.

\begin{table*}[htp]
\centering
\begin{tabular}{l|cc|cccccc|ccccc}
Type            & \multicolumn{2}{c|}{Filtering} & \multicolumn{6}{c|}{Analysis \GRECO} & \multicolumn{5}{c}{Analysis \dragon}                  \\ 
                & Total   & DeepCore  & L3      & L4     & L5    & L6    & L7      & L7 LE     & L3    & L4     & L5        & L6     & L6 LE    \\
\hline
atm. $\mu$      & 991000  & 9180      & 970     & 50.5   & 4.10  & 0.443 & 0.100    & 0.092    & 1310  & 44.7   & 0.163     & 0.0297 & 0.0259   \\
Noise           & 35900   & 8120      & 284     & 12.0   & 1.80  & 0.102 & $<$0.001 & $<$0.001 & 292   & 0.0006 & $<$0.0006 & 0.0003 & $<$0.0003\\
$\nu_e$         & 1.84    & 1.72      & 1.26    & 0.783  & 0.544 & 0.362 & 0.325   & 0.194     & 1.29  & 0.278  & 0.180     & 0.149  & 0.126    \\
$\nu_{\mu}$     & 11.3    & 6.36      & 4.76    & 2.50   & 1.63  & 1.01  & 0.676   & 0.552     & 4.93  & 1.02   & 0.558     & 0.396  & 0.342    \\
$\nu_{\tau}$    & 0.293   & 0.270     & 0.206   & 0.134  & 0.103 & 0.074 & 0.051   & 0.045     & 0.210 & 0.052  & 0.038     & 0.031  & 0.028    \\
MC Total        & 1030000 & 17300     & 1260    & 65.9   & 8.18  & 1.99  & 1.15   & 0.884      & 1608  & 46.1   & 0.94      & 0.61   & 0.52     \\
\hline
Data            & 1150000 & 19100     & 1090    & 68.6   & 7.42  & 1.84  & 0.87   & 0.715      & 1981  & 34.9   & 0.844     & 0.504  & 0.432    \\
\end{tabular}
\caption{\label{tab:TableRate} The event rate in mHz for the common filtering and the subsequent event selection levels for Analyses \greco and \dragon, respectively. After the final selection level, the analyses only include events with energies in the region from 5.6\,GeV to 56\,GeV which is denoted as ``LE''.}
\label{tab:event_rates}
\end{table*}

\section{Higher Level Selection for Sample \GRECO}
\label{app:greco}
This section discusses the progressive stages of the event selection for Analysis \GRECO . Optimized for atmospheric $\nu_{\tau}$ analyses, the event selection focuses on both cascade- and track-like neutrino signatures at $\mathcal{O}$(10)\,GeV, while removing as many background events as possible. Given three years of detector exposure, Analysis \GRECO  expects more than 55,000 neutrino events.

\subsection{Sample \GRECO, Level 3}
\label{app:greco_level3}
In addition to the common selection criteria described in Sec.~\ref{app:common_L3}, an extra criterion is applied at Level~3 based on charges inside the extended DeepCore fiducial volume and that outside the volume.
The total fiducial charge is simply the sum of all charges inside the volume. For charges outside the extended DeepCore fiducial region, an algorithm is performed to search for the largest clusters of hits, and the total veto charge is the sum of all hits in the cluster. Based on the charges deposited in the two regions, a cut is applied to remove potential background muon events.

\subsection{Sample \GRECO, Level 4}
\label{app:greco_level4}
After removing events that are likely caused by random detector noise and atmospheric muons, the Analysis \GRECO event selection uses a boosted decision tree (BDT)~\cite{FREUND1997119} to further reduce atmospheric muon background at Level~4. Six variables are used in training the BDT, including \textit{QR6}, \textit{C2QR6}, and \textit{NAbove200} discussed in Appendix~\ref{app:variables_charge}, as well as \textit{VertexGuessZ} in Appendix~\ref{app:variables_vertex}.

\begin{figure}
    \includegraphics[width=\linewidth]{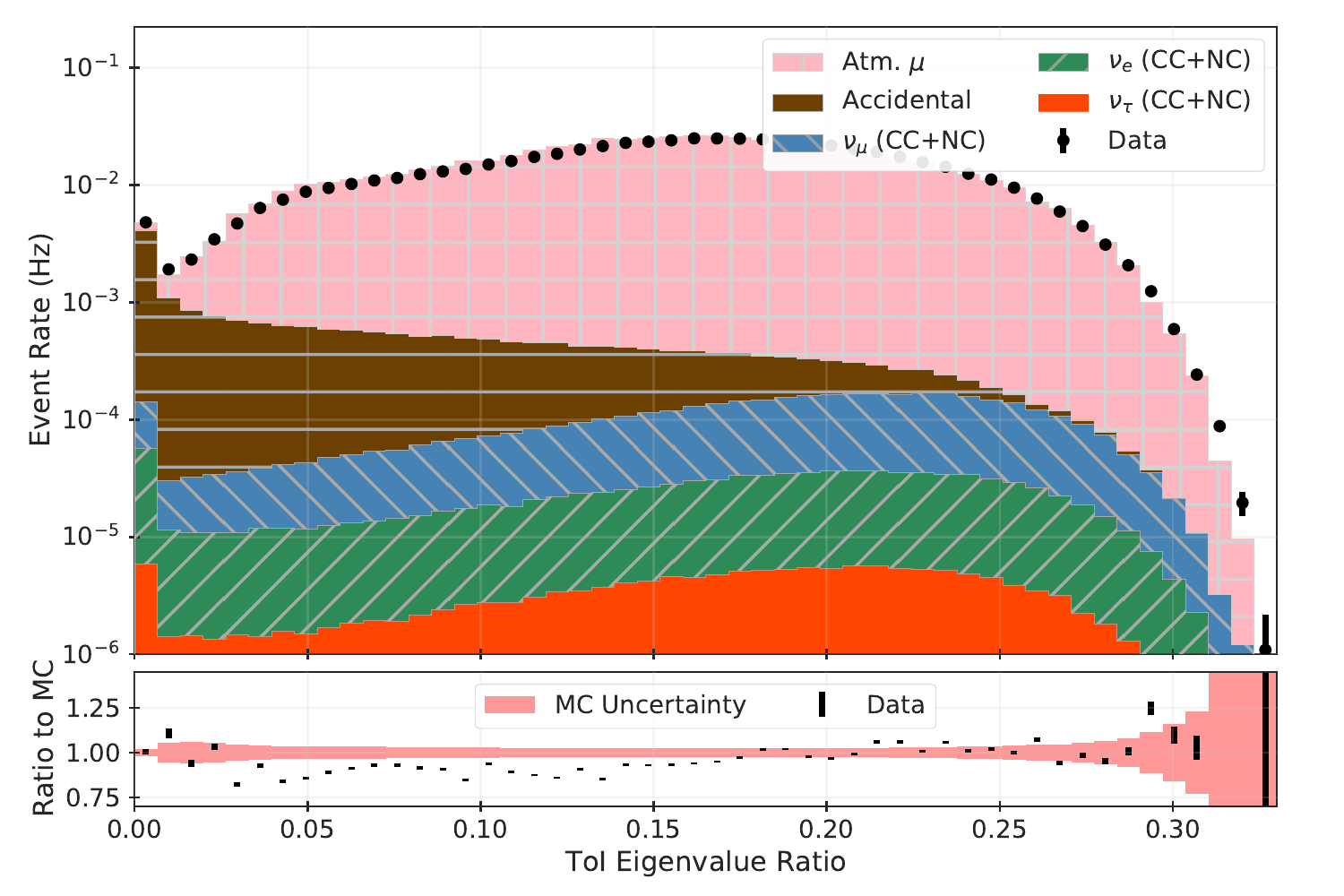}
    \caption{(top) Distribution of \textit{ToIEVal} at Level~4 (before cut applied). Each shaded color represents the stacked histogram from each event type. Black dots represent the data distribution. MC events are weighted by world-averaged best fit oscillation parameters. (bottom) Ratio of distribution from data to that from MC. Black error bars are the statistical fluctuations from data, whereas shaded red areas are the uncertainties from limited MC statistics.}
    \label{appfig:greco_l4_toi}
\end{figure}
The remaining two variables are the estimated speed from a quick \textit{iLinefit} reconstruction (see Appendix~\ref{app:variables_track}) and the rough event topology. The concept of moment of inertia in classical mechanics is adapted to describe the overall shape of an event with $N$ hits, each of which has a charge of $q_i$. The diagonal elements of the moment of inertia are then given by \begin{equation} \begin{split}
I_{x} = \sum_i^N q_i \big ( y_i^2 + z_i^2\big), \\
I_{y} = \sum_i^N q_i \big ( x_i^2 + z_i^2\big), \\
I_{z} = \sum_i^N q_i \big ( x_i^2 + y_i^2\big),
\end{split} \end{equation} where $x_i, y_i, z_i$ are the distance of the $i^{\text{th}}$ hit from String 36 at $z$ = 0\,m. For a spherically-shaped cascade-like event, the numerical values of $I_{x,y,z}$ are similar, while an elongated track-like muon background is more likely to have a smaller lateral distribution of hits than longitudinal. Therefore, a variable based on the tensor of inertia eigenvalue ratio, (\textit{ToIEVal}), is defined as:
\begin{equation}
\text{ToIEVal}=\frac{I_{\text{smallest}}}{I_x+I_y+I_z}.
\end{equation}
\textit{ToIEVal} is included in training the Level~4 BDT, and its distribution is shown in Fig.~\ref{appfig:greco_l4_toi}.

The BDT score distribution is shown in Fig.~\ref{fig:greco_l4_bdtscore}, and a score cut is applied to keep events with a score above 0.04. Compared to Level~3, the muon background from MC estimates after the BDT score cut is reduced from 970\,mHz to 51\,mHz, while $\approx 55$\% of all neutrino events are kept.

\subsection{Sample \GRECO, Level 5}
\label{app:greco_level5}
\begin{figure}
    \includegraphics[width=\linewidth]{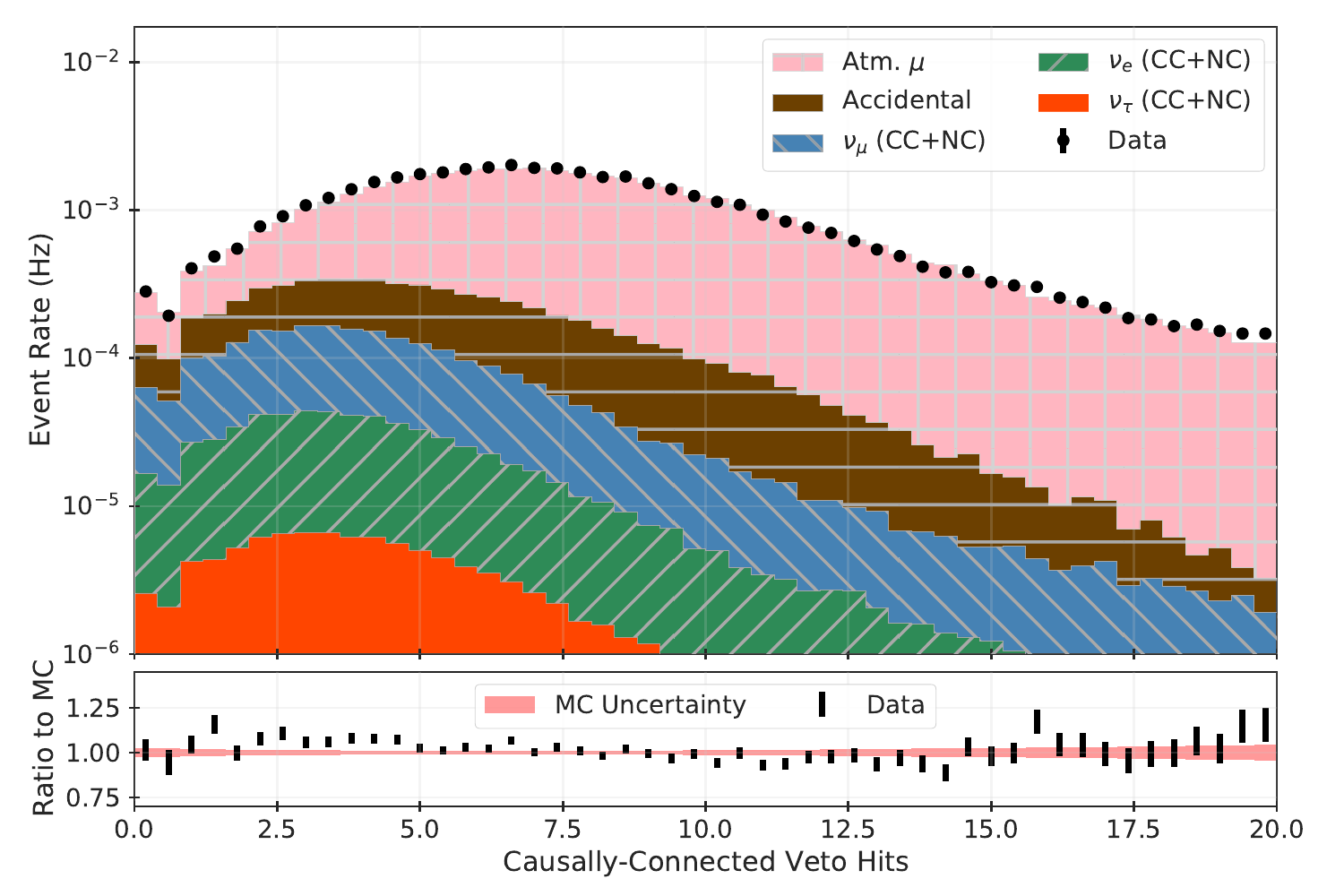}
    \caption{(top) Distribution of \textit{VICH} at Level~5 (before cut applied). Each shaded color represents the stacked histogram from each event type. Black dots represent the data distribution. MC events are weighted by world averaged best fit oscillation parameters. (bottom) Ratio of distribution from data to that from MC. Black error bars are the statistical fluctuation from data, whereas shaded red areas are the uncertainties from limited MC statistics.}
    \label{appfig:greco_l5_vich}
\end{figure}

A second BDT at Level~5 is trained to further reduce background muon contamination. After Level~4 cut is applied, over fifteen times more atmospheric muons and three times more pure-noise events remain compared to the integrated number of events from all neutrino flavors. Six variables are used for training the BDT, four of which are the radial position $\rho$ of the \textit{FirstHLC} (Appendix~\ref{app:variables_vertex}), the \textit{Q1-Q4 separation} and \textit{z-travel} from the CoG algorithm (Appendix~\ref{app:variables_CoG}), and the \textit{VICH} veto variable (Appendix~\ref{app:variables_veto}). The distribution of \textit{VICH} is shown in Fig.~\ref{appfig:greco_l5_vich} as an example.

The remaining two variables are the accumulated time and a simple zenith angle estimation. Because atmospheric muons are likely to travel across the detector, more time is needed, compared to neutrino events, for the hit DOMs to detect photons from the light source. Thus, the time to accumulate 75\% of the charge from a cleaned hit series is included in the Level~5 BDT training. Further, because a majority of muon background events are down going, a fast reconstruction \textit{SPEFit11} (see Appendix~\ref{app:variables_track}) is performed on a cleaned hit series to provide a rough estimate on the zenith angle of the interacting particle.

The Level~5 BDT score distribution is shown in Fig.~\ref{fig:greco_l5_bdtscore}, and a BDT score cut is applied to keep events with a score above 0.04. Compared to Level~4, the background muon rate after the Level~5 BDT score cut is reduced to $\approx 4$\,mHz, which is slightly more than a factor of 10. Moreover, $\approx 85$\% of noise-triggered events are rejected.

\subsection{Sample \GRECO, Level 6}
\label{app:greco_level6}
Previous selection criteria have reduced the background rate to $\approx 6$\,mHz, which is comparable to the total neutrino rate of $>$ 2\,mHz. At this stage, most obvious background muons are rejected, and the remaining atmospheric muons are more difficult to be identified. Therefore, Level~6 includes two straight cuts to reject events caused by random detector noise and two more cuts to identify sneaky atmospheric muons.

\begin{figure}
    \includegraphics[width=\linewidth]{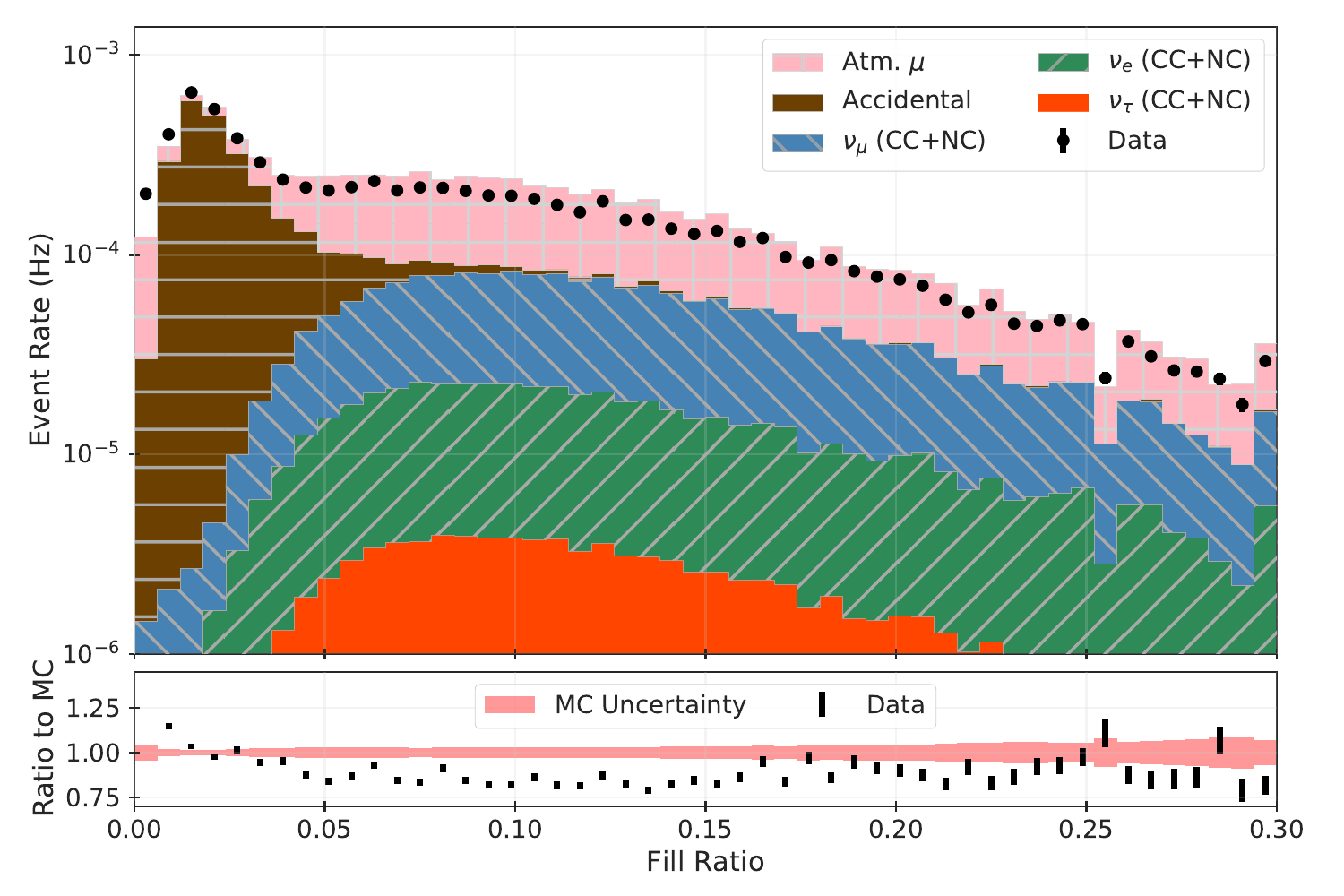}
    \caption{(top) \textit{FillRatio} distribution at Level~6 (before cut applied). Each shaded color represents the stacked histogram from each event type. Black dots represent the data distribution. The vertical line is the cut value of 0.05, events below which are rejected. MC events are weighted by world averaged best fit oscillation parameters. (bottom) Ratio of distribution from data to that from MC. Black error bars are the statistical fluctuation from data, whereas shaded red areas are the uncertainties from limited MC statistics.}
    \label{appfig:greco_l6_fillratio}
\end{figure}
To identify events caused by random detector noise, an algorithm called \textit{Fill-Ratio} is performed to look for the topology of hits in an event around an estimated vertex position given by \textit{FirstHLC} (see Appendix~\ref{app:variables_vertex}). A sphere is defined around the vertex with a radius 60\% larger than the average distance from all cleaned hits. The pattern of hits in an event can be estimated with a Fill-Ratio variable defined by the ratio of number of hit DOMs inside the sphere to the total number of DOMs inside the sphere. Since hits in a pure-noise event are randomly scattered across the detector, such event tends to have a longer mean distance away from its vertex and less hit DOMs inside the sphere. Therefore, pure-noise events often have smaller values of Fill-Ratio compared to physics events, as shown in Fig.~\ref{appfig:greco_l6_fillratio}. A straight cut is placed at 0.05, events above which are kept.

\begin{figure}
    \includegraphics[width=\linewidth]{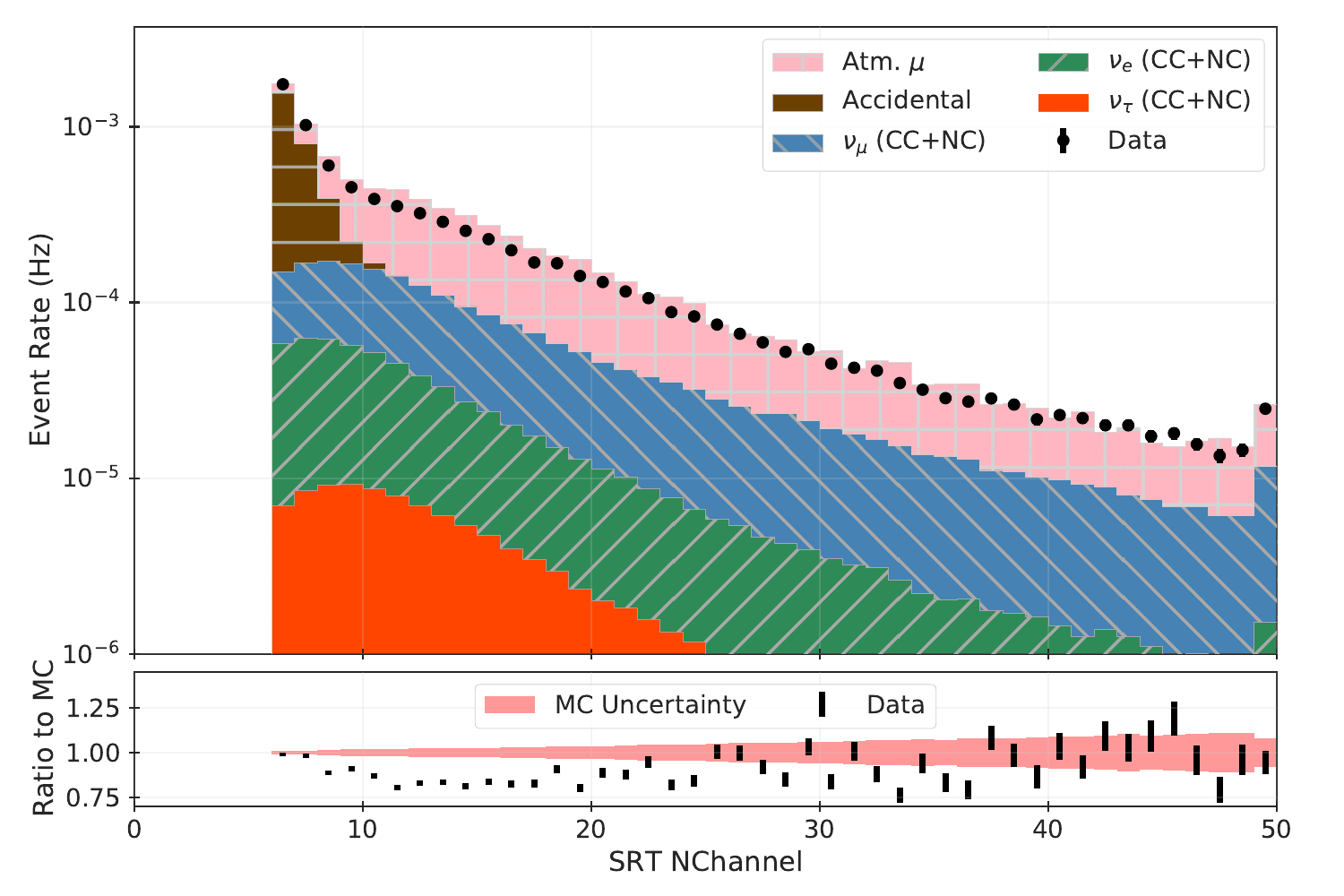}
    \caption{(top) \textit{NChannel} distribution at Level~6 (before cut applied). Each shaded color represents the stacked histogram from each event type. Black dots represents the data distribution. The vertical line is the cut value of 8, events below which are rejected. MC events are weighted by world averaged best fit oscillation parameters. (bottom) Ratio of distribution from data to that from MC. Black error bars are the statistical fluctuation from data, whereas shaded red areas are the uncertainties from limited MC statistics.}
    \label{appfig:greco_l6_nchannel}
\end{figure}
A second cut placed on \textit{NChannel} also help further remove noise-triggered events. \textit{NChannel} is the number of hit DOMs in a cleaned hit series, and its distribution is shown in Fig.~\ref{appfig:greco_l6_nchannel}. Most noise-triggered events have less than eight hit DOMs. Further, the computationally-intensive, likelihood-based reconstruction algorithm described in Sec.~\ref{sec:reco} fits eight parameters. In order to have enough degrees of freedom for a reasonable fit, a direct cut is applied on \textit{NChannel} to remove events with less than eight hit DOMs, reducing the contribution from pure-noise events by a factor of 20.

\begin{figure}
    \includegraphics[width=\linewidth]{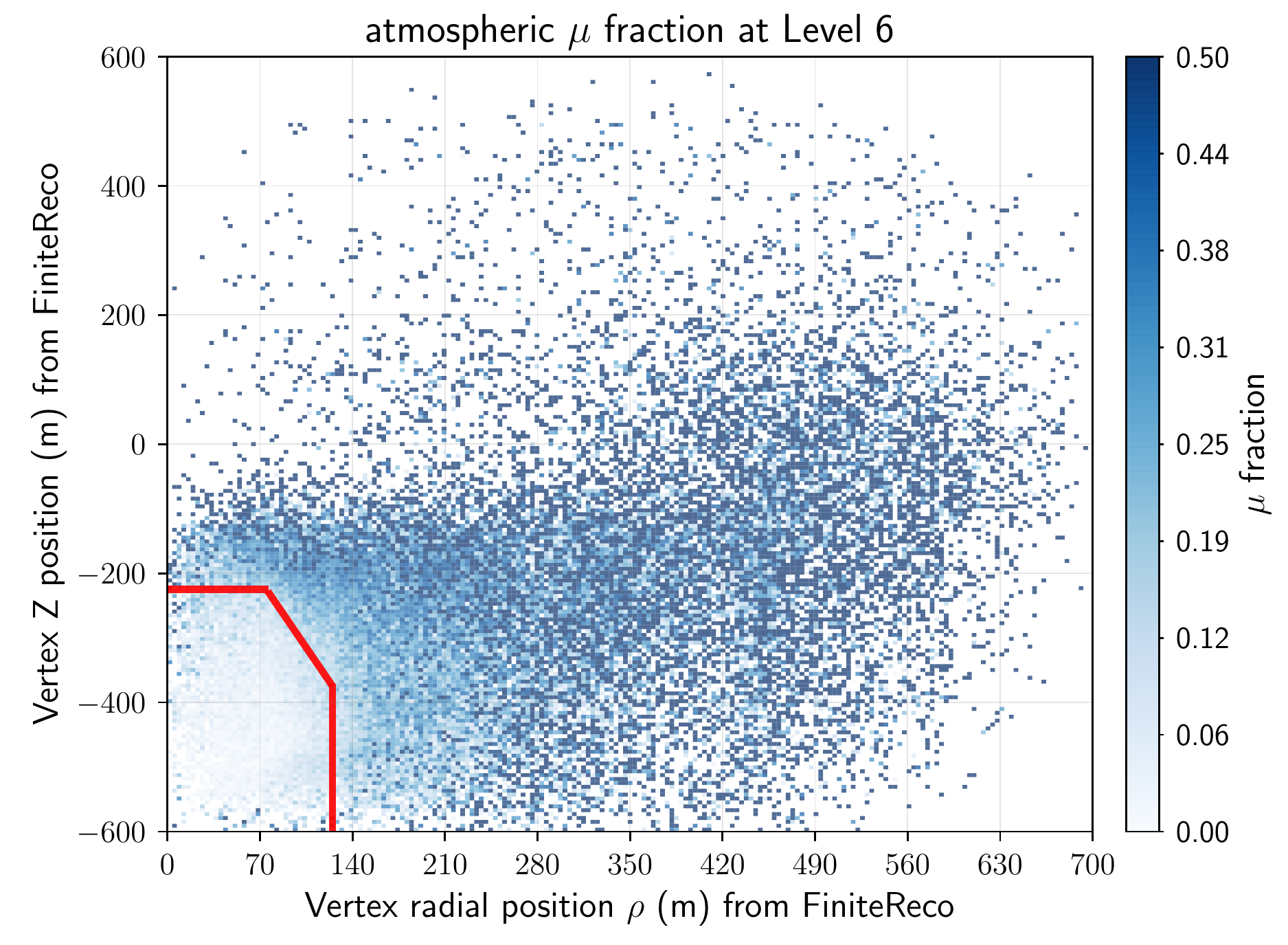}
    \caption{Fractional distribution at Level~6 (before cut applied) in vertex radial $\rho$ and depth $z$ positions by \textit{FiniteReco} for atmospheric muon events. Color axis represents the fraction of atmospheric muon with respect to the total expected MC events. Red lines represent the cut, within which events are kept.}
    \label{appfig:greco_l6_2D}
\end{figure}
For atmospheric muons, two straight cuts are applied. The first cut is applied on the number of \textit{corridor} DOMs as explained in Appendix~\ref{app:variables_veto}. Events with two or more direct hits found along those corridors are rejected. The second cut involves a likelihood-based algorithm known as \textit{FiniteReco}. Given an infinite track event hypothesis, \textit{FiniteReco} reconstructs an event based on the probabilities of the individual DOMs to see a hit or not~\cite{Euler:2014}. It provides a relatively quick estimate on the interaction vertex of an event. Figure~\ref{appfig:greco_l6_2D} shows the fractional 2D distribution of radial $\rho$ and depth $z$ positions from \textit{FiniteReco} for atmospheric muons. Most atmospheric muon events have vertices located 125\,m away from the center of the detector and above $-200$\,m in depth. Therefore, three cuts are applied such that events are required to have a vertex position with $z \ < -225$\,m, $\rho \ <$ 125\,m, and $z \ < \ -3\cdot\rho$. The latter cut removes events at the upper edge of the DeepCore fiducial volume, where atmospheric muons can enter DeepCore through the dust layer.

After Level~6 cuts are applied, more than 92\% and 96\% of atmospheric muon and noise-triggered events are rejected compared to Level~5. Neutrinos contribute more than 66\% of the sample, which opens the opportunity to use a computational-expensive but more comprehensive reconstruction algorithm discussed in Sec.~\ref{sec:reco}.

\subsection{Sample \GRECO, Level 7}
\label{app:greco_level7}
The comprehensive reconstruction discussed in Sec.~\ref{sec:reco} is performed between Level~6 and 7. To reduce the remaining background events and to improve agreement between data and MC, final selection cuts are applied at Level~7.

\begin{figure}
    \includegraphics[width=\linewidth]{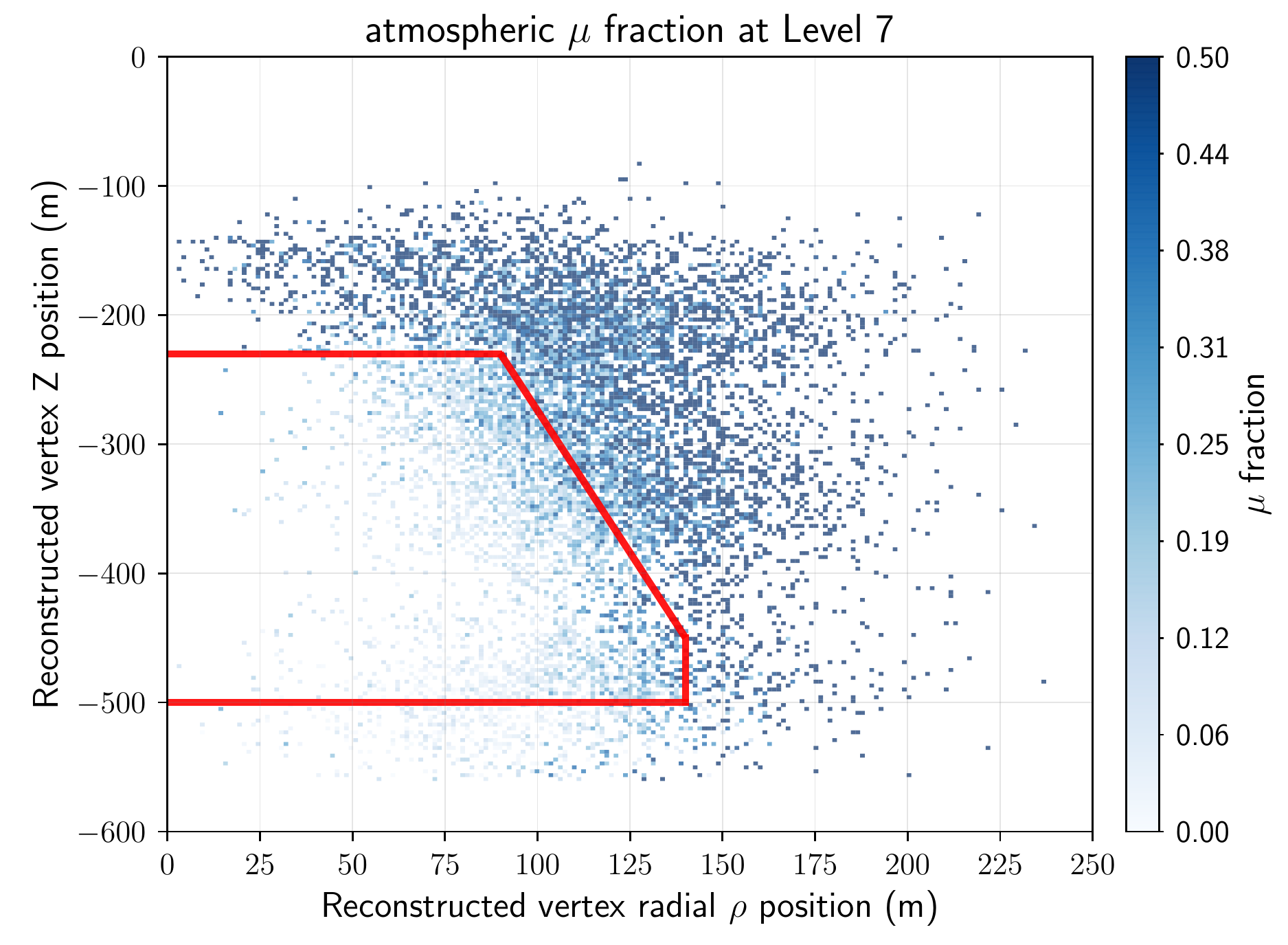}
    \caption{Fractional distribution at Level~7 (before cut applied) in vertex radial $\rho$ and depth $z$ positions by the final reconstruction discussed in Sec.~\ref{sec:reco} for atmospheric muon events. Color axis represents the fraction of atmospheric muon with respect to the total expected MC events per bin. Red lines represent the cut, within which events are kept.}
    \label{appfig:greco_l7_2D}
\end{figure}
The first selection criteria is similar to the containment requirement from the \textit{FiniteReco} vertex positions at Level~6. With a more sophisticated reconstruction method described in Sec.~\ref{sec:reco}, the fitted interaction vertex of an event is better than the estimates from \textit{FiniteReco}. Figure~\ref{appfig:greco_l7_2D} shows the fractional 2D distributions of radial $\rho$ and depth $z$ positions obtained from the final reconstruction. An extra containment condition is added to exclude events below $z$ position of $-500$\,m, and events outside the red lines are rejected.

\begin{figure}
    \includegraphics[width=\linewidth]{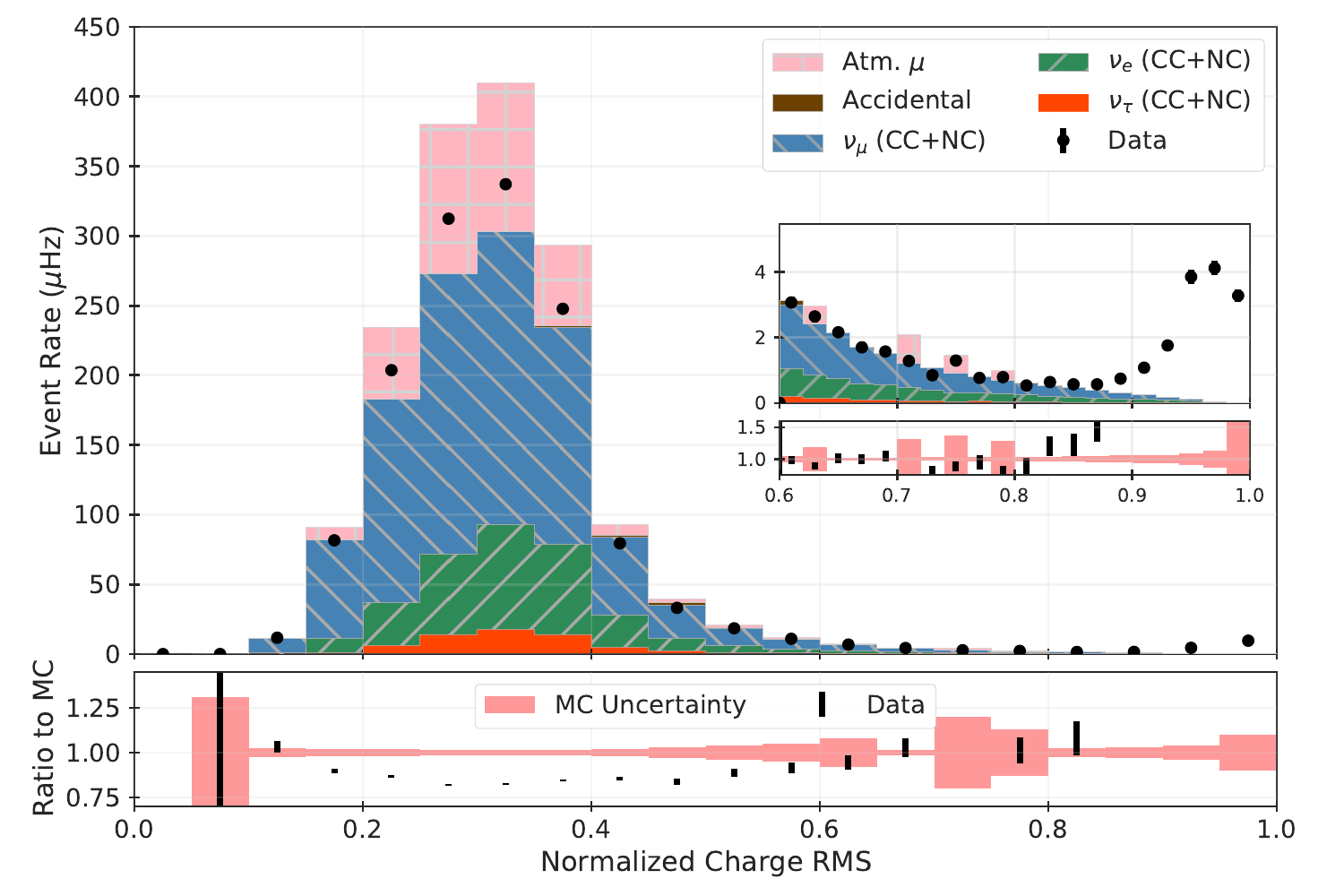}
    \caption{(top) distribution of normalized RMS of total charge at Level~7 (before cut applied). Each shaded color represents the stacked histogram from each event type. Black dots represent the data distribution. The vertical line is the cut value of 0.85, events above which are rejected. MC events are weighted by world averaged best fit oscillation parameters. (bottom) Ratio of distribution from data to that from MC. Black error bars are the statistical fluctuation from data, whereas shaded red areas are the uncertainties from limited MC statistics.}
    \label{appfig:greco_l7_flaringDOM}
\end{figure}

Two final cuts are applied during the development of the \GRECO selection. The first cut is based on a 2D distribution of reconstructed energy per number of hit DOMs and the root mean square (RMS) of photon arrival times from a cleaned hit series. The second cut is related to `flaring' DOMs, which emit light sporadically. The extra light is not simulated in MC, and a disagreement between data and MC is shown from the distribution of normalized RMS of total charges (see Fig.~\ref{appfig:greco_l7_flaringDOM}). Therefore, a cut is placed to remove events in which the normalized RMS of total charges are above 85\%. The above two cuts only remove 5\% of the total events and do not alter any physics results. Nonetheless, they are applied.

Finally, only events within the analysis histogram ranges stated in Sec.~\ref{sec:analysis} are used for the oscillation analyses. This restricts to events from all sky with a reconstructed energy between 5.6\,GeV to 56\,GeV and a reconstructed track length less than 1,000\,m.

In summary, the event rates as a function of event type and selection level is shown in Table~\ref{tab:event_rates}. The neutrino rates are the combination of the NC+CC channels and use the atmospheric neutrino flux predictions from~\cite{Honda:2015fha} with values of $\theta_{23}$ and $\Delta m^2$ from~\cite{Fogli:2012ua}. At the final level, 92\% of the \GRECO sample is neutrino events, while the contamination from atmospheric muon and noise-triggered events are 8\% and 0.1\% respectively.
\section{Higher Level Selection for Sample \DRAGON}
\label{app:dragon}
This section focuses on the event selection method for analysis \DRAGON. It features a set of straight cuts and a boosted decision tree to improve the purity of neutrino events at the final selection level. Based on simulations, about 40,000 neutrinos are expected given three years of detector exposure. The resultant sample is also used for the most recent published measurements of atmospheric neutrino oscillation parameters from the IceCube Collaboration~\cite{Aartsen:2017nmd}.

\subsection{Sample \DRAGON, Level 4}
\label{app:dragon_level4}
After the common Level~3 filtering discussed in Appendix ~\ref{app:common_L3}, a set of straight cuts is applied at Level~4 to further remove events due to random detector noise and atmospheric muons. These selection requirements rely on hit information from a cleaned hit series, charge information in veto regions, and an estimated interaction vertex position of an event.

To ensure that enough information are detected in an event, the first two selection variables are \textit{NChannel} and \textit{RTFiducialQ}. \textit{NChannel} is the number of hit DOMs in a cleaned pulseseries, and only events with at least eight cleaned hits are kept. Further, \textit{RTFiducialQ} is a charge-related variable from an algorithm, which searches for clusters of cleaned hits in the DeepCore fiducial region that satisfy space-time correlations. Events with a minimum of one cluster with at least 7\,p.e.\@ are kept.

Three additional selection criteria are placed based on variables from the \textit{Center of Gravity} (CoG) algorithm discussed in Appendix~\ref{app:variables_CoG}. First, the space-time interval $\Delta s^2$ has some power to distinguish events caused by random detector noise from physics events. Therefore, a cut is applied to keep events where $\Delta s^2$ is between $-(400$\,m)$^2$ and 0\,m$^2$. The remaining two cuts depend on the size of an event, which is estimated by the charge-weighted spread of vertex $z$ position ($\sigma_z$) and photon arrival time ($\sigma_t$) from a cleaned hit series. To be specific, only events with $\sigma_t \ \leq$ 1,000\,ns and 7\,m $\leq \ \sigma_z \ \leq$ 100\,m are kept.

Because an atmospheric muon event can be identified using the charge information in the veto regions, two veto requirements are placed. The first variable counts veto charges using the CoG algorithm (see Appendix~\ref{app:variables_CoG}). This cut is very similar to the total veto charge requirement at Level~3 where the veto region is outside the extended DeepCore volume. At Level~4, the same algorithm is applied to the veto region outside the standard DeepCore fiducial volume defined in Fig.~\ref{fig:detector}. With a slightly larger veto volume, a tighter cut is applied at Level~4 to remove events with a total veto charge greater than 5\,p.e. The second veto charge requirement is applied on the \textit{VICH} variable discussed in Appendix~\ref{app:variables_veto}. With a veto volume defined by the estimated point of interaction, \textit{VICH} looks for potentially causally-related hits in the event-by-event veto region. A cut is placed at 7\,p.e.\@ to reject potential atmospheric muon events.

To increase the purity of $\nu_{\mu}$ CC events, a cut is applied on the number of direct hits. Direct hits are hits due to photons that experience minimal scattering in the ice between its emission point and detection in the DOM. When a muon track passes next to a string, the intersection of its Cherenkov cone with a string results forms a hyperbolic pattern as a function of the photon direct arrival depth and times. The orientation of this pattern is determined by the angle between the string and the passing muon track. The procedure to identify direct hits is explained in~\cite{Aartsen:2014yll}, and only events with at least three direct hits are kept.

Finally, four containment conditions are applied to ensure that the point of interaction in an event is located inside the DeepCore fiducial volume. The first two cuts are based on the radial $\rho_{\text{HLC}}$ and depth $z_{\text{HLC}}$ positions from \textit{FirstHLC} discussed in Appendix~\ref{app:variables_vertex}. Only events with $\rho_{\text{HLC}}$ less than 150\,m and $z_{\text{HLC}}$ position between $-475$\,m and $-200$\,m are kept. In addition, a similar criteria is placed on the $\rho_{\text{Q1}}$ and $z_{\text{Q1}}$ vertex positions from the first quartile (Q1) in the CoG algorithm (see Appendix~\ref{app:variables_CoG}). Only events with $\rho_{\text{Q1}}$ less than 150\,m and $z_{\text{Q1}}$ position between $-475$\,m and $-150$\,m are kept.

After the above selection criteria, the number of atmospheric muon events is reduced by more than 95\% compared to Level~3. Further, the contamination due to random detector noise is also significantly dropped.

\subsection{Sample \DRAGON, Level 5}
\label{app:dragon_level5}
At Level~5, a boosted decision tree (BDT)~\cite{Hocker:2007ht} is trained to further reduce the atmospheric muon background. Eleven variables are included for training the BDT.

\begin{figure}[!htbp]
    \includegraphics[width=\linewidth]{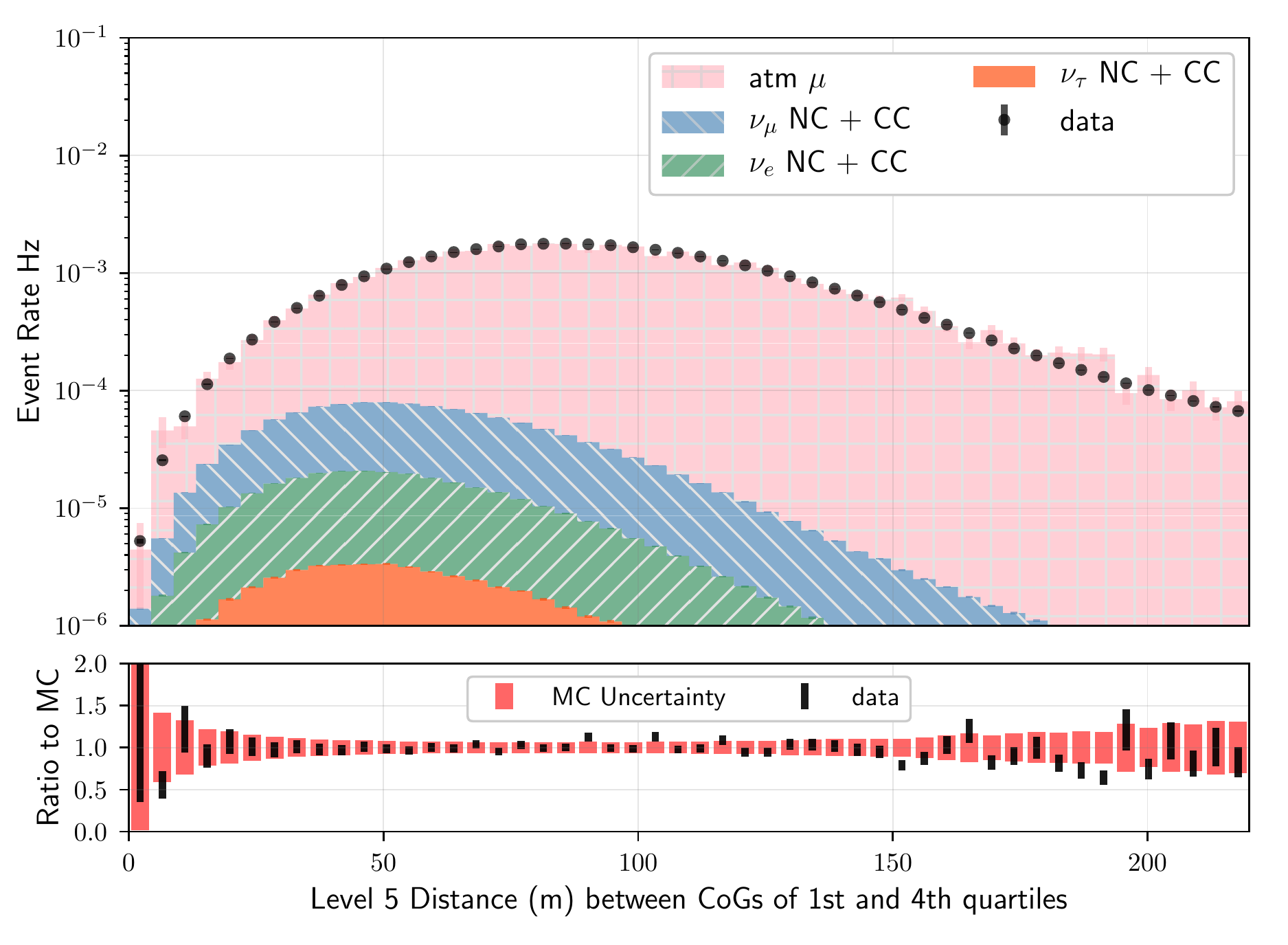}
    \caption{(top) Distribution of \textit{Q1-Q4 separation} at Level~5 (before BDT score cut applied). Each shaded color represents the stacked histogram from each event type. Black dots represents the data distribution. MC events are weighted by world averaged best fit oscillation parameters. (bottom) Ratio of distribution from data to that from MC. Black error bars are the statistical fluctuation from data, whereas shaded red areas are the uncertainties from limited MC statistics.}
    \label{appfig:dragon_l5_separation}
\end{figure}

Three variables for BDT training are obtained from the two quick reconstruction algorithms described in Appendix~\ref{app:variables_track}. They are the reconstructed speed from \textit{iLineFit} and the zenith angles from \textit{SPEFit11} and \textit{iLineFit}.

The next four BDT variables are related to the hit and charge information from a cleaned pulseseries, including the charge ratios \textit{QR3} and \textit{C2QR3} discussed in Appendix~\ref{app:variables_charge}. The number of hit DOMs in the cleaned hit series is also included. Further, the total charge from a cleaned hit series of an event is also used for BDT training. Since an atmospheric muon tends to deposit more charges compared to a neutrino, the background-dominated, high-charge region in the total charge distribution can help identify atmospheric muons from neutrino events.

\begin{figure}[!htbp]
    \includegraphics[width=\linewidth]{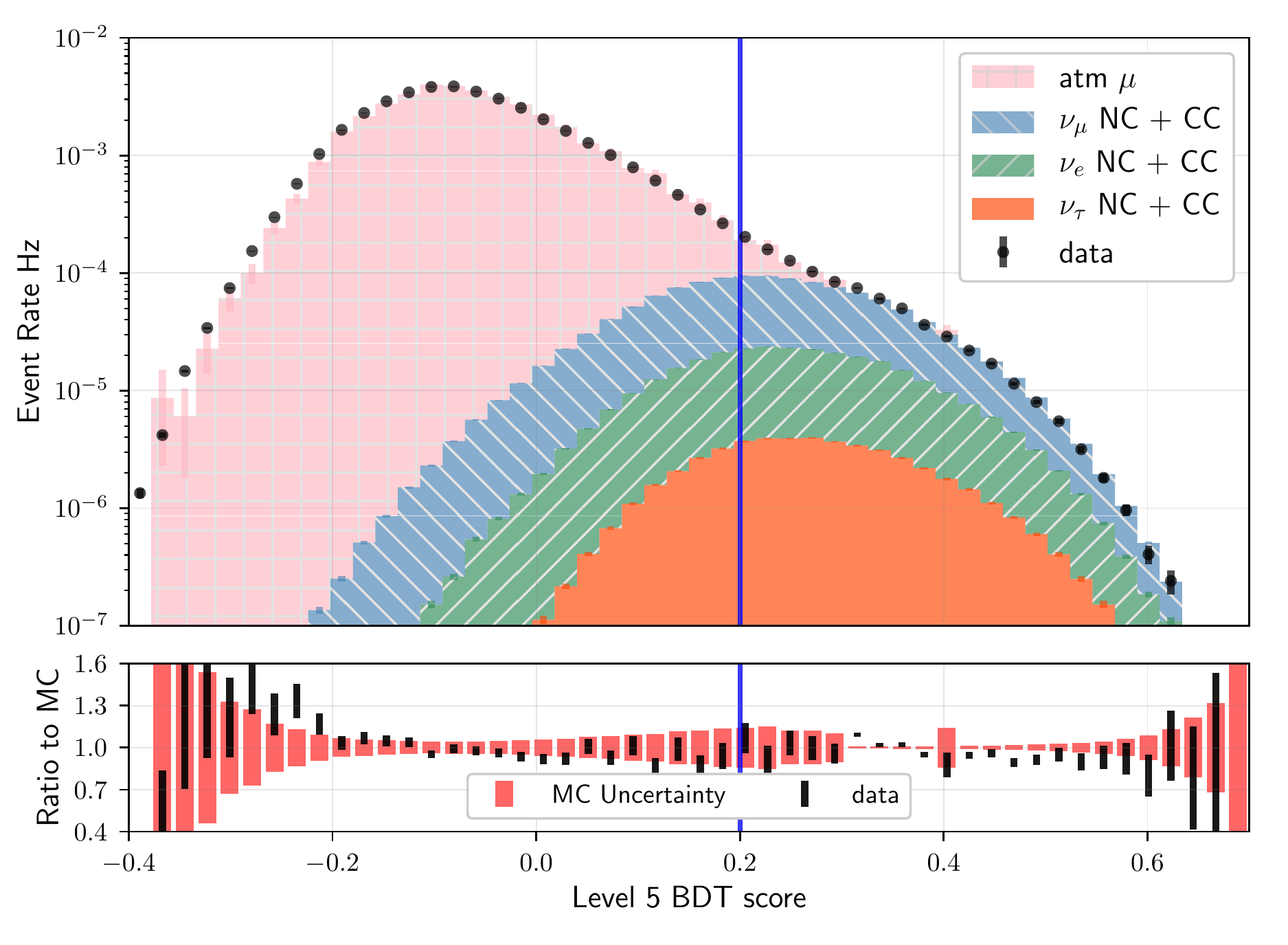}
    \caption{(top) BDT score distribution at Level~5 (before cut applied). Each shaded color represents the stacked histogram from each event type. Black dots represents the data distribution. The blue vertical line is the cut value of 0.2, events below which are rejected. MC events are weighted by world averaged best fit oscillation parameters. (bottom) Ratio of distribution from data to that from MC. Black error bars are the statistical fluctuation from data, whereas shaded red areas are the uncertainties from limited MC statistics.}
    \label{appfig:dragon_l5_bdtscore}
\end{figure}

The last four BDT variables are based on the CoG algorithm discussed in Appendix~\ref{app:variables_CoG}. One of them is the separation between the first and the last quartiles of a cleaned hit series, or \textit{Q1-Q4 separation}. As shown in Fig.~\ref{appfig:dragon_l5_separation}, atmospheric muon background tends to have a longer spatial distance between the two quartiles, compared to neutrino events. The remaining three variables are reused from the previous level. They are the charge-weighted spread of vertex $z$ position ($\sigma_z$) from all cleaned hits and the estimated radial $\rho_{\text{Q1}}$ and depth $z_{\text{Q1}}$ positions of the interaction vertex from the first quartile (Q1) of CoG.

Figure~\ref{appfig:dragon_l5_bdtscore} shows the BDT score distribution, and a cut is applied to accept events with a score above 0.2. Compared to Level~4, 99.9\% of the atmospheric muon background is removed, whereas 58\% of all neutrinos is kept after the BDT score cut is placed.

\subsection{Sample \DRAGON, Level 6}
\label{app:dragon_level6}
In between Level~5 and 6, the comprehensive reconstruction discussed in Sec.~\ref{sec:reco} is performed. At Level~6, two final selection requirements are placed to further improve the quality of the final sample.

\begin{figure}
    \includegraphics[width=\linewidth]{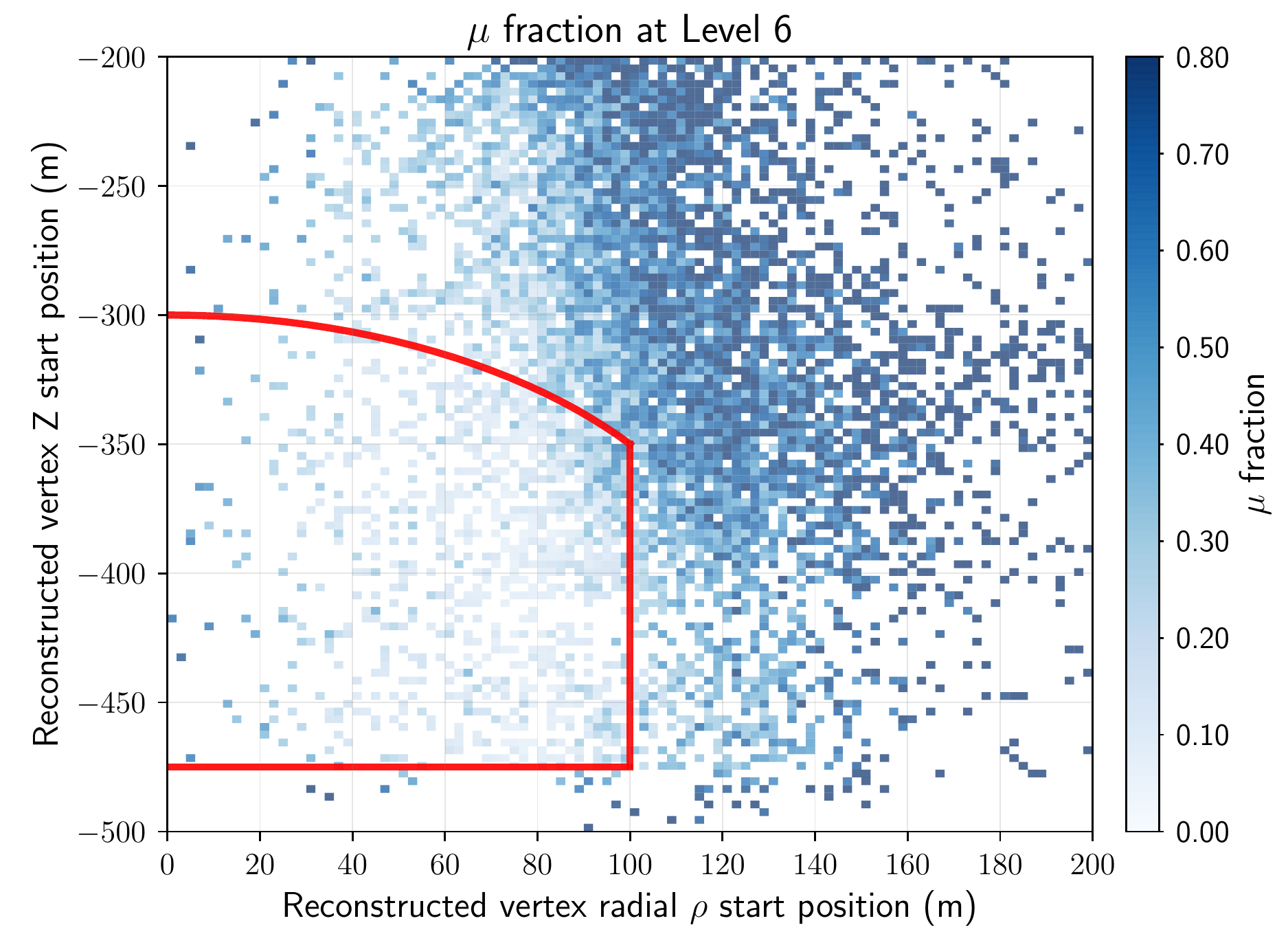}
    \caption{Fraction of $\mu$ events to total at Level~6 (before cut applied) in the starting radial $\rho$ and depth $z$ positions of reconstructed track by the final reconstruction discussed in Sec.~\ref{sec:reco}. Red lines represent the cut, within which events are kept.}
    \label{appfig:dragon_l6_contain}
\end{figure}

\begin{figure}
    \includegraphics[width=\linewidth]{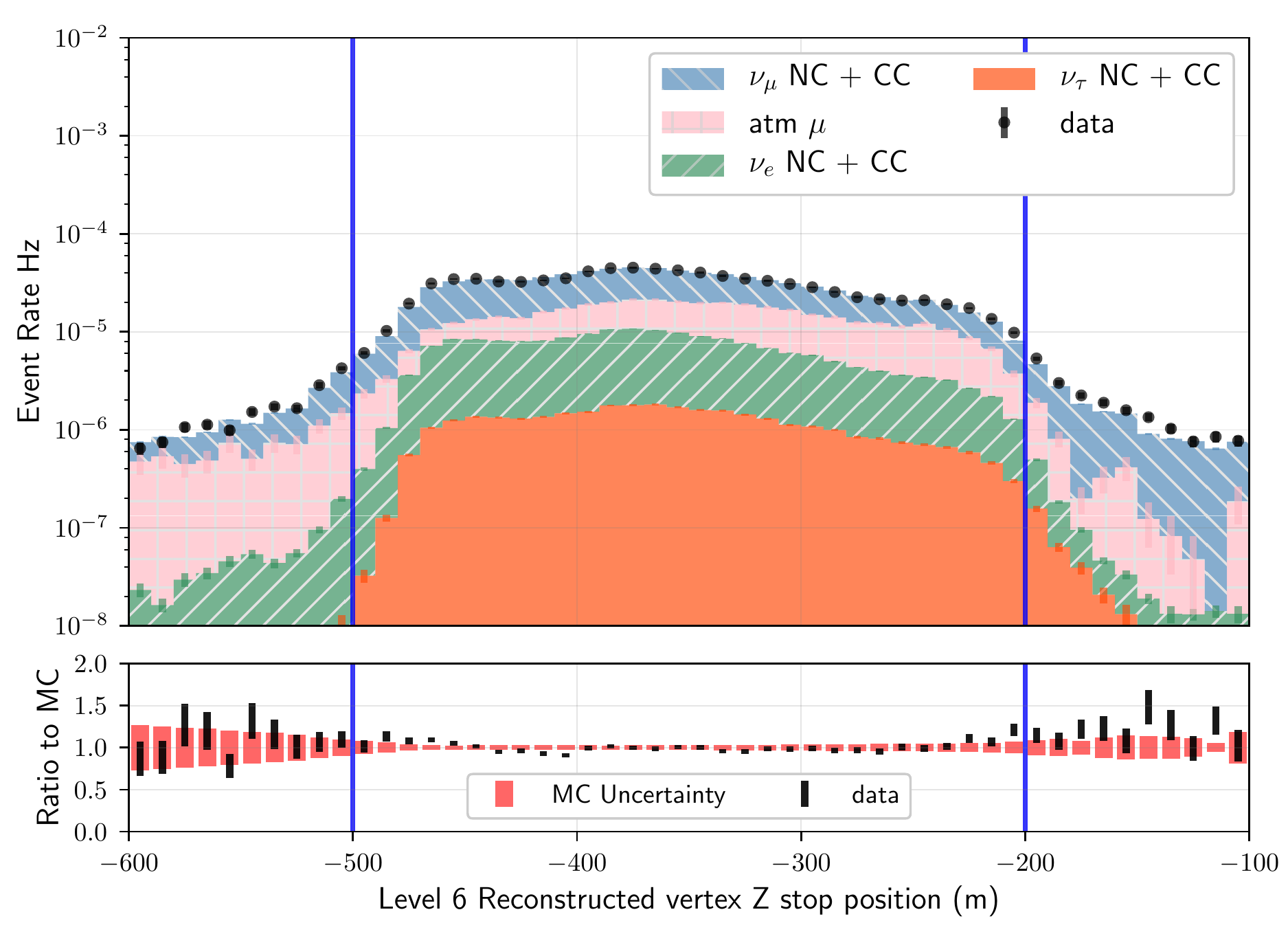}
    \caption{(top) Distribution of stopping $z_{\text{stop}}$ positions of the reconstructed track at Level~6 (before cut applied). Each shaded color represents the stacked histogram from each event type. Black dots represents the data distribution. The vertical lines are the cut values; events between the two vertical lines are kept. MC events are weighted by world averaged best fit oscillation parameters. (bottom) Ratio of distribution from data to that from MC. Black error bars are the statistical fluctuation from data, whereas shaded red areas are the uncertainties from limited MC statistics.}
    \label{appfig:dragon_l6_stop_z}
\end{figure}

First, final containment criteria is required based on the starting and stopping positions of a reconstructed track. The track must starts within the region defined by the red lines in Fig.~\ref{appfig:dragon_l6_contain}, which further rejects atmospheric muons coming through the dust layer. In addition, energy resolution can be improved by ensuring the entire track is contained within the more densely instrumented DeepCore region. Thus, the stopping radial $\rho_{\text{stop}}$ and depth $z_{\text{stop}}$ positions (see Fig.~\ref{appfig:dragon_l6_stop_z}) of the track must satisfy $\rho_{\text{stop}} \leq$ 150\,m and $-500$\,m $\leq \ z_{\text{stop}} \ \leq$ $-200$\,m respectively.

Last, to further reduce the contamination from sneaky penetrating atmospheric muons, the number of \textit{corridor} DOMs, defined in Appendix~\ref{app:variables_veto}, is required to be less than 2. 

Table~\ref{tab:event_rates} shows the event rates for each event type at each selection level. The neutrino rates are the combination of the NC+CC channels and use the atmospheric neutrino flux predictions from~\cite{Honda:2015fha} with world average best fit values for the oscillation parameters. At the final level, the rate of atmospheric muons is reduced by a factor of $\approx 10^8$, and neutrino events contribute to $\approx 95$\% of the \DRAGON sample.
\section{Nuisance and Physics Parameter Impacts in Analysis~\GRECO}
\label{app:systematics}
This section shows the change in expected event rates for all systematic uncertainties included in Analysis~\GRECO. The figures show the percentage change when the corresponding nuisance parameter is shifted off nominal by the amount specified in each subcaption.

\onecolumngrid

\begin{sidewaysfigure}
  \centering
  \subfloat[$\nu_e/\nu_\mu$ Flux Ratio +5\%]{\includegraphics[width=0.4\textwidth]{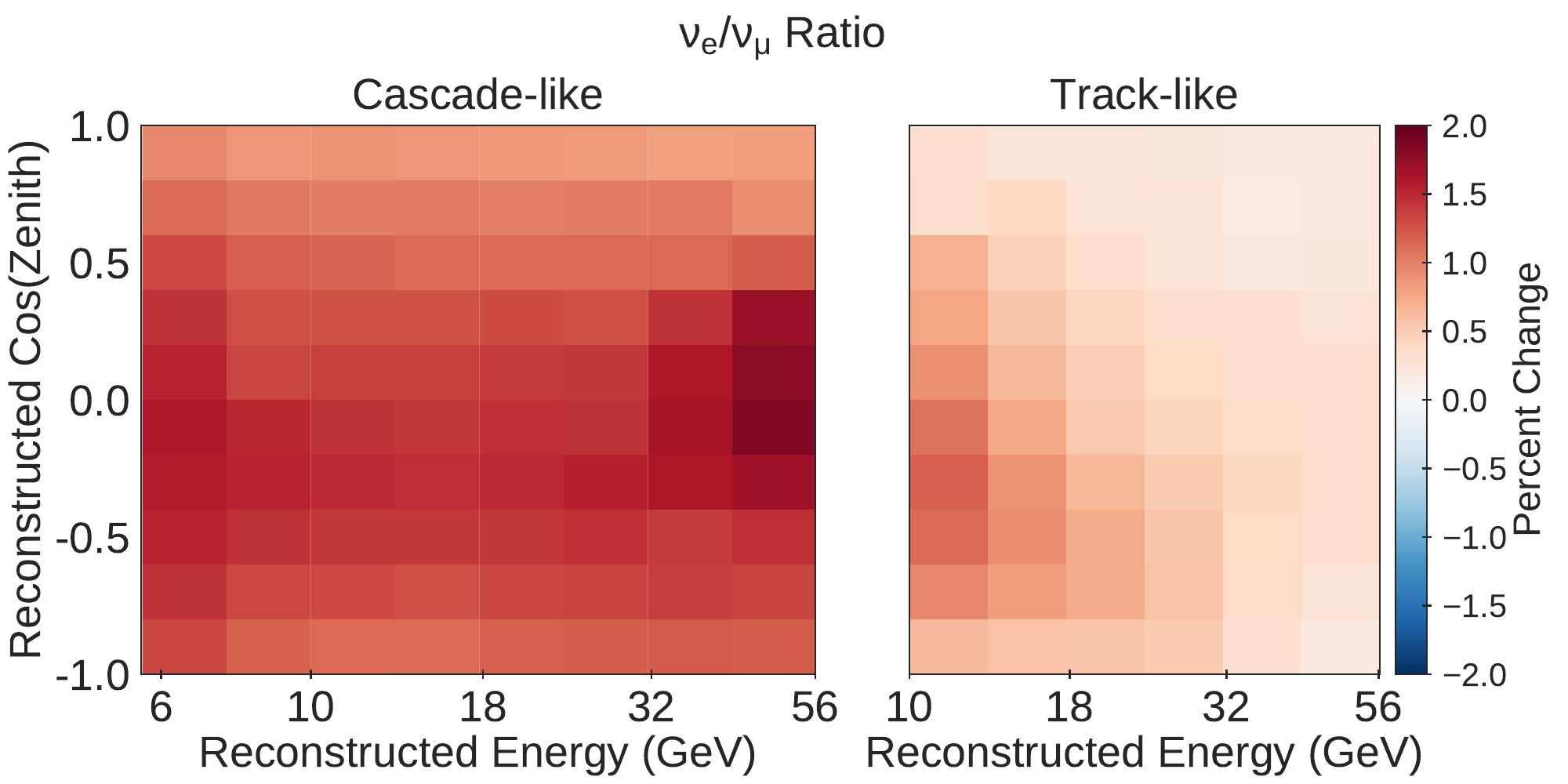}}
  \subfloat[$\nu/\bar{\nu}$ Flux Ratio +1$\sigma$]{
    \includegraphics[width=0.4\textheight]{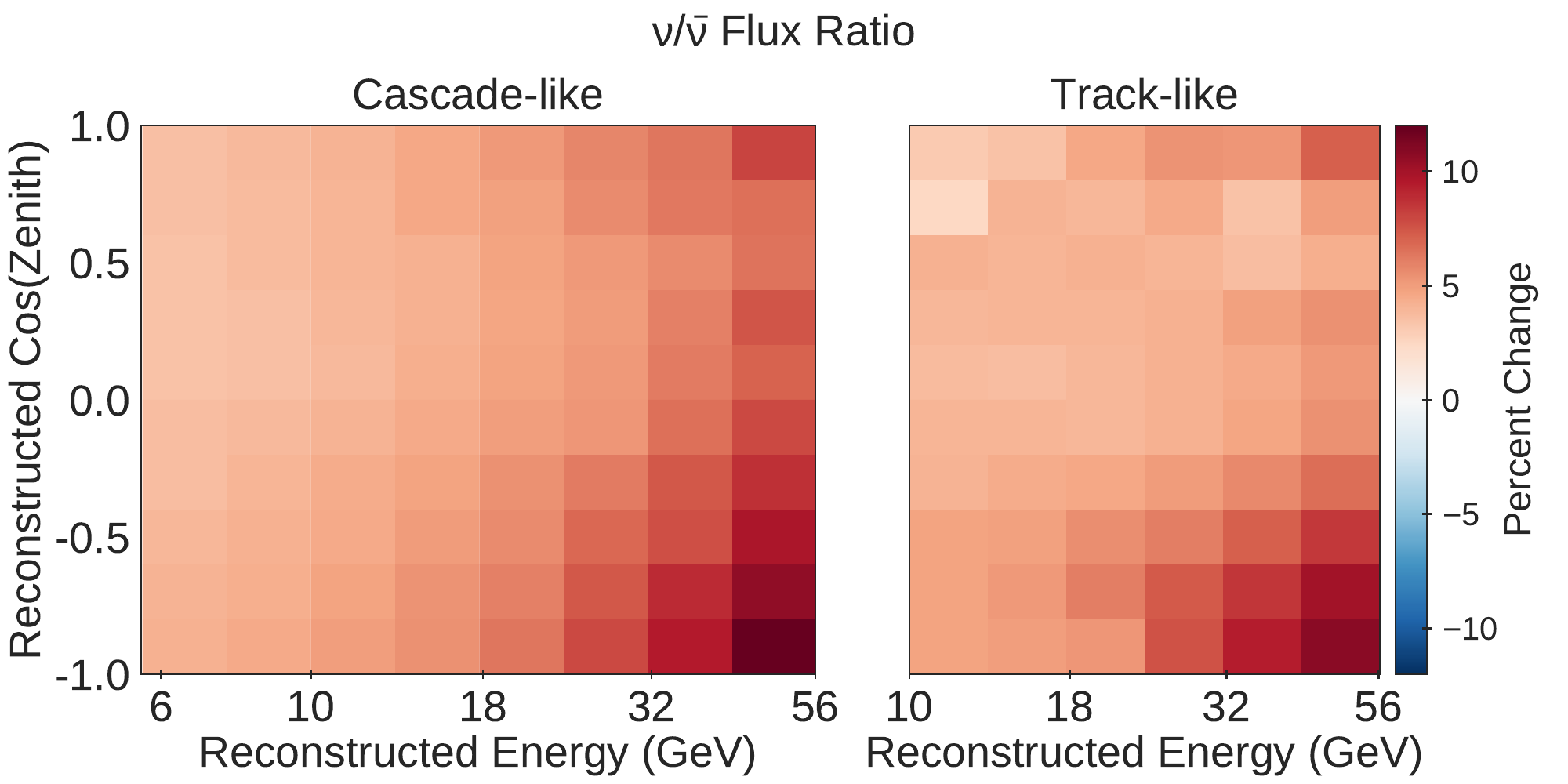}}

  \subfloat[Upward/Horizontal Flux Ratio +1$\sigma$]{
     \includegraphics[width=0.4\textheight]{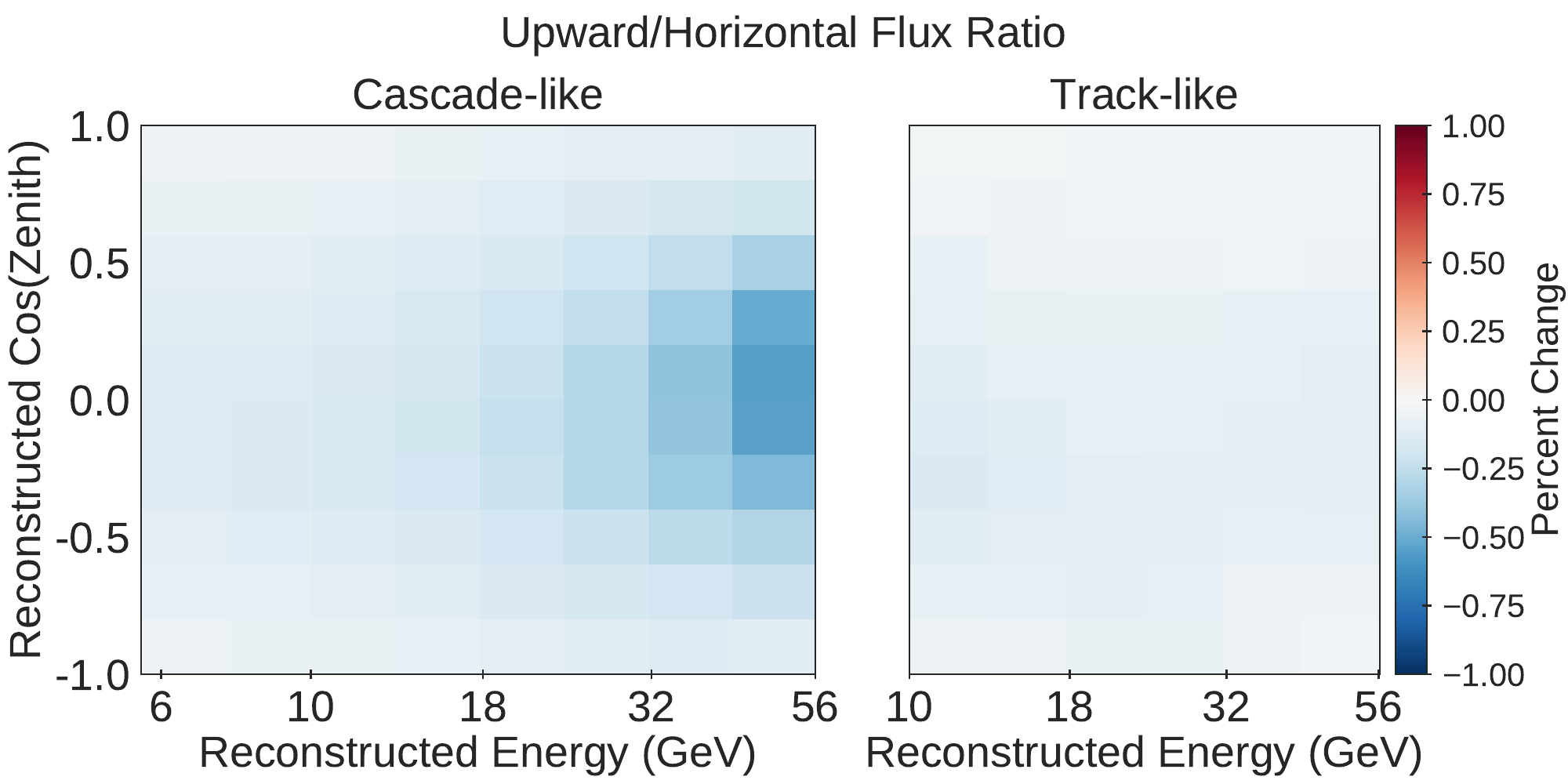}}
  \subfloat[Neutrino Spectral Index +0.1]{
     \includegraphics[width=0.4\textheight]{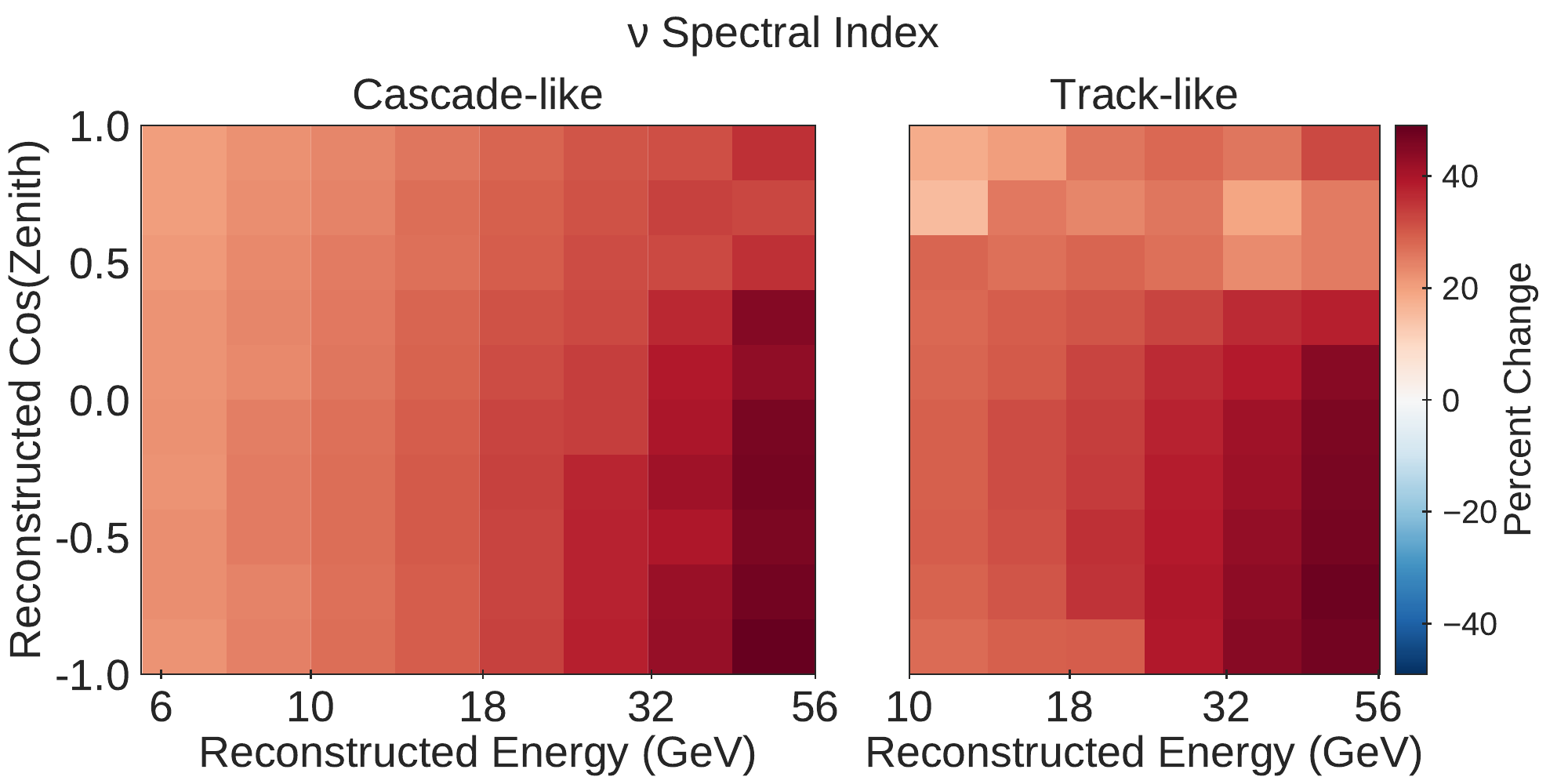}}
     
  \subfloat[$M_{A}^{CCQE}$ +1$\sigma$ (+0.248~GeV)]{
     \includegraphics[width=0.4\textheight]{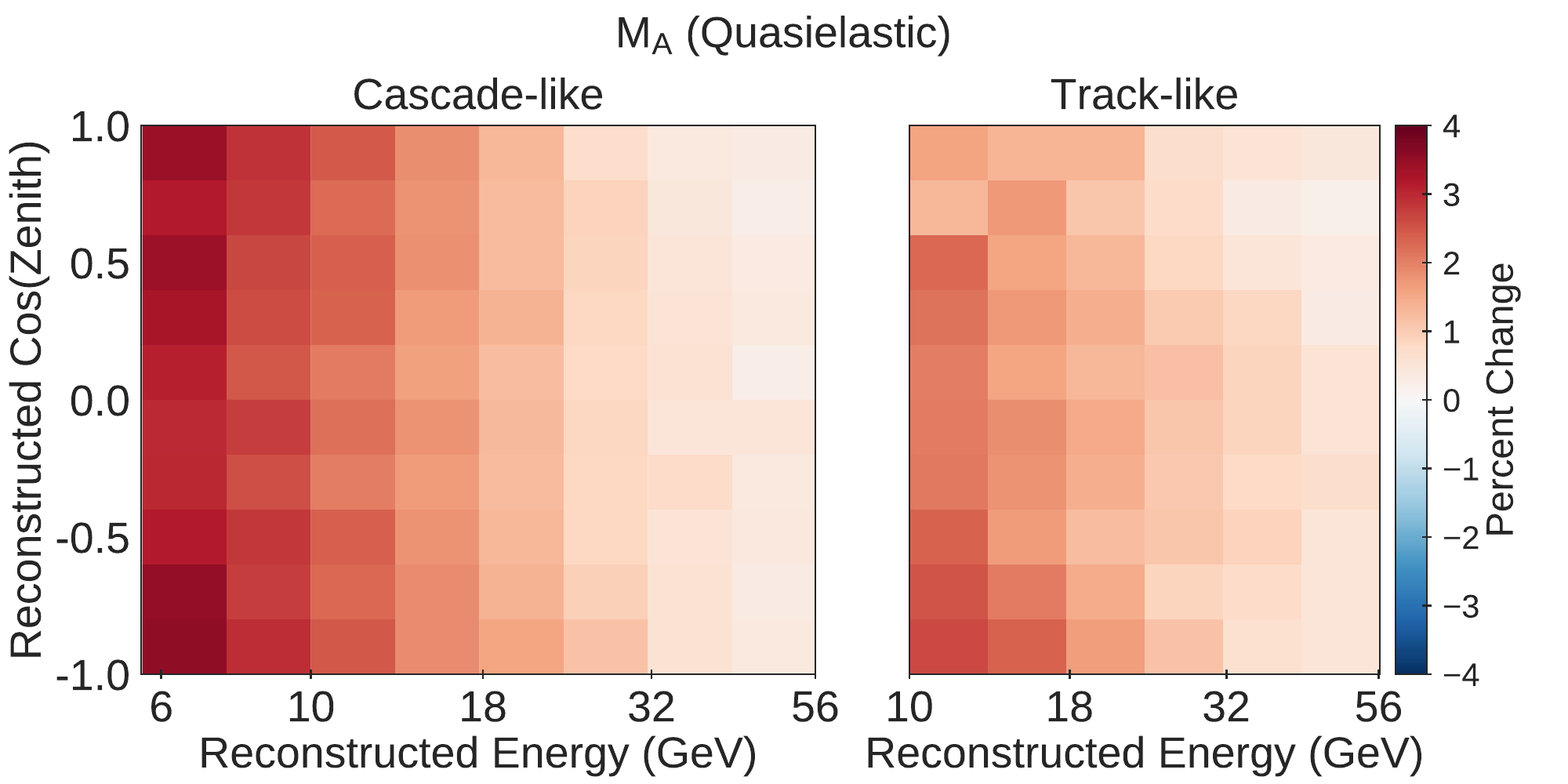}}
  \subfloat[$M_{A}^{RES}$  +1$\sigma$ (+0.22~GeV)]{
     \includegraphics[width=0.4\textheight]{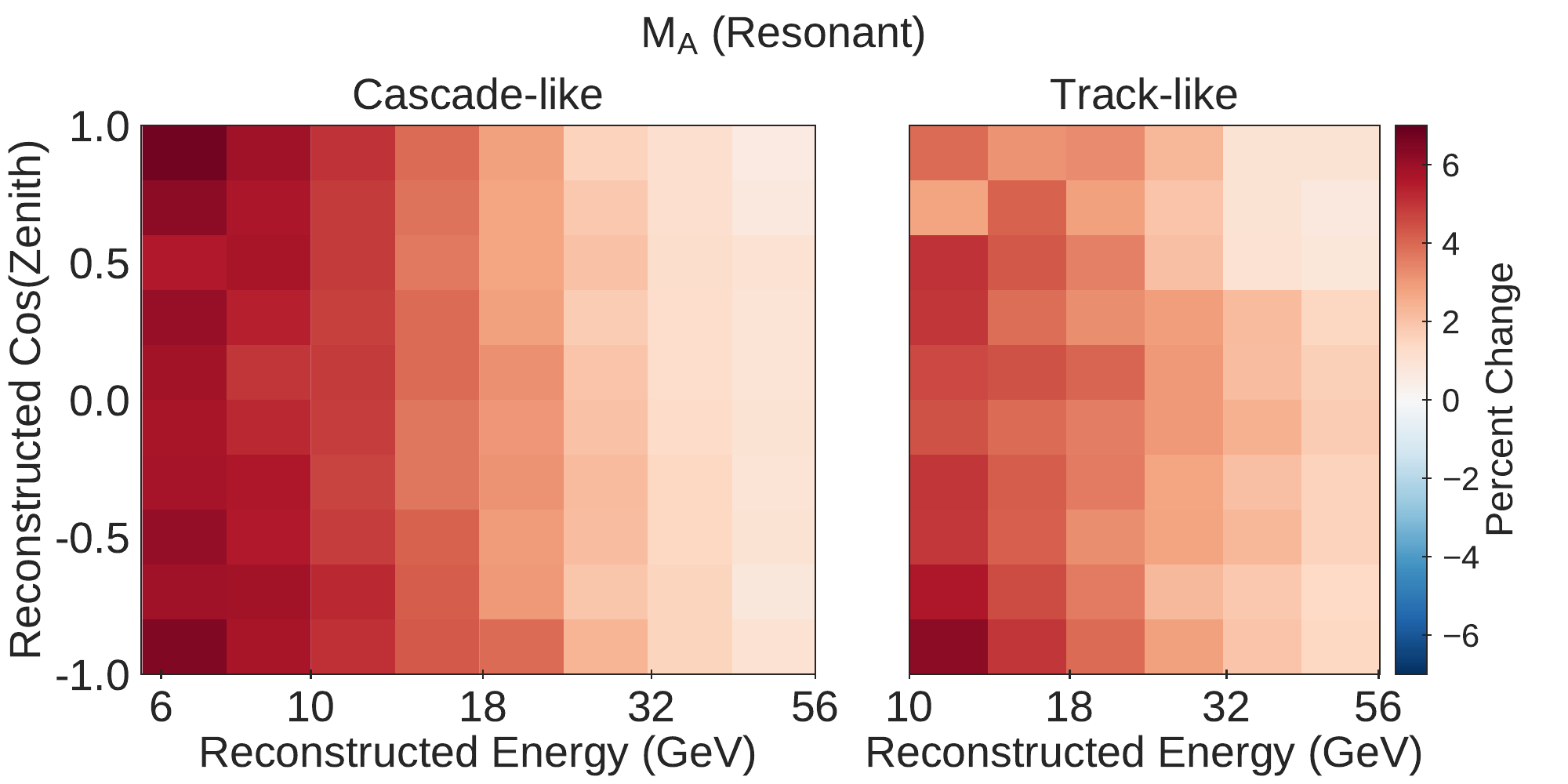}}
     \caption{Changes in the event rates from a subset of the systematic uncertainties included in the analyses.}
\end{sidewaysfigure}

\begin{sidewaysfigure}
  \centering
  \subfloat[$\nu^{NC}$ Normalization +20\%]{
     \includegraphics[width=0.4\textheight]{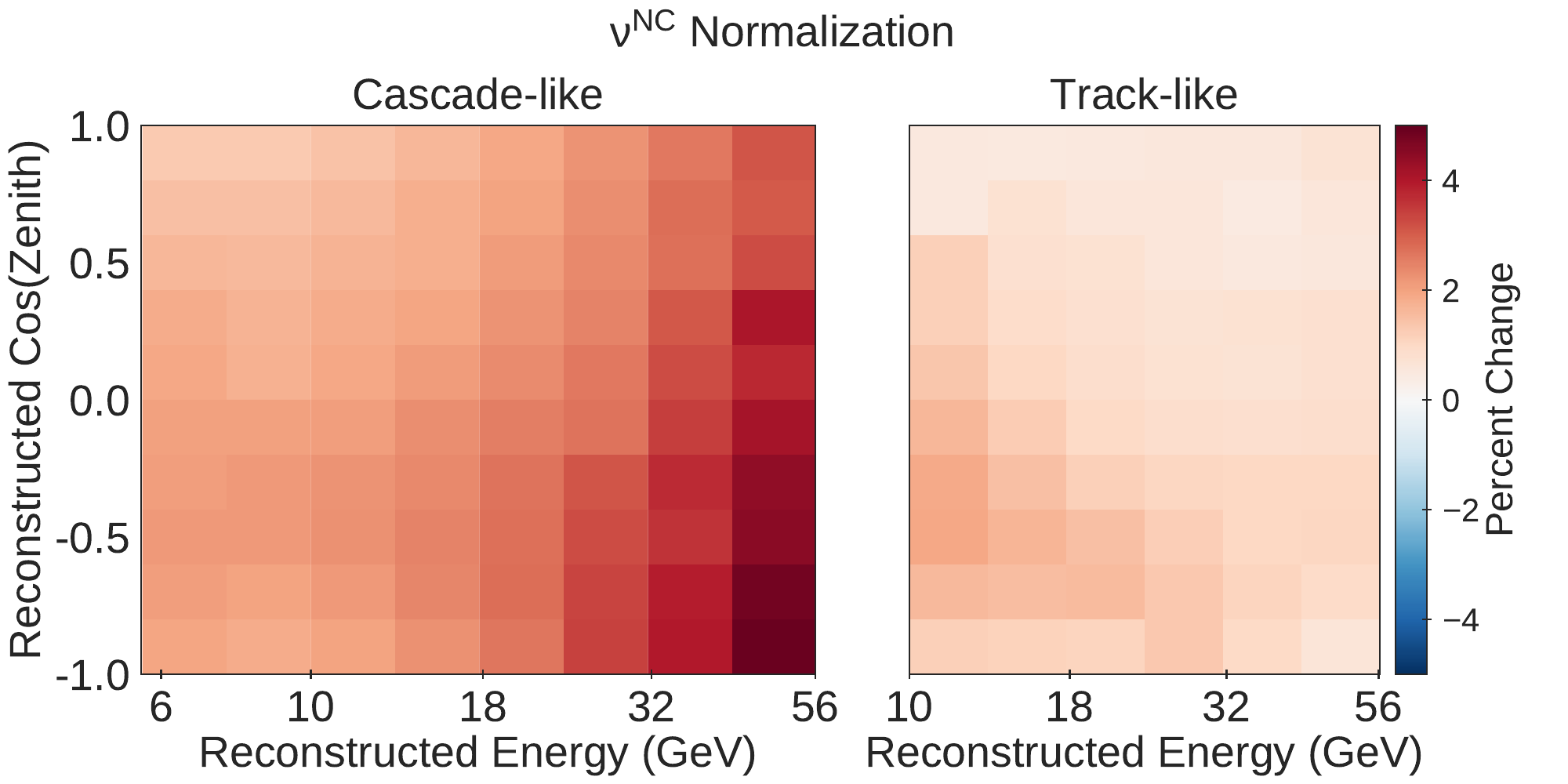}}
  \subfloat[$\theta_{23}$ shifted from 40$^\circ$ to 45$^\circ$]{
     \includegraphics[width=0.4\textheight]{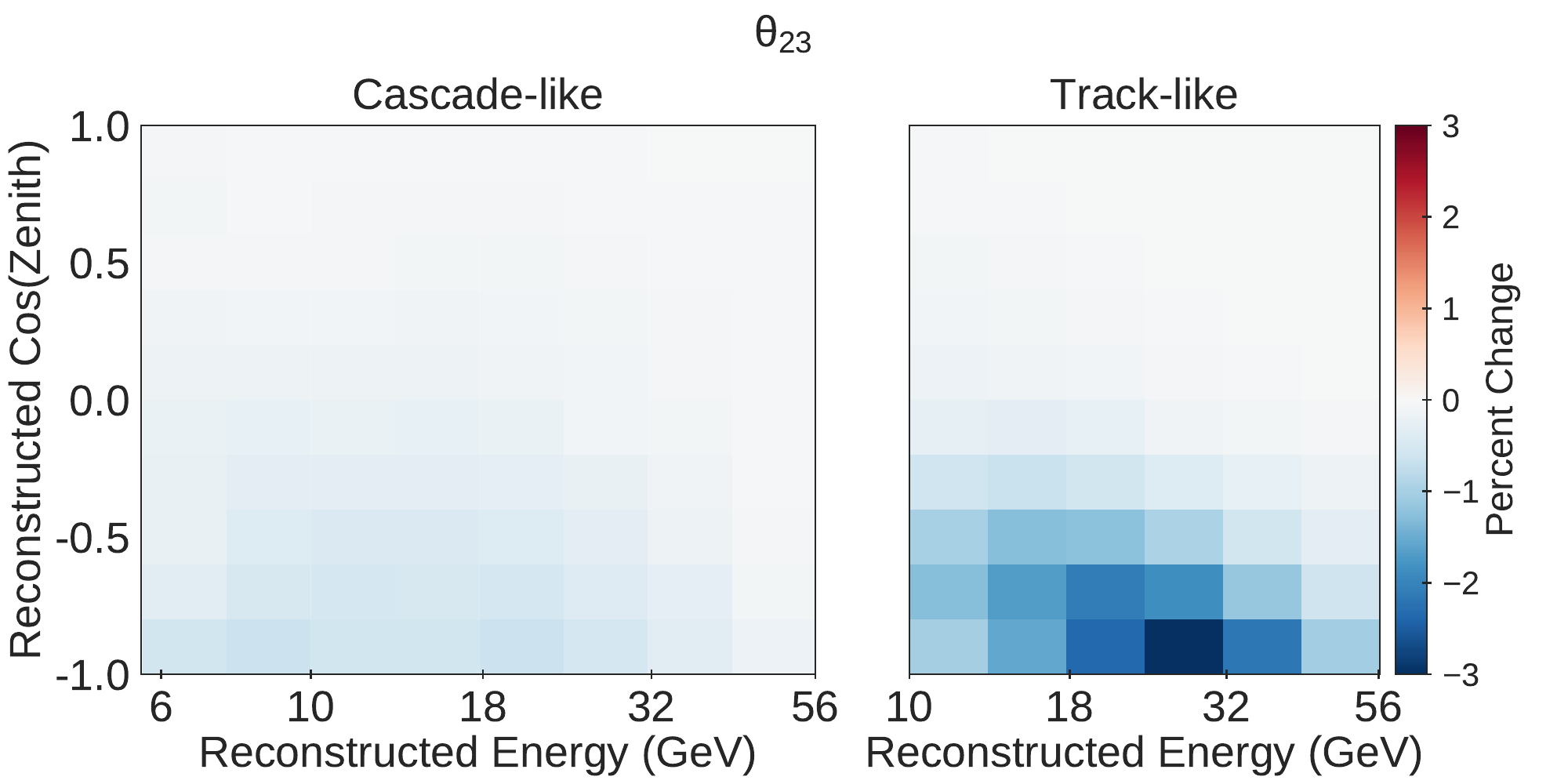}}

  \subfloat[$\Delta m^2_{32}$ shifted from 2.526 to 2.778 ($10^{-3}\rm{eV}^2$) ]{
     \includegraphics[width=0.4\textheight]{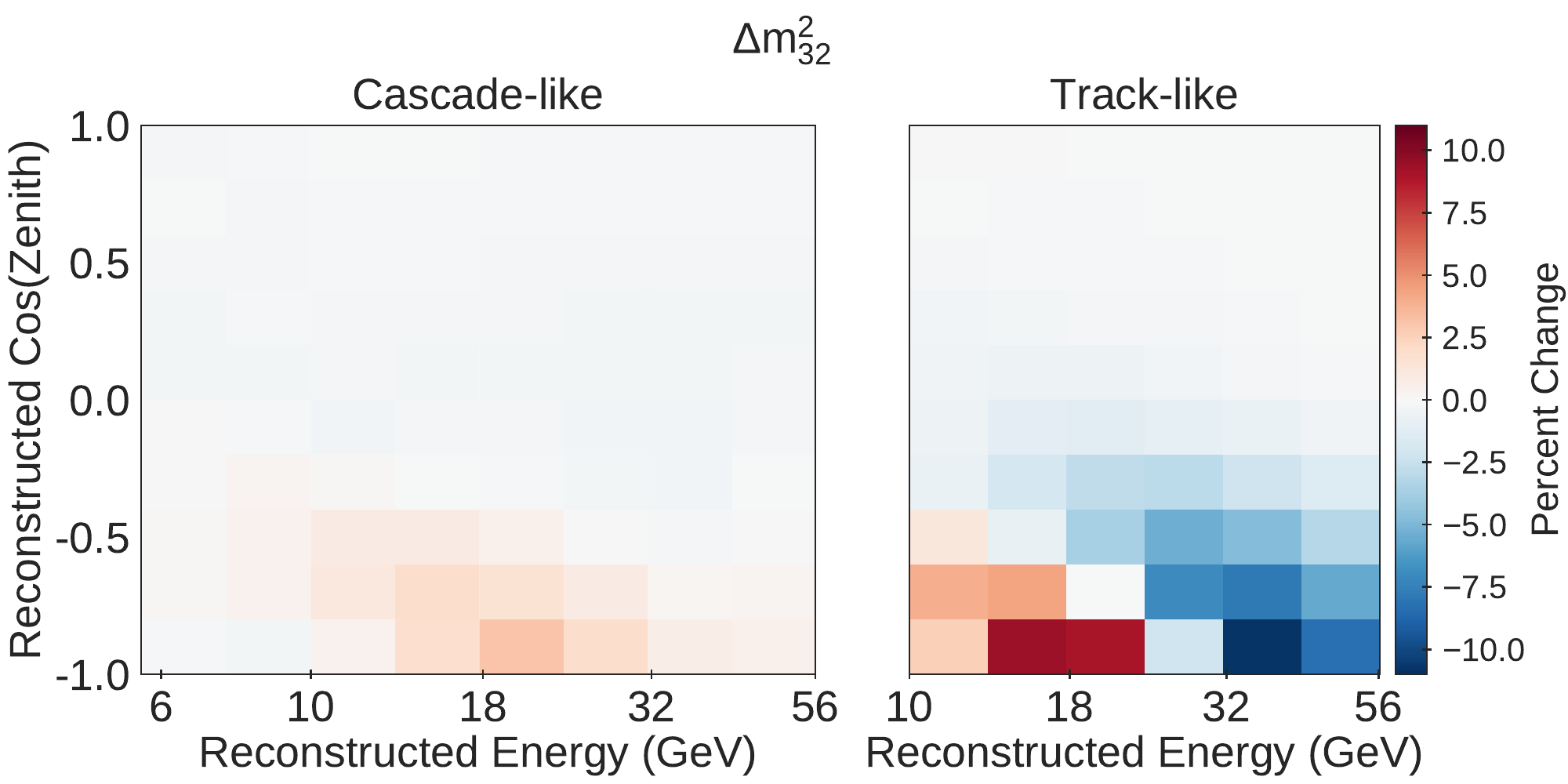}}
  \subfloat[Overall Optical Efficiency +10\%]{
     \includegraphics[width=0.4\textheight]{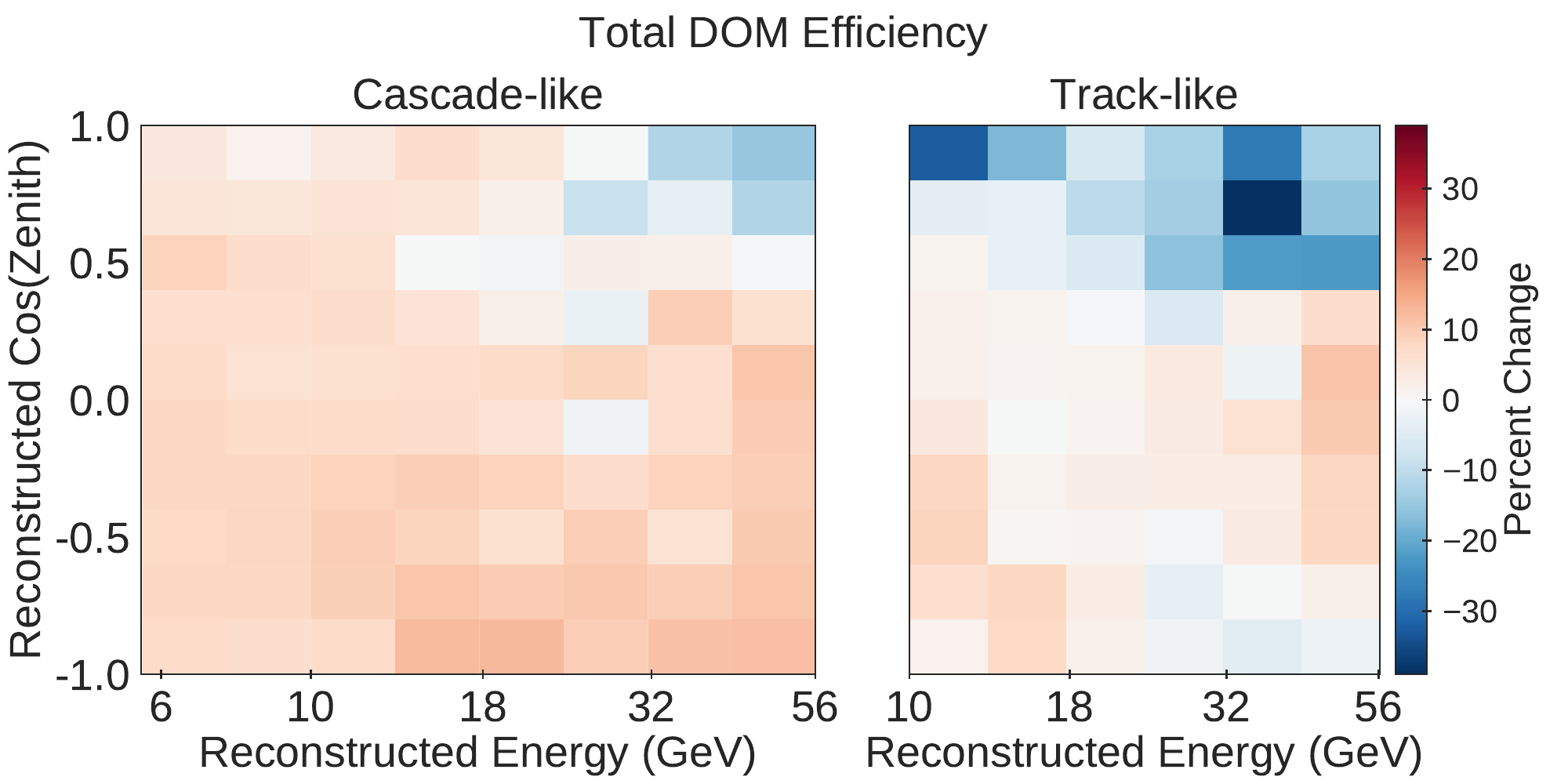}}

  \subfloat[Lateral Optical Efficiency +1$\sigma$]{
     \includegraphics[width=0.4\textheight]{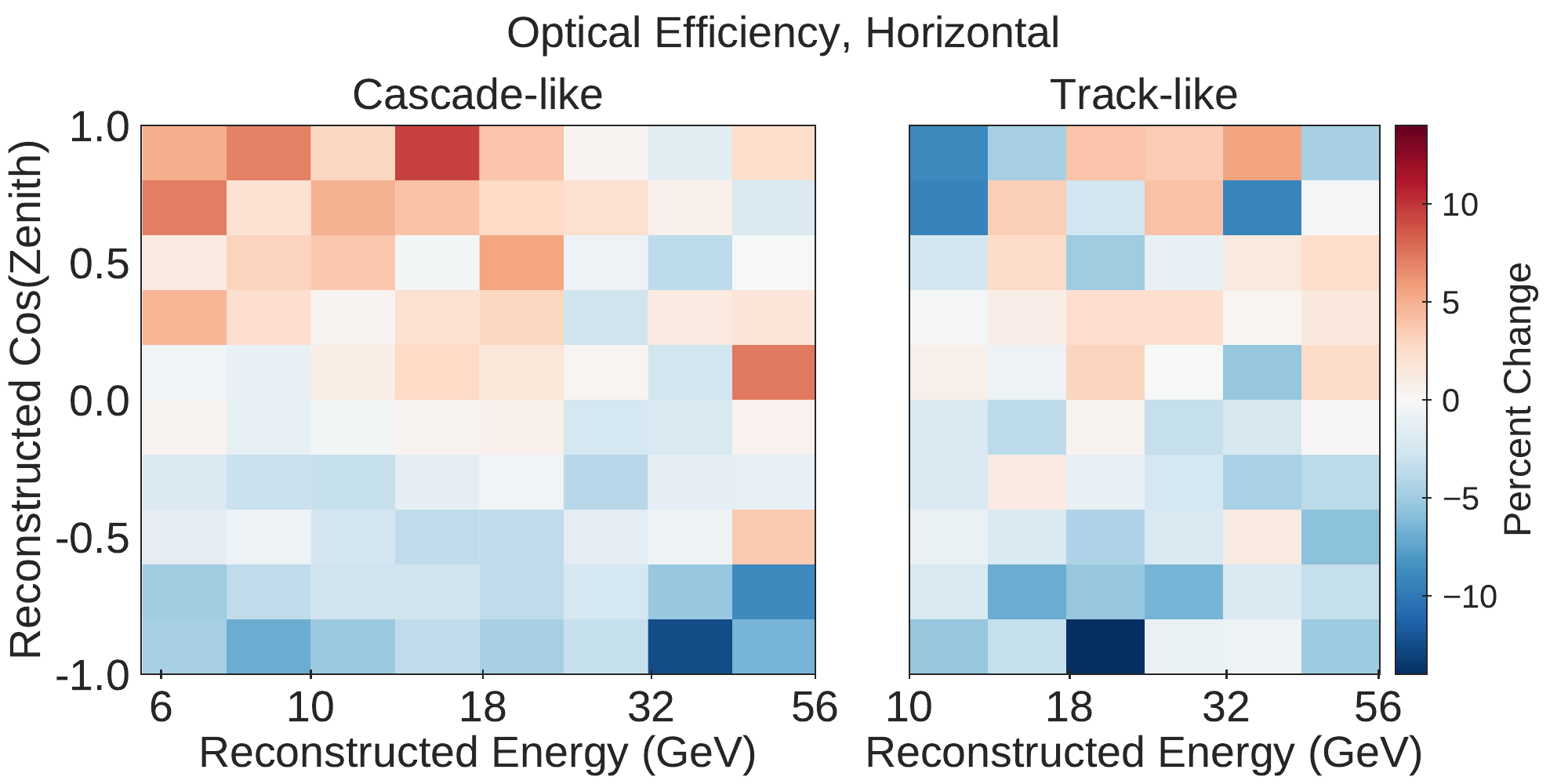}}
  \subfloat[Head-on Optical Efficiency +1 (a.u.)]{
     \includegraphics[width=0.4\textheight]{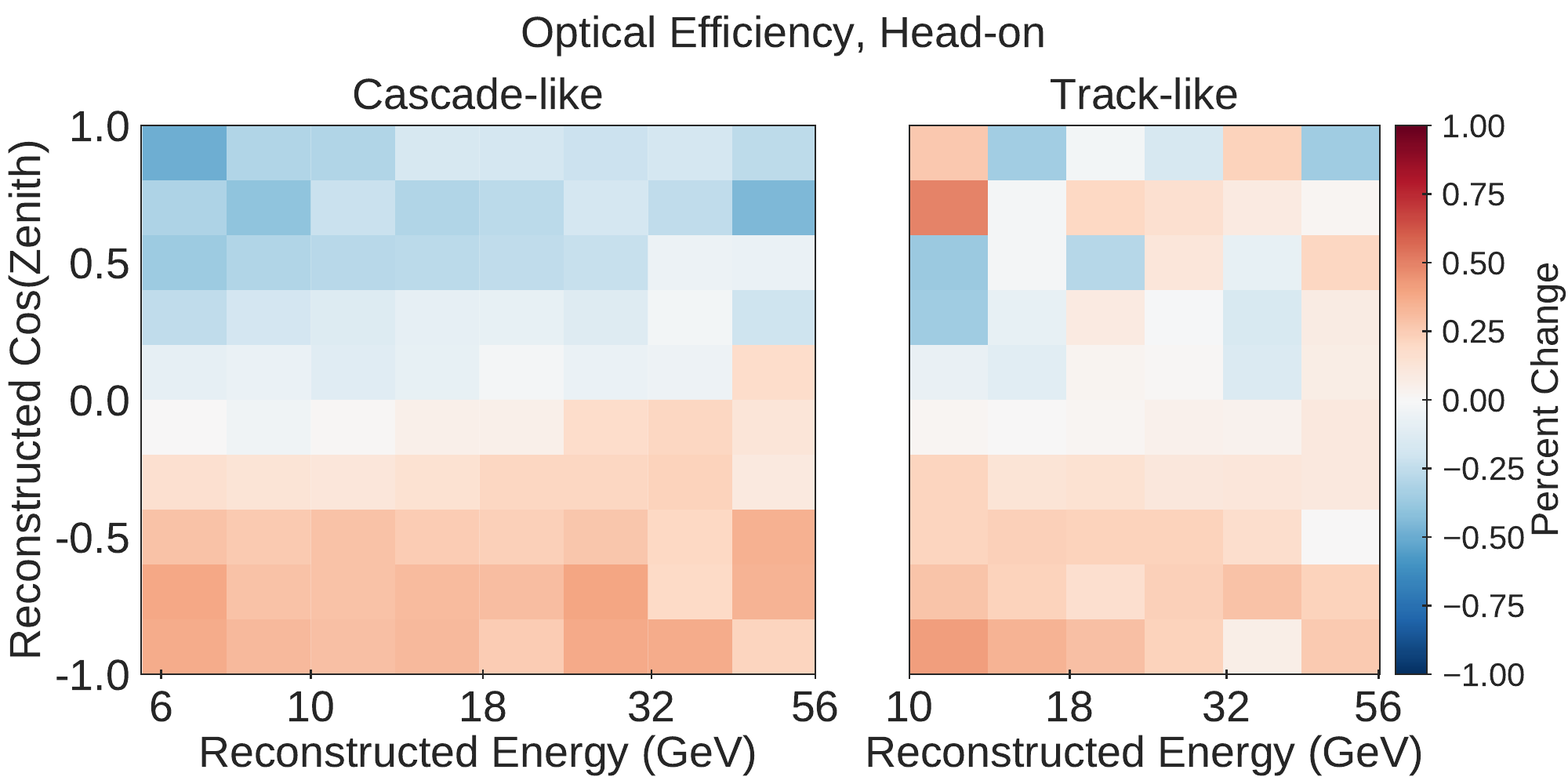}}
     \caption{Changes in the event rates from a subset of the systematic uncertainties included in the analyses.}
\end{sidewaysfigure}

\begin{sidewaysfigure}
  \centering
  \subfloat[Bulk Ice Scattering +10\%]{
     \includegraphics[width=0.4\textheight]{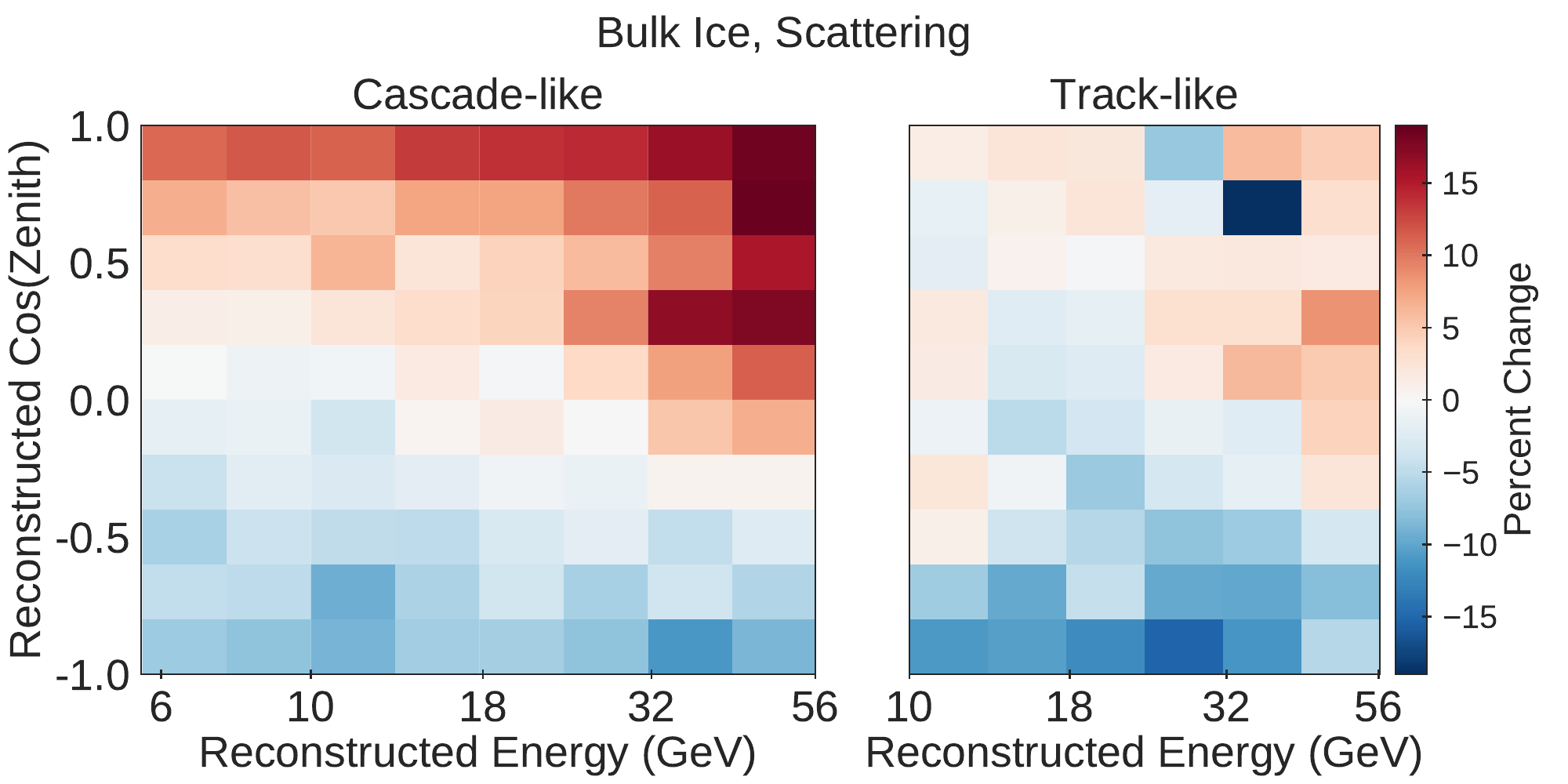}}
  \subfloat[Bulk Ice Absorption +10\%]{
     \includegraphics[width=0.4\textheight]{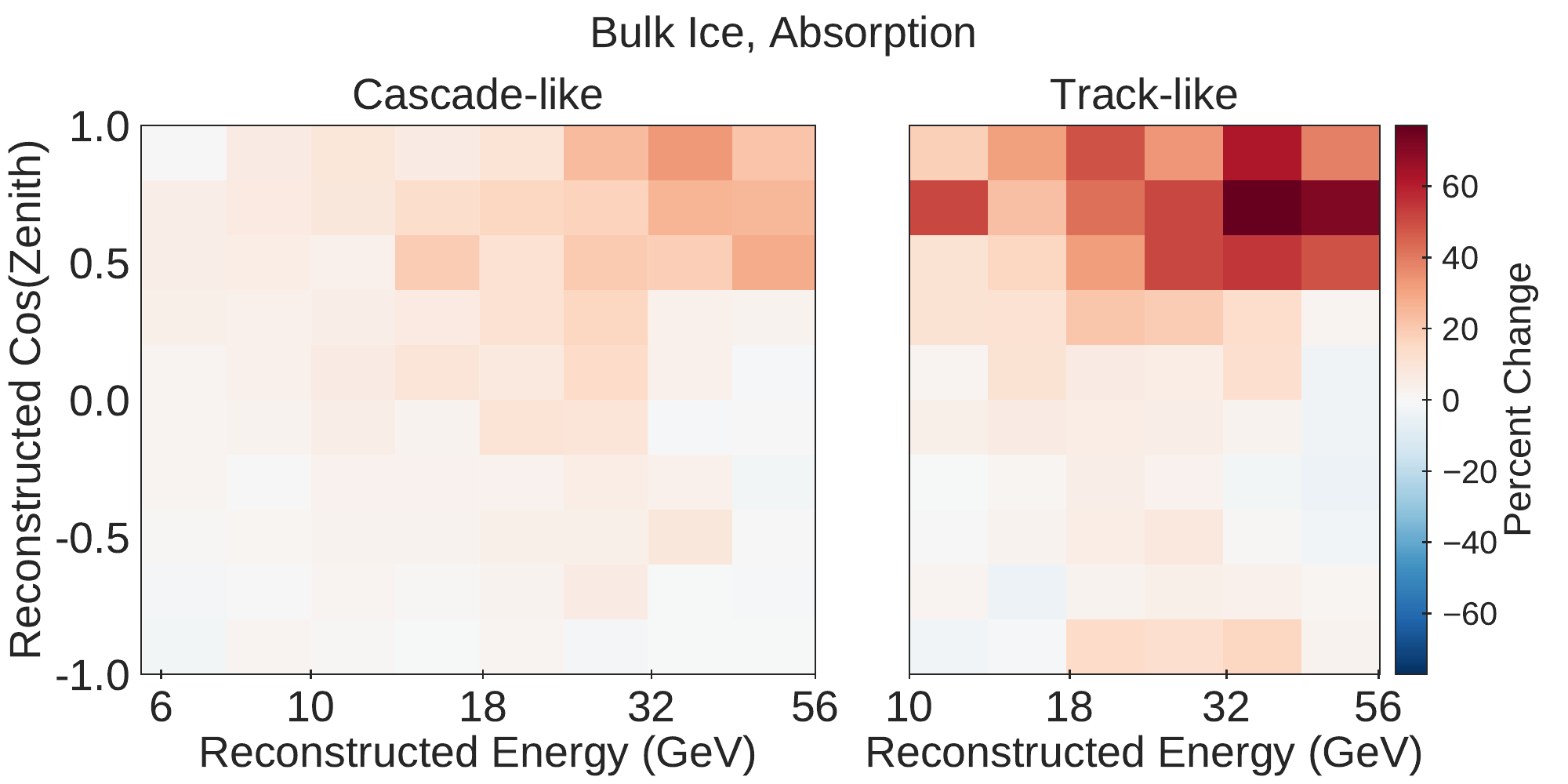}}

  \subfloat[Atmospheric Muon Fraction shifted from 8.1\% to 8.9\%]{
     \includegraphics[width=0.4\textheight]{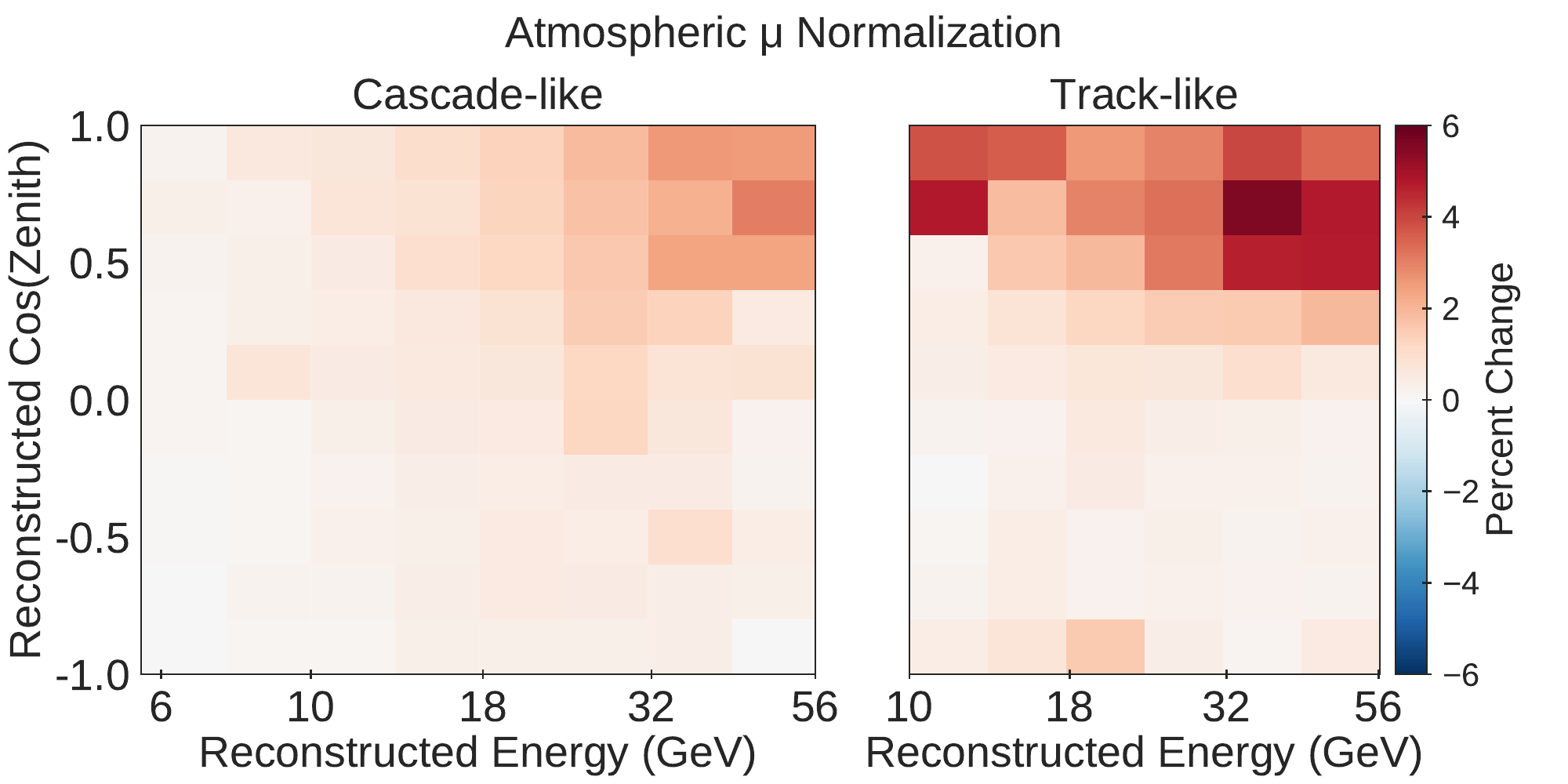}}
  \subfloat[Muon Spectral Index +1$\sigma$]{
     \includegraphics[width=0.4\textheight]{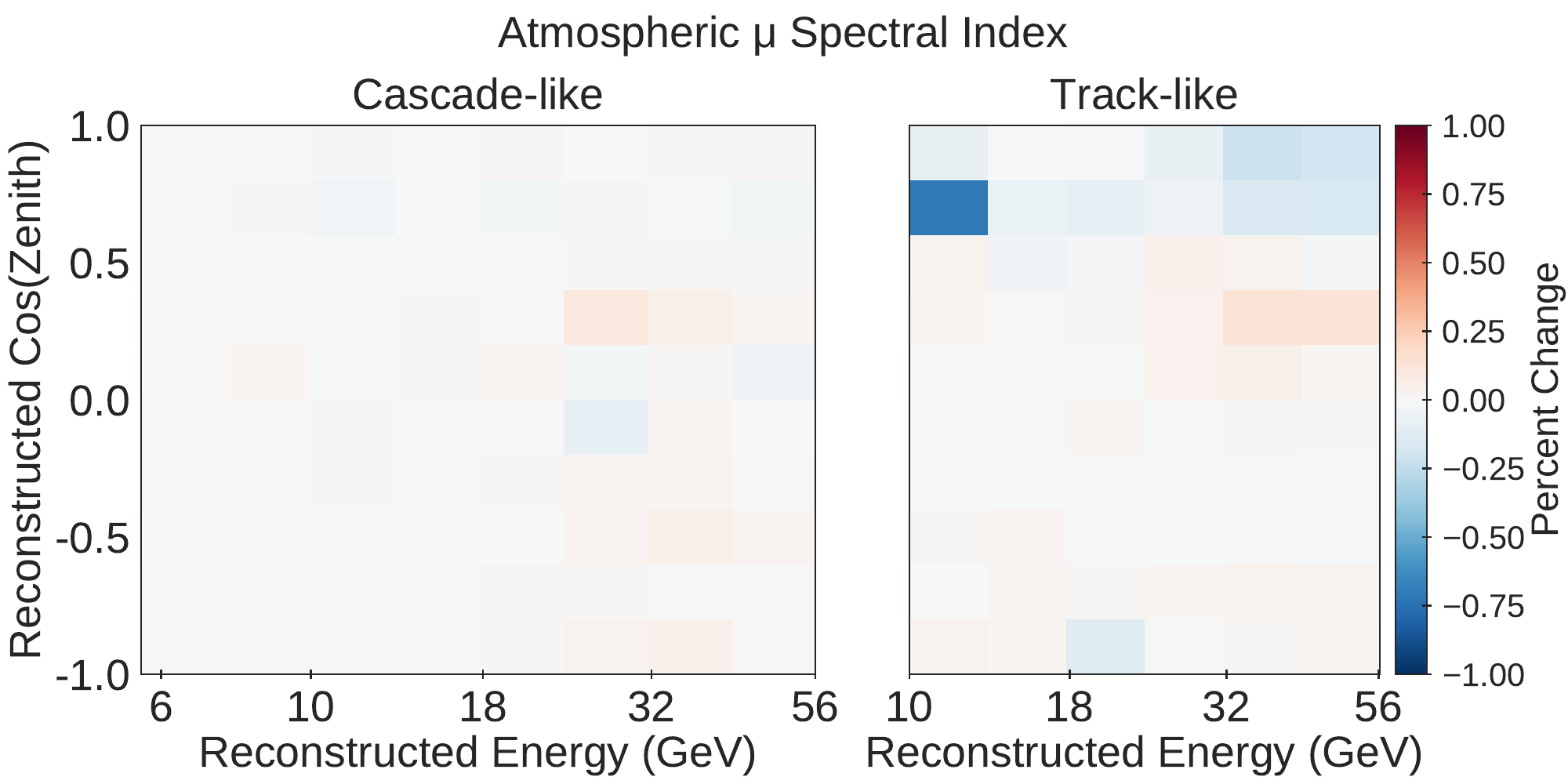}}

  \subfloat[Coincident $\mu$+$\nu$ Fraction +10\%]{
     \includegraphics[width=0.4\textheight]{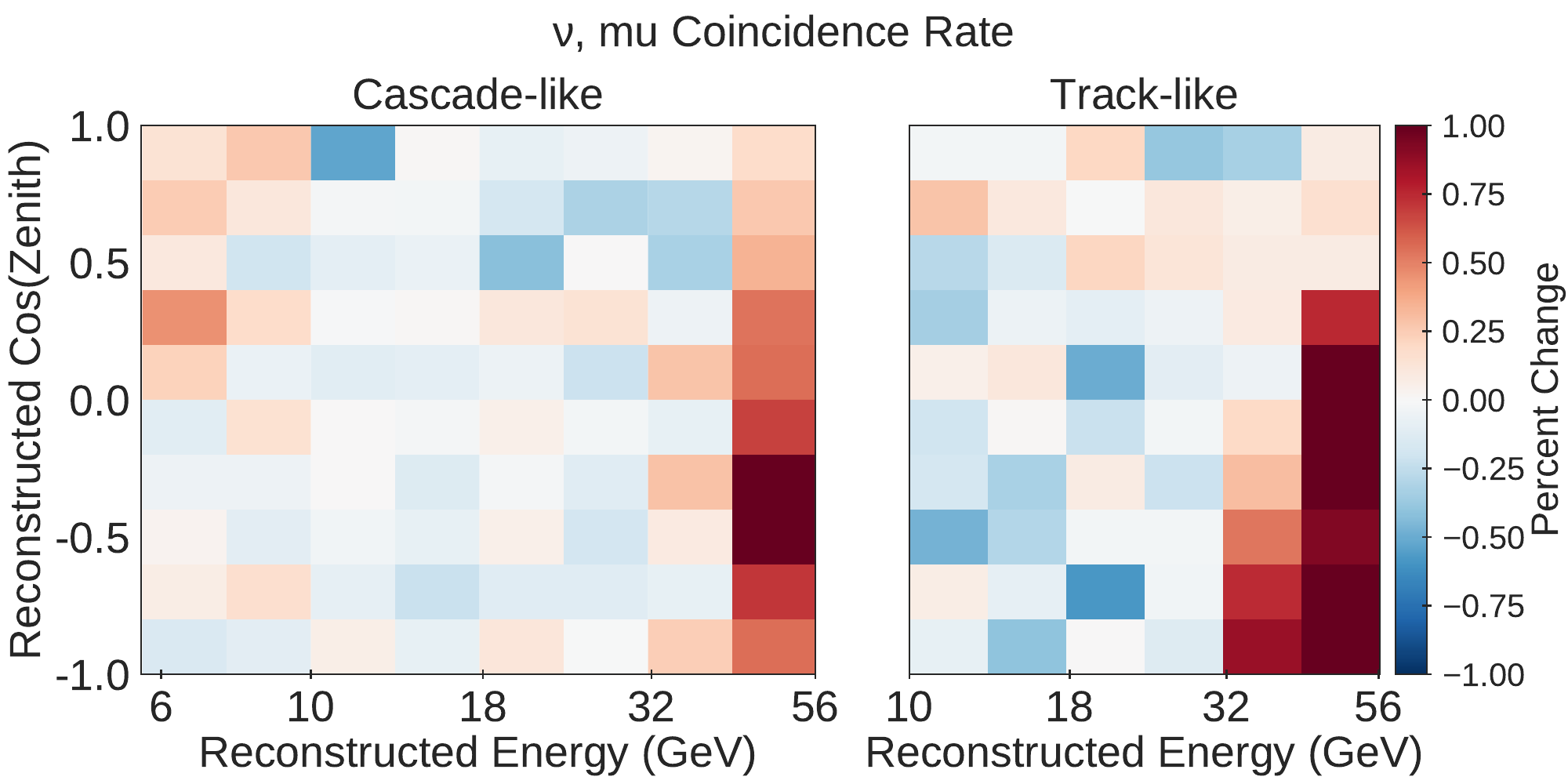}}
     \caption{Changes in the event rates from a subset of the systematic uncertainties included in the analyses.}
\end{sidewaysfigure}

\twocolumngrid

\end{document}